\documentclass{aa}  

\usepackage{graphicx}
\usepackage{amsmath,amssymb}
\usepackage{txfonts}
\usepackage{hyperref}
\usepackage{xcolor}

\begin{document}

   \title{Coupling hydrodynamics with comoving frame radiative transfer}
   \subtitle{III. The wind regime of early-type B hypergiants}

   \author{M. Bernini-Peron\inst{\ref{inst:ari}}
          \and 
          A.A.C.\ Sander\inst{\ref{inst:ari}}
          \and
          F.\ Najarro\inst{\ref{inst:cab}}
          \and
          G.N.\ Sabhahit\inst{\ref{inst:aop}}
          \and
          D.\ Pauli\inst{\ref{inst:kul}}
          \and
          R.R.\ Lefever\inst{\ref{inst:ari}}
          \and
          J.S.\ Vink\inst{\ref{inst:aop}}
          \and
          V. Ramachandran \inst{\ref{inst:ari}}
          \and
          L.M.\ Oskinova\inst{\ref{inst:up}}
          \and
          G.\ Gonz{\'a}lez-Tor{\`a}\inst{\ref{inst:ari}}
          \and
          E.C. Sch\"osser\inst{\ref{inst:ari}}
          }

   \institute{
           {Zentrum f{\"u}r Astronomie der Universit{\"a}t Heidelberg, Astronomisches Rechen-Institut, M{\"o}nchhofstr. 12-14, 69120 Heidelberg\label{inst:ari}\\
        \email{matheus.bernini@uni-heidelberg.de}}
    \and
           {Departamento de Astrof\'{\i}sica, Centro de Astrobiolog\'{\i}a, (CSIC-INTA), Ctra. Torrej\'on a Ajalvir, km 4,  28850 Torrej\'on de Ardoz, Madrid, Spain\label{inst:cab}}
     \and
           {Armagh Observatory and Planetarium, College Hill, Armagh BT61 9DG, N. Ireland\label{inst:aop}}
     \and 
           {Institute of Astronomy, KU Leuven, Celestijnenlaan 200D, 3001 Leuven, Belgium\label{inst:kul}}
           \and
{Institut f\"ur Physik und Astronomie, Universit\"at Potsdam, Karl-Liebknecht-Str. 24/25, 14476 Potsdam, Germany\label{inst:up}}
   }

   \date{Received February 13, 2025; accepted March 18, 2025}
 
  \abstract
    {B hypergiants (BHGs) are rare but important for our understanding of high-mass stellar evolution. While they occupy a similar parameter space as B supergiants (BSGs), some BHGs are known to be luminous blue variables (LBVs). Their spectral appearance with absorption and emission features shares similarities with the hotter Of/WNh stars. Yet, both their wind physics and their evolutionary connections are highly uncertain. 
}
    {
    In this study we aim to understand (i)  the stellar atmospheric and wind structure, (ii) the wind-launching and wind-driving mechanisms, and (iii) the spectrum formation of early-type BHGs. As an observational prototype, we use $\zeta^1$\,Sco (B1.5Ia+), which has a broad spectral coverage from the far-UV to the mid-IR regime.
    } 
    {
    Using the stellar atmosphere code PoWR$^\textsc{hd}$, we calculated the first hydrodynamically consistent atmosphere model in the BHG wind regime. These models inherently connect stellar and wind properties in a self-consistent way. They also provide insights into the radiative driving of the calculated wind regimes and enable us to study the influence of clumping and X-rays on the resulting wind properties and structure.
    }
    {
    Our hydrodynamically consistent atmosphere model nicely reproduces the main spectral features of $\zeta^1$\,Sco and represents a new framework of quantitative spectroscopy. The obtained mass-loss rate is higher than for BSGs of similar spectral types. However, despite the spectral morphology, the wind optical depth of BHG atmospheres is still considerably below unity, making them less of a transition type than the Of/WNh stars. To reproduce the spectrum, we need mild clumping with subsonic onset ($f_\infty = 0.66$, $\varv_\mathrm{cl} = 5$\,km\,s$^{-1}$). The wind shows a shallow-gradient velocity profile that deviates from the widely used $\beta$ law. Even beneath the critical point, the wind is mainly driven by \ion{Fe}{III} opacity.
    }
   {
    Our investigation suggests that despite more mass loss, early-type Galactic BHGs have winds that are relatively similar to late-type BSGs. Their winds are not sufficiently optically thick that we would characterize them as ``transition-type'' stars, unlike Of/WNh, implying that emission features arise more easily in cooler than in hotter stars. The spectral BHG appearance is likely connected to atmospheric inhomogeneities already arising beneath the sonic point. To reach a spectral appearance similar to known LBVs, BHGs need to be either closer to the Eddington limit or have higher wind clumping than inferred for $\zeta^1$\,Sco.
    }
    {}


   \maketitle

\section{Introduction}
\label{sec:intro}

B hypergiants (BHGs) are a rare type of evolved massive stars. To date, slightly fewer than 40 objects are known in the whole Local Group  (see Table\,\ref{tab:all_BHGs}). Farther away, B stars, with luminosities comparable to BHGs, are detected at high redshifts ($z \gtrsim 6.2$) via gravitational lensing \citep[][]{Welch+2022}.  However, only five BHGs have so far been properly analyzed with quantitative spectroscopy, and the physical mechanisms responsible for their stellar wind driving are not yet understood. 

In the Hertzsprung-Russell diagram, BHGs are located near B supergiants (BSGs). However, their usually high luminosities place them closer to luminous blue variables (LBVs). Spectroscopically, they differ from the ``typical'' BSGs by the presence of at least H$\beta$ in emission {\citep[pure emission or a P-Cygni profile; see, e.g.,][]{Lennon+1992}}. However, it is also common to observe many more H and He lines appearing as P-Cygni profiles \citep[e.g.,][]{Clark+2012,Walborn+2015,Mahy+2016}.


In recent years, several pioneering efforts have been made to model radiative-driven winds with hydrodynamically consistent atmosphere models (i.e., no longer relying on the use of ad hoc velocity fields like the $\beta$ law). These studies significantly advanced our understanding of Wolf-Rayet (WR) stars \citep{Sander+2020, Poniatowski+2021, Sander+2023}, O stars \citep{Krticka-Kubat2017, Sander+2017, Sundqvist+2019, Bjorklund+2021}, and early-B stars \citep{Sander+2018, Krticka+2021, Bjorklund+2023}. A few wind investigations were also performed for the later B-star regime \citep[e.g.,][]{Petrov+2016, Vink2018, Venero+2024}, albeit with more approximate wind dynamics. Yet, for BHGs and LBVs, which are closer to the Eddington limit and may play crucial roles in high-mass stellar evolution, there is a severe lack of dedicated investigations.
In this study we present the first hydrodynamically consistent atmosphere modeling of an early-type BHG\footnote{Given that the vast majority of BHGs have spectral types later than B1 (see Table\,\ref{tab:all_BHGs}), we consider B1 and B1.5 types as "early" in our context.}. 

Using PoWR$^\mathrm{HD}$ \citep[see][]{Sander+2017}, we can predict the mass-loss rate and velocity field directly from a given set of stellar parameters. To validate and benchmark our dynamically consistent model, we compared the obtained synthetic spectra with observed spectra of $\zeta^1$~Sco (B1.5Ia+) from the UV to the IR. 
$\zeta^1$\,Sco is currently the brightest BHG in the sky \citep[$V = 4.78$,][]{Zacharias+2005} and was chosen as a prototype as it is the Galactic BHG with the broadest spectral coverage. We further compared our results with previous work using ``traditional'' (i.e., non-hydrodynamically consistent) atmosphere models \citep[e.g.,][]{Crowther+2006, Clark+2012}.

In Sect.\,\ref{sec:atmos_model} we briefly describe PoWR and the assumptions adopted in our modeling. In Sect.\,\ref{sec:observ_data} we present and discuss the observational data and compare them to our model spectra in different wavelength regions. In Sect.\,\ref{sec:star-prop} we discuss the obtained stellar properties, and in Sect.\,\ref{sec:wind-prop} we discuss the wind properties, focusing on the dynamics, mass-loss rates, and wind clumping. In Sect.\,\ref{sec:trans} we discuss how BHGs compare to Of/WNh stars and how their winds are presumably transitioning to the ``optically thick wind regime.''
We present our conclusions in Sect.\,\ref{sec:conclu}.

\section{Hydrodynamically consistent atmosphere model}
\label{sec:atmos_model}

For our investigation, we made use of the state-of-the-art PoWR stellar atmosphere code PoWR \citep[e.g.,][]{Graefener+2002, Sander+2015}. More specifically, we used PoWR$^{\rm HD}$ branch \citep{Sander+2017,Sander+2018,Sander+2023}, which solves the momentum equation consistently with the radiation field, temperature stratification, and the ionization/excitation rate equations at each layer of the atmosphere/wind. During the model iteration the velocity/density field is updated so that the acceleration terms (from radiative and pressure forces) are equal to the deceleration terms (due to gravity and inertia) in each atmosphere layer (see \citealt{Sander+2017,Sander+2018, Sander+2023} for a detailed explanation of the implementation).

One key aspect of consistent stellar atmosphere codes is that the wind structure and terminal velocity $\varv_\infty$ is neither an ad hoc parametrization (as the traditionally used $\beta$ law) nor an input property. Rather, it is calculated directly from the fundamental properties and inner-boundary conditions of the model (e.g., radius, luminosity, and mass). Likewise, the mass-loss rates, $\dot{M}$ (or alternatively the mass, $M$) is also computed consistently. However, given the 1D and stationary nature of stellar atmosphere codes, physical effects such as clumping and X-ray emission still need to be described parametrically.

\paragraph{Clumping description.}
Both theory and observation reveal that the winds of hot stars are inhomogeneous.
The stellar atmospheres are expected to have regions of over-densities (clumping) and rarefaction in their outflows. Clumping affects the formation of lines in various direct and indirect ways. For instance, (i) the strength of recombination lines, such as Balmer lines, dependent on $\rho^2$, and (ii) scattering resonance lines, such as UV P-Cygni profiles, are affected by changes in the ionization structure caused by the alterations in the density structure.
We applied the exponential parametrization of the clumping structure introduced by \cite{Hillier+2003},
\begin{equation}
  \label{eq:volfill}
    f(r) = f_\infty + (1 - f_\infty) \exp\left(-\frac{\varv(r)}{\varv_\mathrm{cl}}\right)
,\end{equation}
for the volume filling factor $f$, which is the inverse of the clumping factor ($f_\mathrm{cl} = 1/f$) if clumps are assumed to be optically thin and the inter-clump medium is void. The same has previously be used for BHGs in \cite{Clark+2012}. In this description, the clumping profile depends only on two parameters $f_\infty$ and $\varv_\mathrm{cl}$, where the former controls the degree of inhomogeneity at an ``infinite'' distance, and the latter is the onset characteristic velocity, which controls how abruptly clumping will start growing in the wind, essentially working as a scale height.

\paragraph{X-rays.}
As discussed in \cite{Bernini-Peron+2023, Bernini-Peron+2024}, the inclusion of X-rays in cool BSG models ($T_\mathrm{eff} \sim 21$\,kK) is necessary to produce the so-called superionization \citep[e.g.,][]{Cassinelli+1981, Odegard-Cassinelli1982, Krticka-Kubat2016}, characterized by the presence of \ion{N}{V}, \ion{C}{IV}, and \ion{Si}{IV} lines in the UV with well-developed P-Cygni profiles, which are not expected to appear at such low effective temperatures. Conversely, the extra source of ionization is also necessary to quench the profiles of \ion{C}{II} and \ion{Al}{III}, which would be saturated otherwise. The neglect of X-rays would likely yield to an underestimation of the mass-loss rate, as the saturation of \ion{C}{II} and \ion{Al}{III} would then have to be done by assuming thinner winds.
For BHGs, this is not an option, as they also show well-developed \ion{N}{V}, \ion{C}{IV}, and \ion{Si}{IV} profiles, indicating that superionization needs to happen in their atmospheres despite their comparably slow winds ($\varv_\infty < 500$\,km\,s$^{-1}$). Hence, we included X-rays in our models. 

PoWR implements X-rays as free-free emission (bremsstrahlung), following the formalism of \cite{Baum+1992}.
While not including specific line contributions, this formalism is sufficient to provide the necessary radiation field at high frequencies to produce superionization, via either the Auger process or direct ionization due to a hot thermal component. Isolated BHGs have not yet been detected in X-rays. Thus, superionization provides the only means to constrain the X-rays in the winds of these stars and investigate their effect on ionization structure and wind driving.

\paragraph{Atomic data.}
For models without hydrodynamic consistency, usually, only the ions that produce measurable signatures and/or sufficient blanketing are included in atmosphere calculations. In the context of B stars, these are H, He, C, N, O, Mg, Al, Si, P, S, and Fe. Following \cite{Sander+2018}, we also explored elements that could possibly contribute to the wind driving. Besides the previously mentioned elements, we additionally included Na, Ne, Cl, K, Ca, and Ar.  However, our test calculations revealed that including these elements did not significantly alter the derived wind properties nor the output spectrum.

In PoWR, Fe is treated conflated with Sc, Ti, V, Cr, Mn, Co, and Ni \citep[see][for details]{Graefener+2002}. Thus, the abundance of the ``generalized iron'' is given by the sum of the listed iron-group elements. As Fe is by far the most abundant among them \citep[see, e.g.,][]{Asplund+2009}, and we do not aim for precise Fe abundance determination in this study, we refer to the ``generalized iron'' simply as Fe in the rest of this work.
The list of elements, ionization levels, and the respective number of transitions are shown in Table~\ref{tab:atomic}.

\section{Observational data}
\label{sec:observ_data}

We compared our synthetic spectra with archival observed data of the prototypical star $\zeta^1$\,Sco in the UV, optical, and IR. The UV observations were obtained using archival data from the Far Ultraviolet Spectroscopic Explorer (FUSE), whose observation PI is T.P.\,Snow (program ID: P116), and from the International Ultraviolet Explorer (IUE). In the optical, we chose to compare with the spectrum obtained with the Echelle SPectrograph for Rocky Exoplanets and Stable Spectroscopic Observations (ESPRESSO) by the PI A.\,de\,Cia (program ID: 0102.C-0699) due to its higher resolution among the available spectra. The near-IR spectrum was obtained with X-Shooter at the Very Large Telescope (VLT). We also used spectrum from the Infrared Space Observatory (ISO) obtained via the ISO SWS Atlas \citep[][]{Sloan+2003} in the mid- and far-IR. 
$\zeta^1$\,Sco presents some line variability across the whole spectrum, which we show in Appendix \ref{app:specvar}.
As the available observed spectra have an absolute flux incompatible with the retrieved magnitudes, we normalized the optical and near-IR spectra by dividing the observed spectra by a curve obtained from linearly interpolating individual points in the continuum.
The UV range was normalized using our elected best model's synthetic continuum. 

The flux-calibrated FUSE, IUE, and ISO spectra were used to build the spectral energy distribution (SED) for constraining the luminosity. Given the available multi-epoch IUE data, we used an averaged spectrum for the SED. Additionally, we also used the magnitudes from the optical to the far-IR: The Johnson-Cousins magnitudes are retrieved from \cite{Paunzen+2022} and \cite{Zacharias+2005}. The \textit{Gaia} magnitudes used are those listed in \textit{Gaia} Data Release 3 \citep{GaiaCollab+2016,GaiaCollab+2023}. We adopted the distance from \cite{BailerJones+2021}. The employed IR magnitudes are the JHK bands from the Two Micron All-Sky Survey (2MASS) and bands from the Wide-field Infrared Explorer (WISE) \citep[both sets retrieved from][]{Cutri+2013}. In Table\,\ref{tab:phot} we list the nominal values of all employed magnitudes. 

\section{Spectroscopic validation of the model}
\label{sec:spec_valid}

To create a dynamically consistent atmosphere model that reasonably reproduces the spectrum of $\zeta^1$\,Sco,
we started with the derived parameters from \citet{Clark+2012}. 
We then varied the different stellar properties, initially the luminosity ($L$), the inner-boundary temperature\footnote{The inner boundary is defined in terms of the Rosseland-mean optical depth $\tau$, which in our calculations was set to $\tau = 50$. Hence, $T_\ast =  T_\mathrm{eff}(\tau = 50)$.} ($T_\ast$), and the stellar mass ($M$). After finding the best suitable parameter set, we then adjusted the helium mass fraction ($X_\mathrm{He}$), microturbulent velocity ($\varv_\mathrm{turb}$), clumping, and X-rays. When necessary, the initial parameter set was also readjusted. As we do not aim at a precise abundance determination, we initially adopted the values of \cite{Clark+2012} for the He and CNO content. We tested adjustments within their respective error bars, but this did not lead to significant improvements in the spectral reproduction. The elemental abundances of all other metals are fixed to the values of \cite{Asplund+2009}. 

By design, hydrodynamically consistent models have a lower number of free parameters and it is not possible to tweak the stellar and wind parameters separately to improve the spectral fit. Instead, as we discuss in more detail in a related work on very massive stars \citep{Sabhahit+2025}, changes in one input parameter can have multiple affects on the spectrum. The main purpose of the current work is therefore not to reproduce every diagnostic in the observed spectrum and fine-tune the stellar parameters, but capture the overall spectral appearance of a BHG like $\zeta^1$\,Sco and test if our physical assumptions in this regime are sufficient to describe the wind of an early-type BHG. A more detailed description of our spectral analysis procedure and the employed diagnostics is given in Appendix \ref{app:diagnostics}.

\subsection{Ultraviolet spectrum}
\label{sec:uvspec}

\begin{figure}
\centering
\includegraphics[width=1.0\linewidth]{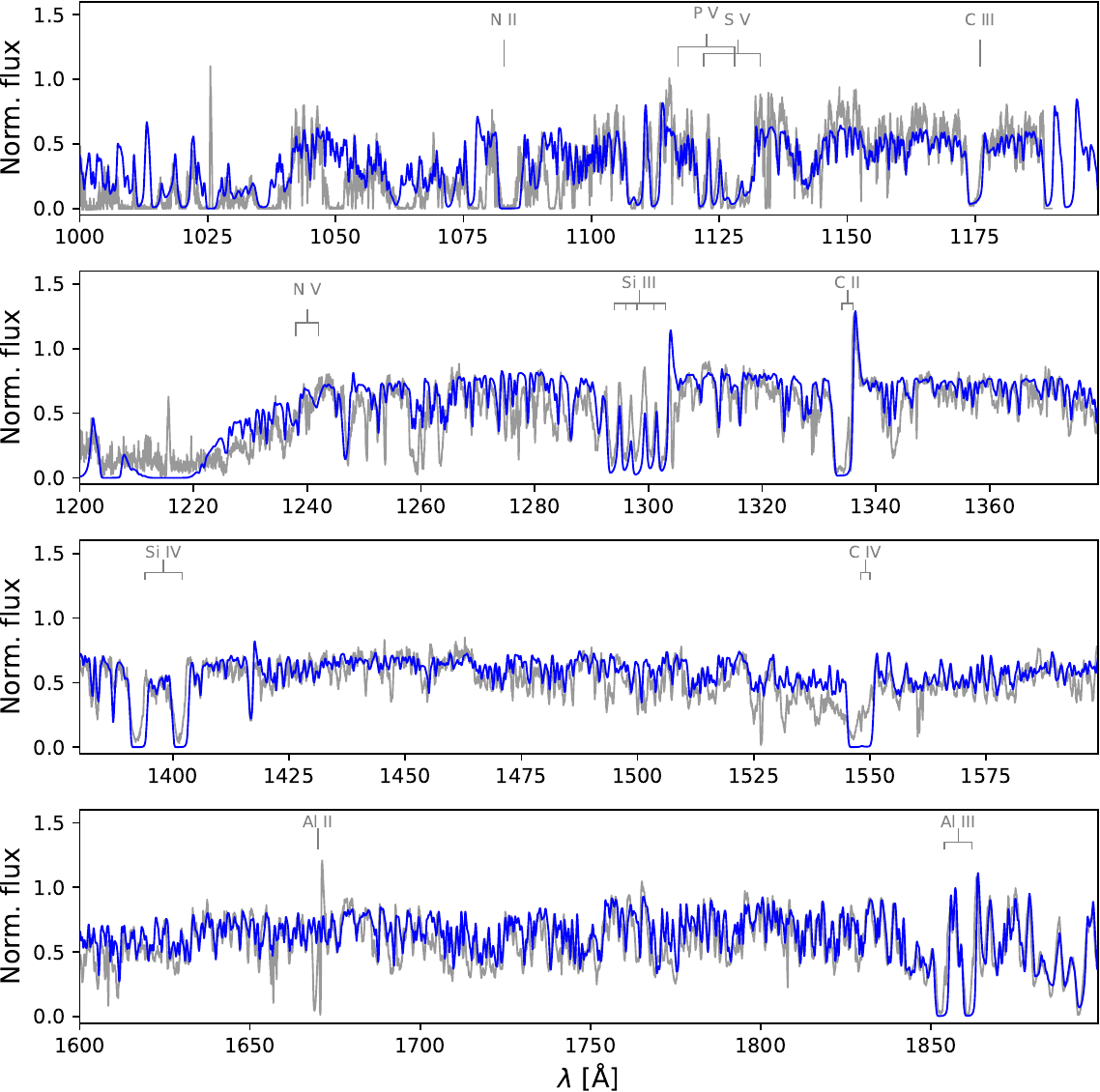}
  \caption{
  Comparison of our final hydrodynamically consistent model (blue line) with the far-UV spectra (FUSE and IUE; gray lines) of $\zeta^1$\,Sco. 
  }
    \label{fig:Z1Sco-UV-fit}
\end{figure}

The overall UV spectrum of $\zeta^1$\,Sco remains unchanged over time. We inspected IUE data from 1980 to 1995 and no significant changes in most of P Cygni profiles were observed. Small changes could be seen in the saturation level of $\ion{Si}{IV}$~$\lambda1400$ and $\ion{Al}{III}$~$\lambda1670$. This points to a quite stationary global outflow, at least closer to the terminal velocity regime. However, we find that \ion{N}{V}~$\lambda1240$, which in B star models requires X-rays, changes its shape and strength meaningfully. In Fig.\,\ref{fig:Z1Sco-UV-fit} we show the comparison between our final PoWR$^\textsc{hd}$ model and observations from FUSE and IUE (average spectrum).

To obtain a spectrum that matches the absorption troughs of the UV P Cygni lines, we applied a maximum wind microturbulence of $\xi_\mathrm{max} = 180$\,km\,s$^-1$ in the computation of the observer's frame spectrum. Otherwise, the obtained terminal velocity $\varv_\infty$ from the PoWR$^\textsc{hd}$ solution would not produce wide enough P-Cygni profiles. We discuss this aspect further in Sect. \ref{sec:wind-prop}.

Despite the general similarities between the UV spectra of BSGs and BHGs (e.g., with respect to the presence of superionization), there are also noticeable differences:
A prominent example is the prominent presence of \ion{Al}{II}~$\lambda$1670 and \ion{Mg}{II}~$\lambda2800$ in the few available UV spectra of B1Ia+ and B1.5Ia+ hypergiants\footnote{In the Milky Way, only $\zeta^1$\,Sco, HD190603, and P\,Cyg have available high-quality UV spectrum.}, which in BSGs appears only for B8 and later spectral types \citep[e.g.,][]{Walborn+1987}. This is likely a consequence of the higher wind densities in BHGs. However, we were not able to reproduce these low ionization features.

\begin{figure}
\centering
\includegraphics[width=1.0\linewidth]{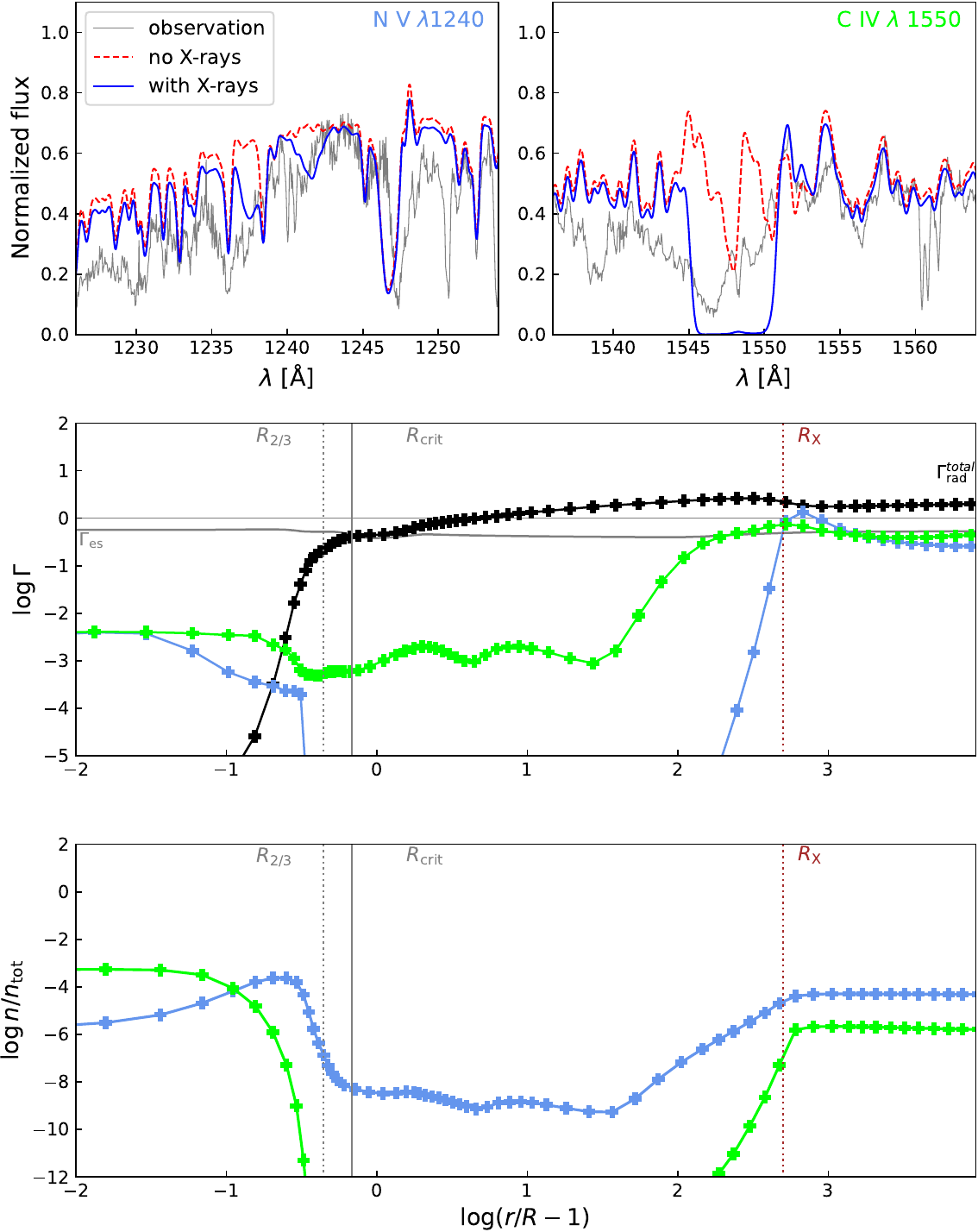}
  \caption{
  \textit{Upper panels:} Comparison of models' spectra with and without X-rays (dashed red and solid blue curves, respectively). \textit{Middle panel}: Contribution of \ion{Fe}{III}, \ion{C}{IV}, and \ion{N}{V} for the wind acceleration (black, green, and blue lines, respectively). The lines with thick (thin) crosses indicate the model with (without) X-rays. \textit{Lower panel}: Population number of \ion{N}{V} and \ion{C}{IV} following the same color and symbol code as the middle panel. In the middle and lower panels, the dashed blue line indicates the photosphere ($R_{2/3}$) and the solid gray line indicates the critical radius ($R_\mathrm{crit}$). The onset of X-rays ($R_\mathrm{X}$) is represented by the dashed brown line. 
  }
    \label{fig:xR-on-off}
\end{figure}

To reproduce the observed superionization P-Cygni profiles in the UV, we assumed a ``hot plasma component,'' that is, X-rays with $T_\mathrm{X}=0.35$\,MK and an onset radius of $R_\mathrm{X} = 500$\,$\mathrm{R}_*$. This means that X-rays only affect the outermost wind layers, where the velocity is similar to $\varv_\infty$. Our parameters mark the best compromise between reproducing \ion{C}{IV}\,$\lambda$1550 and \ion{N}{V}\,$\lambda$1240. Other lines were not meaningfully affected. Our final model has $L_\mathrm{X}/L = -8.5$, in line with the upper limit from \cite{Berghoefer+1997} of $L_\mathrm{X}/L < -7.4$ (0.2 to 2.4\,keV).
In Fig.\,\ref{fig:xR-on-off} we show the impact of the X-ray emission on the model spectrum. As evident, the inclusion of X-rays made \ion{N}{V} and \ion{C}{IV} appear as P-Cygni profiles, a direct consequence of the change in the population numbers. In contrast to BSGs \citep[see][]{Bernini-Peron+2023, Bernini-Peron+2024}, the profiles of \ion{C}{II}, \ion{Si}{IV}, and \ion{Al}{III} in BHGs seem barely affected by X-rays. In other wavelength regimes (optical and IR), the inclusion of X-rays has no meaningful impact.

The X-ray parametrization of our model is very hard to explain under the assumption of shock-induced X-rays. $\zeta^1$\,Sco has a slow wind, which would not have enough kinetic energy to produce X-rays via high-velocity collisions in the plasma \citep{Cohen+2014}. Moreover, the variability observed in \ion{C}{IV}\,$\lambda$1550 and \ion{N}{V}\,$\lambda$1240, more intense than in the other lines, would imply a very drastic variation in the X-ray emission over a very extended wind region. Hence, it is possible that the additional ionization source parametrized as X-ray emission simply mimics other physical conditions.

A plausible alternative to explain the superionization is the presence of a ``hot component'' in the wind, which would not necessarily have X-ray temperatures. Recent multidimensional simulations of O and WR stars further predict a distribution of velocities and temperatures for the wind material \citep[see][]{Moens+2022, Debnath+2024}. Although not tested yet, one could speculate that the observed superionization is the manifestation of the hotter part of such a distribution. This might then also explain the different behavior and morphology of \ion{C}{IV}\,$\lambda$1550 and \ion{N}{V}\,$\lambda$1240.

\subsection{Optical}

The optical region of BHG spectra is characterized by a rich set of absorption lines, P-Cygni lines, and emission lines. Thereby, it encodes important information about the photosphere as well as the inner layers of the stellar wind \citep{Clark+2012}. However, as the Balmer absorption lines are blended with wind P-Cygni features, the determination of surface gravities ($\log g$) via the evaluation of pressure-broadened wings is challenging or in some cases simply impossible. Therefore, one has to rely on higher-order series members, such as H$\epsilon$ and H$\zeta$ -- whose widths are slightly underestimated by our models. Additionally, the combined imprint of photospheric and wind features marks a particular challenge in the context of dynamically consistent modeling. The spectral appearance is very sensitive to changes in the stellar properties, which here do not only affect the photospheric features but also the wind parameters and resulting spectral imprints (see also Appendix \ref{app:diagnostics}). 

\begin{figure}
\centering
\includegraphics[width=1.0\linewidth]{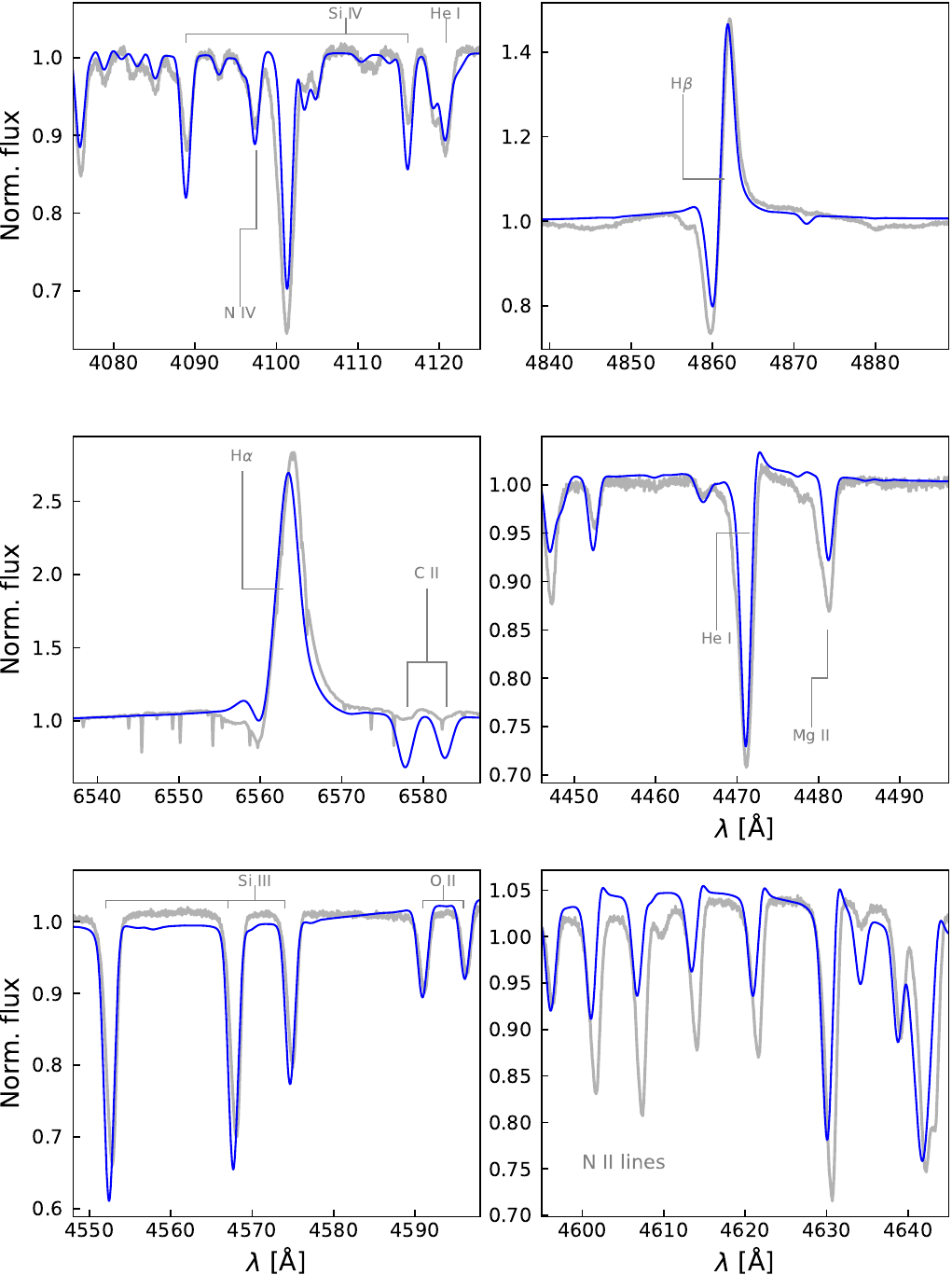}
  \caption{
    Comparison of our final PoWR$^\textsc{hd}$ model (blue line) with the observed optical spectrum (ESPRESSO) of $\zeta^1$\,Sco (gray line).
  }
    \label{fig:optical-fitting}
\end{figure}

In Fig.\,\ref{fig:optical-fitting} we show a comparison between the observed and synthetic spectrum, focusing on specific diagnostic lines in the optical regime. Given the abovementioned sensitivity, the calculated model atmosphere does an excellent job in reproducing the overall spectral features that are characteristic of $\zeta^1$\,Sco and other early BHGs, namely the Balmer lines and some \ion{He}{i} lines in P Cygni. The model further reproduces also the bulk of the metal lines across the optical range well.
In Fig.\,\ref{fig:opt-var} we compare our model with all the available optical spectra from the European Southern Observatory (ESO) archive. It is noticeable that $\zeta^1$\,Sco has a lot of line variability, associated with changes in the innermost layers of the wind. While we cannot account for variability explicitly in a stationary approach, our model captures the overall line behavior within the observed range.

\subsection{Infrared}

\begin{figure}
\includegraphics[width=1.0\linewidth]{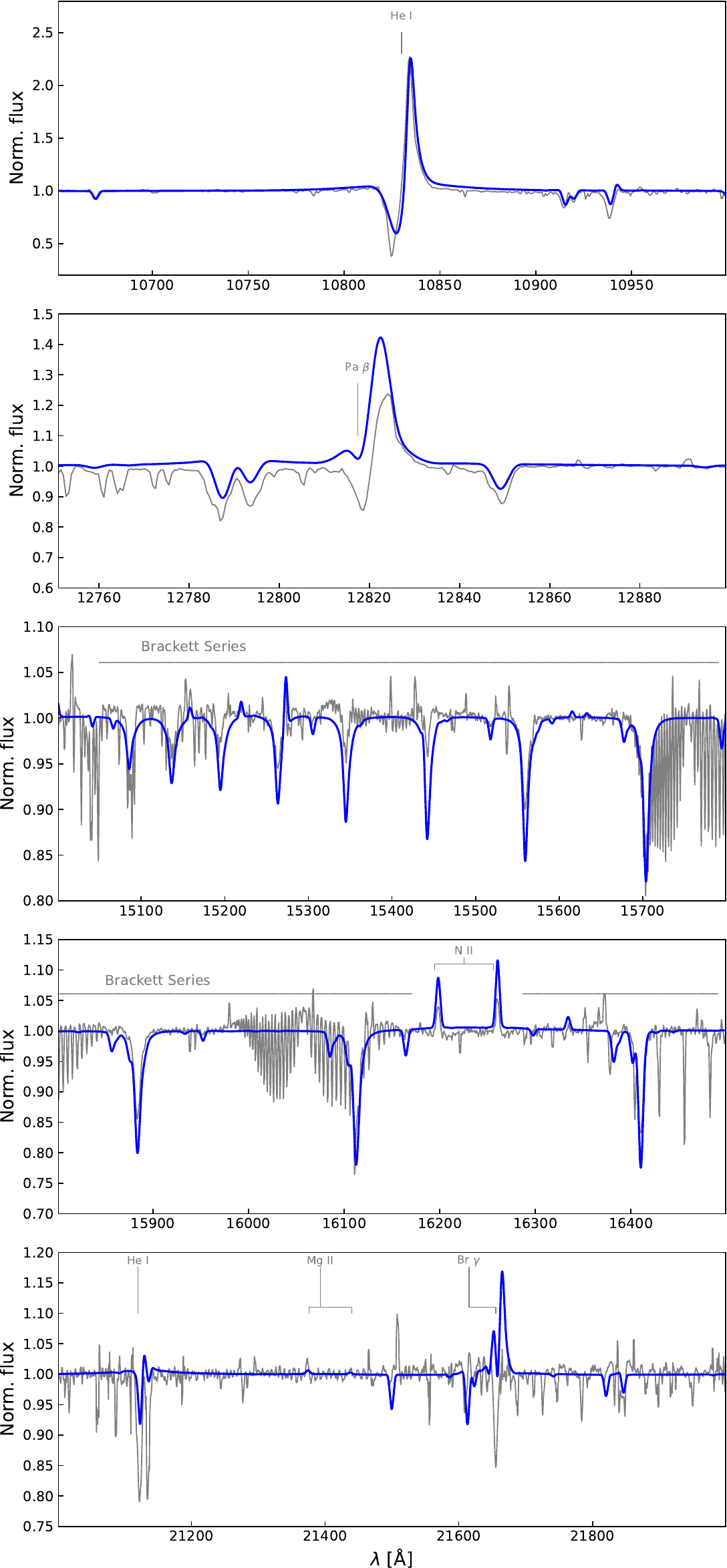}
  \caption{
    Comparison of our best-fit model (blue line) with the near-IR spectral features (X-Shooter; gray lines).}
    \label{fig:Z1Sco-nIR-fit}
\end{figure}

The IR and lower energy spectral regions contain information about the wind density $\rho(r)$. Namely, the continuum is dominated by free-free processes, which scale with $\rho^2$ \citep[e.g.,][]{Barlow-Cohen1977,Najarro+2009, RubioDiez+2022}.
In the IR, one also finds lines such as \ion{He}{I}\,$\lambda$10830 that can be used as an indicator of the terminal velocity in hot supergiants and hypergiants (see Fig.\,\ref{fig:Z1Sco-nIR-fit}). Our model closely reproduces the width of this specific line, indicating that our derived $\varv_\infty$ is realistic.

In contrast to the UV, but similar to the optical, we observe more pronounced line variability in the IR regime, especially in wind-affected H and He lines. This again suggests changes in the photospheric properties as well as in the base of the wind.

\begin{figure}
\includegraphics[width=1.0\linewidth]{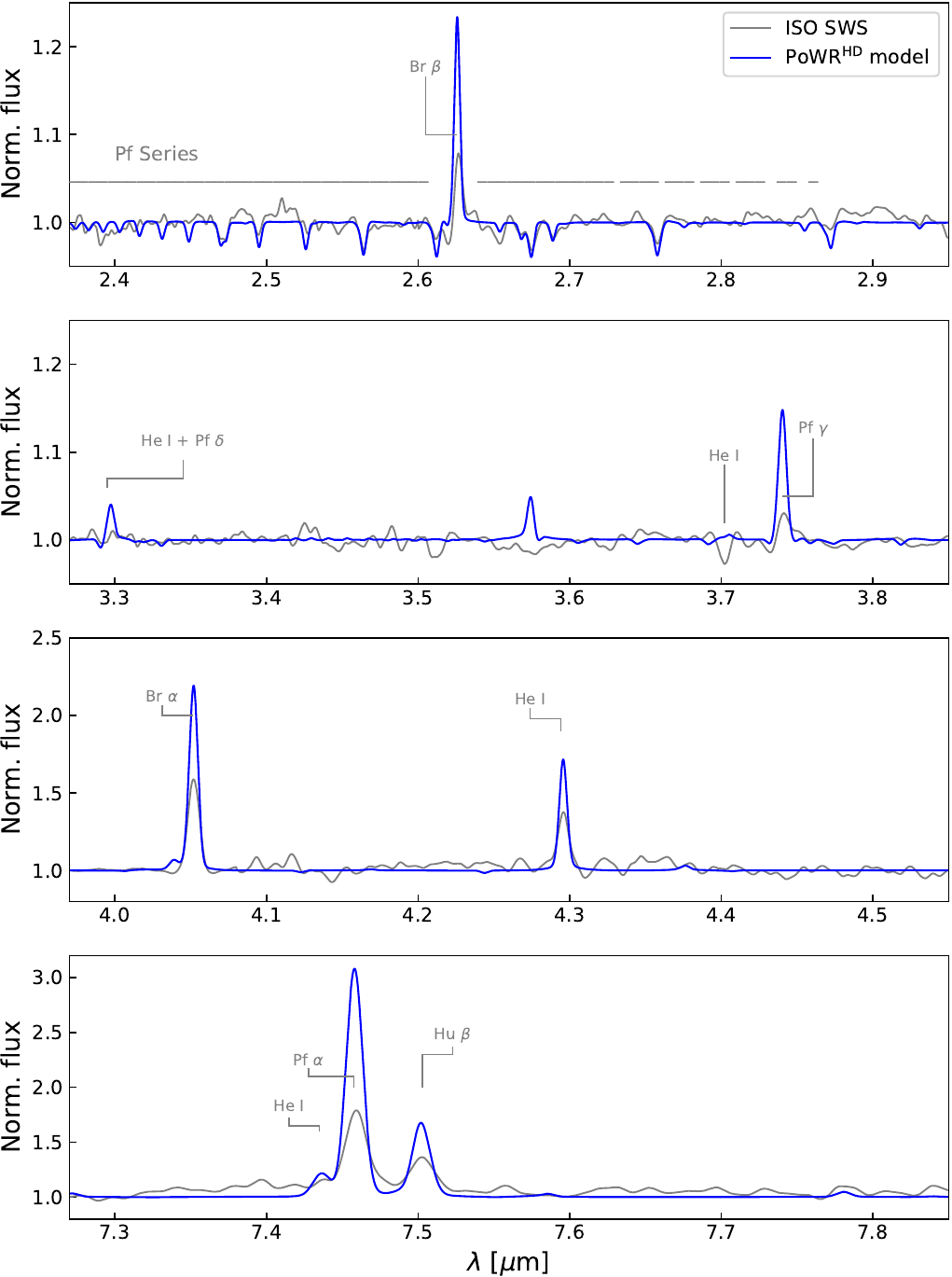}
  \caption{
    Comparison of our best-fit model (blue lines) with the mid-IR spectral features (ISO; gray lines). 
  }
    \label{fig:Z1Sco-ISO-fit}
\end{figure}

In the mid-IR regime, covered by the ISO observations shown in Fig.\,\ref{fig:Z1Sco-ISO-fit}, we find mostly low-energy transitions of H and He. In this range, the vast majority of the predicted emission lines are stronger than what is observed, similar to what we obtain in the near-IR. Nonetheless, the higher order of the Paschen series in the near-IR and the Pfund series in the mid-IR are in good agreement between the model and the observation.

However, we obtain a clear trend of overestimated emission of the Brackett, Paschen, and Pfund lines, indicating too much recombination in the outer wind. This may indicate that the mass-loss rates are slightly too high or that the clumping structure in the outer wind is not well reproduced. As discussed by \citet{Najarro+2011} in the context of LBVs, a better fit in the IR can be obtained considering a volume-filling factor that returns to unity in the outer wind. In our specific case, applying this clumping law does not produce a sufficient reduction of the emission to account for the discrepancy. Still, when considering models with lower clumping, we are able to solve the discrepancy, at the cost of spoiling the fits in other regions as another hydrodynamical solution is obtained (see the discussion in Sect. \ref{sec:clumping}). This implies that the findings by \citet{Najarro+2011} of a decrease in clumping are qualitatively in line with our modeling results, but the specific description of the clumping decrease in the outer wind is insufficient for our particular case.

In contrast to the spectrum of P\,Cyg \citep{Najarro+1996}, we do not observe forbidden lines in $\zeta^1$\,Sco, which would have  been evidence of circumstellar material.

\section{Comparison to previous works}
\label{sec:star-prop}

Overall, the properties of our final PoWR$^\textsc{hd}$ match relatively well the properties deduced in quantitative spectroscopic analyses by \cite{Crowther+2006} and \cite{Clark+2012}, performed with CMFGEN \citep{Hillier-Miller1998} and using atmosphere models without hydrodynamical consistency.
In Table\,\ref{tab:star-prop} we list the basic stellar properties we obtained for our final model and compare them to the aforementioned studies.
As discussed in Sect.\,\ref{sec:spec_valid} and Appendix \ref{app:diagnostics}, the stellar mass obtained with PoWR$^\textsc{hd}$ may be underestimated and affected by a certain degeneracy with the assumed clumping, as we discuss further in Sect.\,\ref{sec:clumping}. Nonetheless, our obtained $L/M$ ratio is essentially the same as obtained by \cite{Clark+2012} -- differing by less than $2\%$. 

\begin{table}
\caption{\label{tab:star-prop} Stellar properties of $\zeta^1$\,Sco deduced from our final PoWR$^\textsc{hd}$ model.}
\centering
\begin{tabular}{lccr}
\hline\hline
Property & This work & Cr06 & Cl12 \\
\hline
$T_*^{\tau\,\mathrm{max}}$ [kK] & 21.5$_{50}$ & -- & 19.8$_{100}$\\
$T_\mathrm{eff}^{\tau = 2/3}$ [kK] & 17.7 & 18.0 & 17.2 \\
$\log L/\mathrm{L_\odot}$  & 5.96 & 6.10 & 5.93  \\
$\log g^{\tau\,\mathrm{max}}$ [cgs] & 2.35$_{50}$ & -- & 2.20$_{100}$  \\
$\log g^{\tau = 2/3}$ [cgs] & 2.02 & 2.20 & 1.97 \\
$\varv \sin i$ [km\,s$^{-1}$] & 45 & 74 & 45 \\
$\varv_\mathrm{mac}$ [km\,s$^{-1}$] & 50 & -- & 50 \\
$M$ [$\mathrm{M_\odot}$] & 39  & 72 & 36\\
$R_*^{\tau\,\mathrm{max}}$ [$\mathrm{R_\odot}$] & 69$_{50}$  & -- & 77$_{100}$ \\
$R^{\tau = 2/3}$  [$\mathrm{R_\odot}$] & 100 & 112  & 103 \\
$\Gamma_\mathrm{e}$ & 0.51 & 0.35 & 0.47 \\
$\varv_\mathrm{turb}$ [km\,s$^{-1}$] & 14 & 15 & 15\\
$\xi_\mathrm{max}$ [km\,s$^{-1}$] & 180 & 50 & 90 \\
$X_\mathrm{H}$ & 0.60 & 0.55& 0.55\\
$X_\mathrm{He}$ & 0.38 & 0.43 & 0.43 \\
$X_\mathrm{C} \times 10^\mathrm{3}$ & 2.1 & 0.1 & 1.0 \\
$X_\mathrm{N} \times 10^\mathrm{3}$ & 6.0 & 1.7 &  6.0 \\
$X_\mathrm{O} \times 10^\mathrm{3}$ & 4.9 & 1.8 &  3.0 \\
\hline
$\varv_\infty$ [km\,s$^{-1}$] & 300 & 390 & 390 \\
$\log \dot{M}$ [$\mathrm{M_\odot}\,\mathrm{yr}^{-1}$] & -5.27 & -5.22 & -5.79 \\
\hline
\end{tabular}
\tablefoot{
For comparison, we also show the results from \citet[Cr06]{Crowther+2006} and \citet[Cl12]{Clark+2012}. The subscripts on $T_*^{\tau\,\mathrm{max}}$, $\log g^{\tau\,\mathrm{max}}$ and $R_*^{\tau\,\mathrm{max}}$ refer to the maximum Rosseland mean optical depth $\tau_\mathrm{max}$ of each study.
}
\end{table}
Despite some mismatches between the observed and synthetic spectra, the obtained stellar properties are in good agreement with previous determinations in the literature. Namely, the differences from our obtained $T_\mathrm{eff}$ and $\log g$ compared to literature values are well within the typical errors associated with different methods of quantitative spectroscopy \citep[see][]{Sander+2024}. 

Despite the high luminosity and dense wind, the Eddington parameter of our model is ``only'' $\Gamma_\mathrm{e} = 0.51$. 
Our finding aligns with what \citeauthor{Clark+2012} reported for the late BHG Cyg\,OB2\,\#12 (B3Ia+), which, despite possibly surpassing the Humphreys-Davidson limit \citep{Humphreys-Davidson1979}, also has $\Gamma_\mathrm{e} = 0.40$. Their value for $\zeta^1$\,Sco is $\Gamma_\mathrm{e} = 0.47$, very close to our findings. When considering the total $\Gamma_\text{rad}$ in the layers below the critical point ($\Gamma_\mathrm{rad}^\mathrm{in}$), which takes into account the total line opacity, we find $\Gamma_\mathrm{rad}^\mathrm{in} \sim 0.9$. This illustrates that due to the line and bound-free opacities, $\zeta^1$\,Sco is actually quite close to the Eddington limit. Our model also shows a small super-Eddington regime in the subsonic region, which might indicate the presence of a turbulent layer \citep[see][]{Debnath+2024} and could possibly explain some of the observed line profile variations. Currently, there is no self-consistent treatment of the resulting turbulent pressure in 1D atmosphere, but it can be included in a parameterized way \citep[e.g.,][]{Gonzalez-Tora+2025}. In this work we adopted a value of $\varv_\text{turb} = 14\,\mathrm{km\,s^{-1}}$, which is similar to \citet{Clark+2012} but actually might be an underestimation. In Appendix \ref{app:diagnostics} we briefly discuss the impact of varying $\varv_\text{turb}$ in the model calculations.

\section{Wind dynamics}
\label{sec:wind-prop}

\subsection{Mass-loss rate and wind acceleration}
\label{sec:wind-drive-ions}

\begin{table}
\caption{\label{tab:mdots} Mass-loss rates and clumping parameters inferred for $\zeta^1$\,Sco from different studies.}
\centering
\begin{tabular}{l|cccr}
\hline\hline
Study & $\dot{M}$ & $f_\infty$ & $\dot{M} \sqrt{f_\infty^{-1}}$ & $\dot{M}_\mathrm{t}$ \\
 -- & $\mathrm{M_\odot\,Myr^{-1}}$ & -- & $\mathrm{M_\odot\,Myr^{-1}}$ & $\mathrm{M_\odot\,Myr^{-1}}$ \\
\hline
This  & $5.3$ & 0.66 & 6.4 & 23.3 \\
Cr06  & $6.0$ & 1.00 & 6.0  & 12.9\\
Cl12  & $1.6$ & 0.06 & 6.5  & 18.9\\
RD22  & $<$$6.2$ & 0.40 -- 1.00 & $<$$6.9$ & -- \\
\hline
\end{tabular}
\tablefoot{
In the last columns the clumping-corrected mass-loss rates and transformed mass-loss rates are shown.
}
\end{table}

Our obtained PoWR$^\textsc{hd}$ solution has a mass-loss rate of $\dot{M} = 10^{-5.27}\,\mathrm{M_\mathrm{\odot}\,yr}^{-1} = 5.3\,\mathrm{M_\mathrm{\odot}\, Myr^{-1}}$. This result lies within the range found in previous studies by \cite{Crowther+2006}, \cite{Clark+2012}, and \cite{RubioDiez+2022}, albeit with some variations by factors of more than 3 (see Table\,\ref{tab:mdots}).
However, as also evident from Table\,\ref{tab:mdots}, when we consider $\dot{M} \sqrt{f_\infty^{-1}}$, we see an agreement within 10\% between the different studies for this quantity. 

On the other hand, when we consider the transformed mass-loss rates $\dot{M}_\mathrm{t} = \dot{M}\cdot f_\infty^{-1/2}\cdot (10^3\, \mathrm{km\,s^{-1}}/\varv_\infty)\cdot (10^6\,\mathrm{L_\odot}/L)^{3/4}$, introduced by \cite{Grafener-Vink2013}, we obtain bigger disagreements. Compared to \cite{Crowther+2006} a difference of a factor of 2 is obtained, whereas for \cite{Clark+2012} the difference is about 20\%. The reason for our obtained high $\dot{M}_\mathrm{t}$ is primarily linked to our obtained lower terminal velocity.

\begin{figure}
\centering
\includegraphics[width=1.0\linewidth]{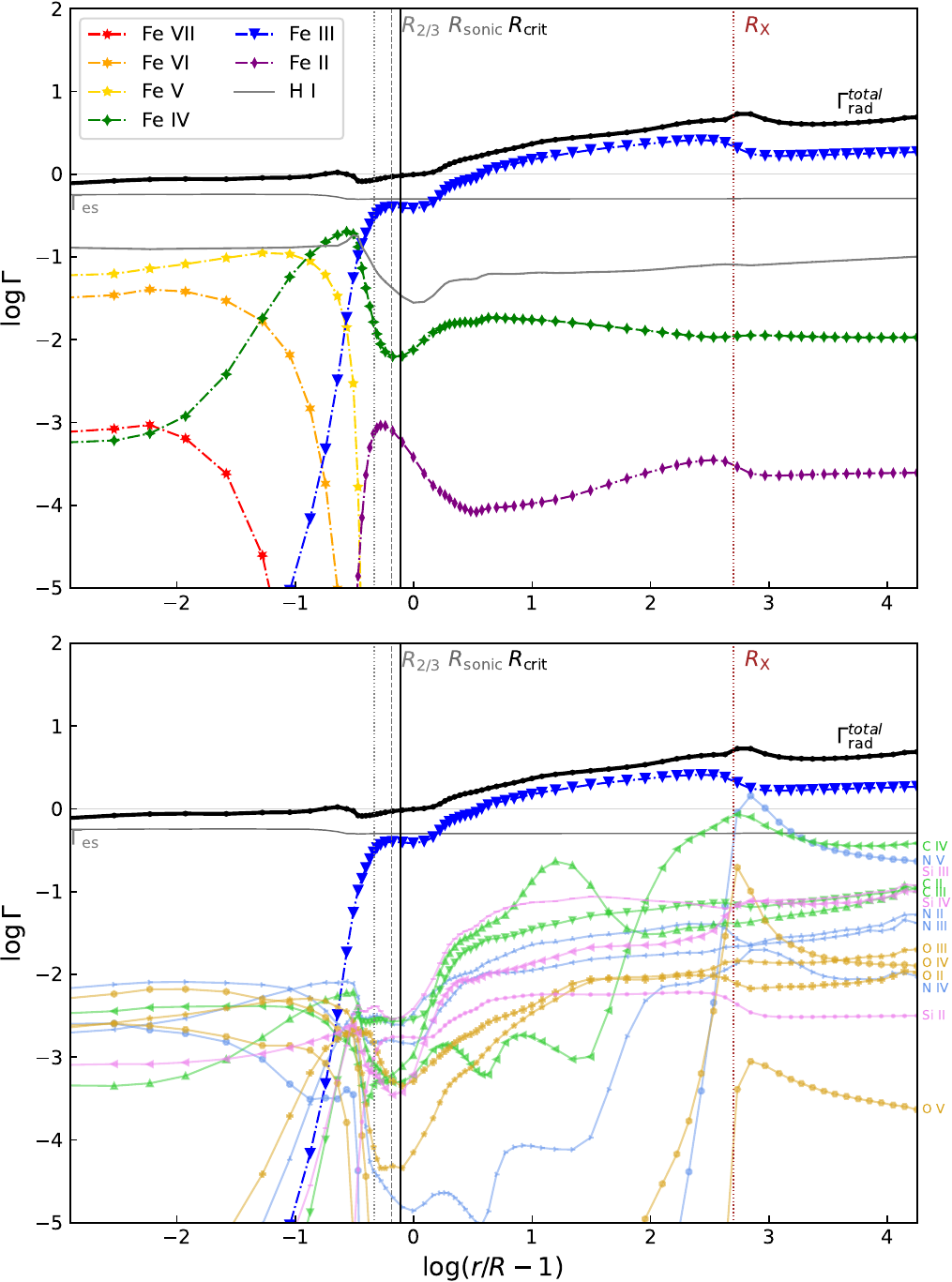}
  \caption{
    Wind-driving contribution of different ions to the total wind acceleration, $\Gamma_\mathrm{rad}^\mathrm{total}$. The upper panel focuses on the contribution of different Fe ions. The acceleration by electron scattering ($\Gamma_\mathrm{es}$) and by \ion{H}{I} are shown. The lower panel shows the breakdown of the contributions of CNO and Si ions. \ion{Fe}{III} and the electron scattering acceleration, the most important contributors to wind launching, are also shown for comparison.  The vertical lines represent the photospheric, sonic, and critical radii (dotted gray, dashed gray, and thin black lines, respectively). The dotted brown line marks the point where X-ray emission is switched on.
  }
    \label{fig:acc-G-alpha}
\end{figure}

In Fig. \ref{fig:acc-G-alpha} we show the wind acceleration normalized by gravity for different ions in our dynamically consistent wind model for $\zeta^1$\,Sco. The acceleration for each Fe ionization stage is shown in the upper panel. In the subsonic regime, \ion{Fe}{iv} and higher ionization stages dominate the acceleration. However, toward the sonic point, \ion{Fe}{iii} takes over and the contribution of \ion{Fe}{iv} drops to $\sim$1\%. Notably, at the critical point $R_\mathrm{crit}$, the contribution of \ion{Fe}{iii} to the total radiative acceleration, being comparable with the electron scattering acceleration, is the most important. Hence, this is the ion that effectively launches the stellar wind. In the outer wind, the \ion{Fe}{iii} also maintains dominance.

This behavior is similar to the hydrodynamically consistent model of the early-BSG donor of Vela\,X-1 (B0.5Ia) computed by \cite{Sander+2018} using the same code. In their models, \ion{Fe}{III} is the ion that effectively launches the wind whereas \ion{Fe}{IV} is secondary. However, in their case, the contribution of \ion{Fe}{IV} is about 10\% near $R_\mathrm{crit}$ due to the higher temperature in the B0.5Ia star. In the case of $\zeta$\,Pup (O4Iaf$^+$), analyzed and discussed in \cite{Sander+2017} following the same methodology, Fe is also the leading element but the \ion{Fe}{IV} ionization stage is the dominant source of acceleration at the critical radius. 

The leading acceleration ion is important in the context of the so-called bi-stability jump. The idea of a bi-stable solution for the mass-loss rates stemmed from \cite{Pauldrach_Puls1990}, who computed the wind solution for P\,Cyg (B1Ia+/LBV), whose temperature was estimated $\sim$21\,kK in their study. Later on, studies such as \cite{Lamers+1995} and \cite{Markova_Puls2008} found drops in the terminal velocity roughly at that temperature, reinforcing the scenario of a sudden change in the wind properties due to the recombination of \ion{Fe}{IV} to \ion{Fe}{III} within a narrow range of stellar temperatures around $\sim$22\,kK. Utilizing the observed terminal velocity change, this led to a significant jump in the mass-loss rates in the Monte Carlo calculations by \citet{Vink+1999}, which also entered the widely used \citet{Vink+2001} mass-loss recipe including its recent update \citep{Vink-Sander2021}. For stars with winds driven (mainly) by \ion{Fe}{iii}, the mass-loss rates obtained with the scheme from \citet{Vink+1999} are about an order of magnitude larger than those of winds driven by \ion{Fe}{iv}. A similar conclusion was reached in the analysis by \citet{Petrov+2016}. However, in the past two decades, empirical studies on BSGs across that region \citep[e.g.,][]{Crowther+2006, Bernini-Peron+2024, Verhamme+2024} and new theoretical calculations \citep[cf.][]{ Krticka+2021, Bjorklund+2023} challenge the scenario of a sharp $\dot{M}$ increase associated with the dominance of \ion{Fe}{III} as the main wind driver. Our modeling results place $\zeta^1$\,Sco on the cool side of the (hot) bi-stability jump. Ignoring the second bi-stability jump, the recipe from \citet{Vink+2001} yields $\log \left(\dot{M}\,[M_\odot\,\mathrm{yr}^{-1}]\right) = -4.42$, while our obtained value of $-5.27$ is almost an order of magnitude lower. Interestingly, both the descriptions from \citet{Bjorklund+2023} and \citet{Krticka+2024} yield results that are significantly too low, namely $-5.93$ and  $-5.94,$ respectively. Assuming our mass is actually an underestimation, the result from \citet{Bjorklund+2023} would even be lower. However, neither the studies by \citet{Bjorklund+2023} nor \citet{Krticka+2024} were designed for BHGs. Nonetheless, we can conclude that despite all similarities to BSGs, the mass-loss rates of BHGs are higher than recently predicted, though still considerably lower than estimated by \citet{Vink+2001}.

\subsection{Wind velocity field}
  \label{sec:windvelofield}

BHGs and LBVs are characterized by their slow winds, which can often be slower than 300\,km\,s$^{-1}$ and, from most of the $\beta$-values required in earlier spectral analysis, are expected to have a quite shallow velocity gradient. These characteristics make it hard to explain the superionization by shock-induced X-rays and defy the typical m-CAK picture of a velocity field with $\beta$ values around and below unity \citep[e.g.,][]{Cure-Araya2023}.

\paragraph{Wind terminal and turbulent velocities.}

The stellar properties used as input for the hydrodynamically consistent atmosphere model yield a terminal velocity of $\varv_\infty = 300$\,km\,s$^{-1}$. This value is about 90\,km\,s$^{-1}$ lower than what previous non-hydrodynamically consistent spectroscopic analyses find \cite[e.g.,][]{Crowther+2006, Clark+2012}. 
As mentioned in the discussion of the UV results (cf.\ Sect.\,\ref{sec:uvspec}), require a microturbulent velocity of $\xi_\mathrm{max} = 180$\,km\,s$^{-1}$ in the formal integral to reconcile the model with the observation.
\citet{Crowther+2006} adopted $\xi_\mathrm{max} = 50$\,km\,s$^{-1}$, whereas \citet{Clark+2012} considered a value of $\xi_\mathrm{max} = 75$\,km\,s$^{-1}$. While their values are lower than ours, they are higher than what is typically used in quantitative spectroscopy of OB stars in general ($\xi_\mathrm{max}  \sim 0.1 \varv_\infty$) and higher than the velocity dispersion obtained in line-deshadowing instability (LDI) hydrodynamic simulations of a BSG wind by \citet{Driessen+2019}. While our derived terminal velocity might be slightly underestimated, this generally higher microturbulence could point to a more fundamental difference between the winds of BSGs and BHGs, potentially related to the previously discussed turbulent pressure arising from the proximity of the hot iron bump to the Eddington limit.

\paragraph{Validity of the $\beta$ law approximation.}

To first order, the resulting velocity field in the supersonic regime is qualitatively similar to a $\beta$ law with a value of $\beta \gtrsim 3$ as this would lead to a comparably gradual acceleration of the wind. However, important differences appear when examining the field in detail, especially the layers further in the wind. Deviations from the $\beta$ law are also found in previous studies with hydrodynamically consistent models for O-type stars and BSGs \citep[e.g.,][]{Sander+2017, Sander+2018, Bjorklund+2021}.

\begin{figure}
\centering
\includegraphics[width=1.0\linewidth]{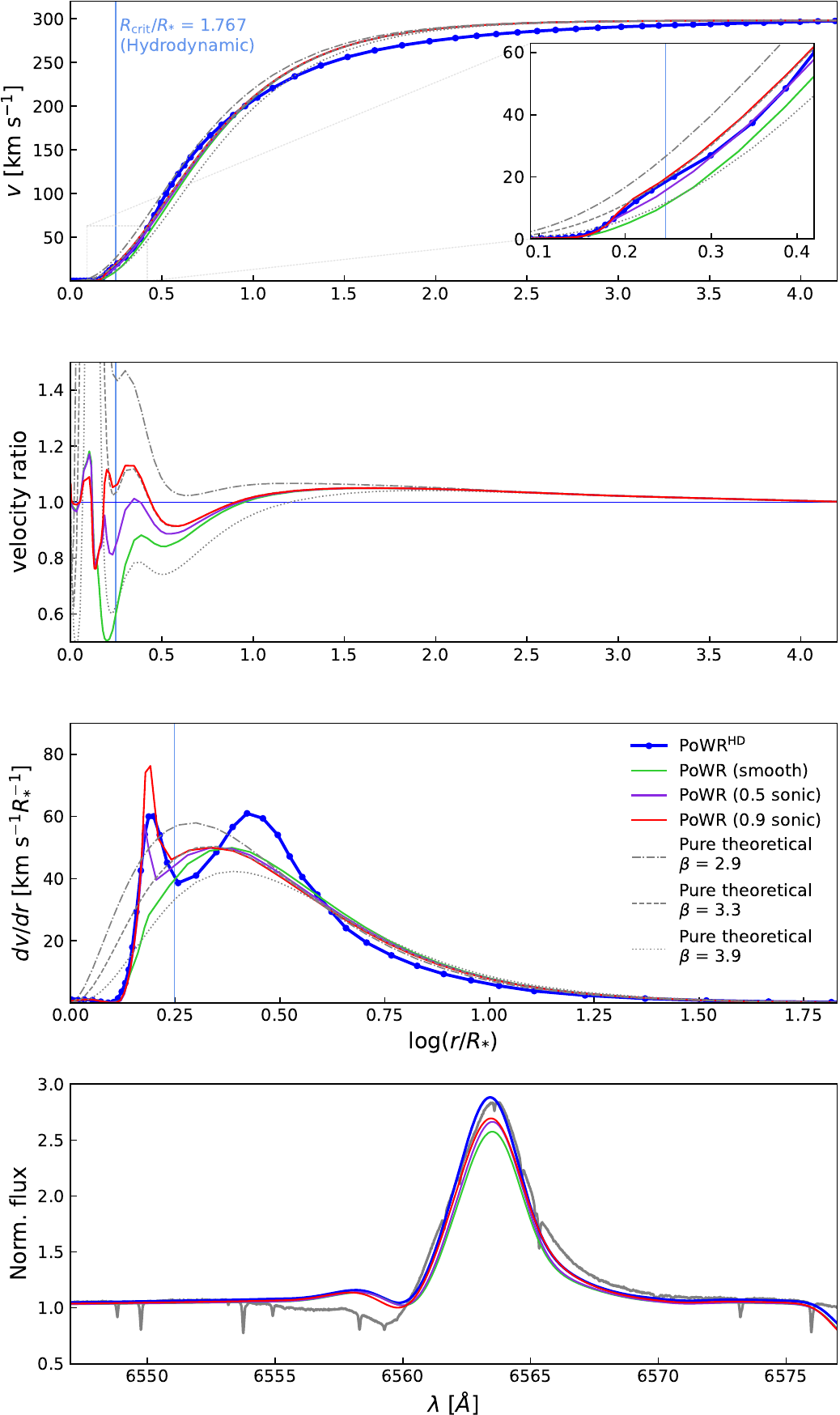}
  \caption{Comparison of hydrodynamically consistent, pure $\beta$ law profiles with classical PoWR models with hydrostatic profiles. \textit{Top panel}: Velocity profile as a function of the stellar radius. The gray lines indicate the pure theoretical $\beta$ law profiles with $\beta =$ 2.9, 3.3, and 3.9 (dot-dashed, dashed, and solid lines, respectively). The bold blue line is the hydrodynamically consistent PoWR$^\textsc{HD}$ model, whereas the green (continuous velocity gradient connection), purple (forced connection at $\varv = 0.5 \varv_\mathrm{sound}$), and red (forced connection at $\varv = 0.9 \varv_\mathrm{sound}$) profiles indicate the classical ($\beta$ law + hydrostatic integration) PoWR models with $\beta = 3.3$. The light blue vertical line indicates the critical point of the PoWR$^\textsc{HD}$ model. \textit{Second panel}: Same as the top panel but dividing all the velocity profiles by the hydrodynamical solution, and hence highlighting where velocity profiles differ more. \textit{Third panel}: Comparison of the velocity gradient ($d\varv/dr$) profiles. \textit{Bottom panel}: Spectroscopic comparison between the PoWR models (following the same color code). The observed optical spectrum (ESPRESSO; light gray) is also shown for reference.}
    \label{fig:z1sco-beta}
\end{figure}

In Fig.\,\ref{fig:z1sco-beta} we compare the velocity fields of a hydrodynamically consistent model with a set of models that use a hydrostatic stratification connected to a $\beta$ law, hereafter referred as ``classical PoWR models.'' All the models have the same stellar properties. Their work ratio\footnote{The ratio between the power deposited by the radiation field plus gas expansion and the total mechanical power of the wind \citep[see, e.g.,][]{Graefener-Hamann2005}.} is also about unity, implying a global, but not necessarily local consistency with respect to the force balance. The classical models differ among themselves in terms of how the connection between the (quasi-)hydrostatic and the wind profile is described. We calculated one model with a smooth subsonic-to-supersonic gradient connection ($(d\varv/dr)_\mathrm{hys} = (d\varv/dr)_\mathrm{\beta}$) and models where the velocity of the connection point $\varv_\mathrm{con}$ is fixed to different fractions of the sound speed
$\varv_\mathrm{sonic}$ (0.9, 0.7, and 0.5). The choice of $\beta = 3.3$ was obtained by employing nonlinear least squares to fit a $\beta$ law to the obtained hydrodynamically consistent solution.

The comparison reveals that the subsonic and transition regions agree remarkably well between the hydrodynamically consistent PoWR$^\textsc{hd}$ and the classical PoWR models with $\varv_\mathrm{con} = 0.9\,\varv_\mathrm{sonic}$. Their velocity gradients also match considerably well. Such an agreement is to be expected for the inner subsonic layers as long as the atmosphere is nearing (quasi-)hydrostatic equilibrium, which is ensured even for PoWR models without full dynamical consistency \citep{Sander+2015}. 
However, when forcing a smoother connection by imposing equal gradients at the transition point (cf.\  Fig.\,\ref{fig:z1sco-beta}), the differences increase. The smooth-connection model also produces spectra that differ more from the observations (see, e.g., the last panel of Fig.\,\ref{fig:z1sco-beta}, where the H$\alpha$ of the smooth model is the weakest). When inspecting this model, we find that the resulting connection point is significantly more inward than in the  ``abrupt'' transition models and connects to the hydrostatic gradient during its large increase. Considering both the hydrodynamic and the $0.9\,\varv_\mathrm{sonic}$ models, we can conclude that the additional subsonic peak in the gradient seems to mark an important feature for correctly predicting the spectrum.

In the supersonic regime, the velocity fields start to differ more distinctively, which is also noticeable in the velocity gradient. Up to $\sim$10$\,R_*$, the wind accelerates faster than the $\beta$ law profile but afterward the trend reverses. This result suggests that winds of BHGs have a shallower acceleration than the typical $\beta$ laws applied to OB stars (i.e., $0.5 \gtrsim \beta \gtrsim 1.0$). This is in alignment with a series of investigations in the past decades that systematically infer $\beta$\,$\gtrsim$\,$2.0$ for late BSGs and BHGs.
Many studies modeling LBVs, which have denser winds, also find $\beta \sim 3$ \citep[e.g.,][]{Najarro+2009, Kostenkov+2020} -- though values of $\beta$$\sim$1 have also been found \citep[e.g.,][]{Najarro+2009, Maryeva+2022}. 

\begin{figure}
\centering
\includegraphics[width=1.0\linewidth]{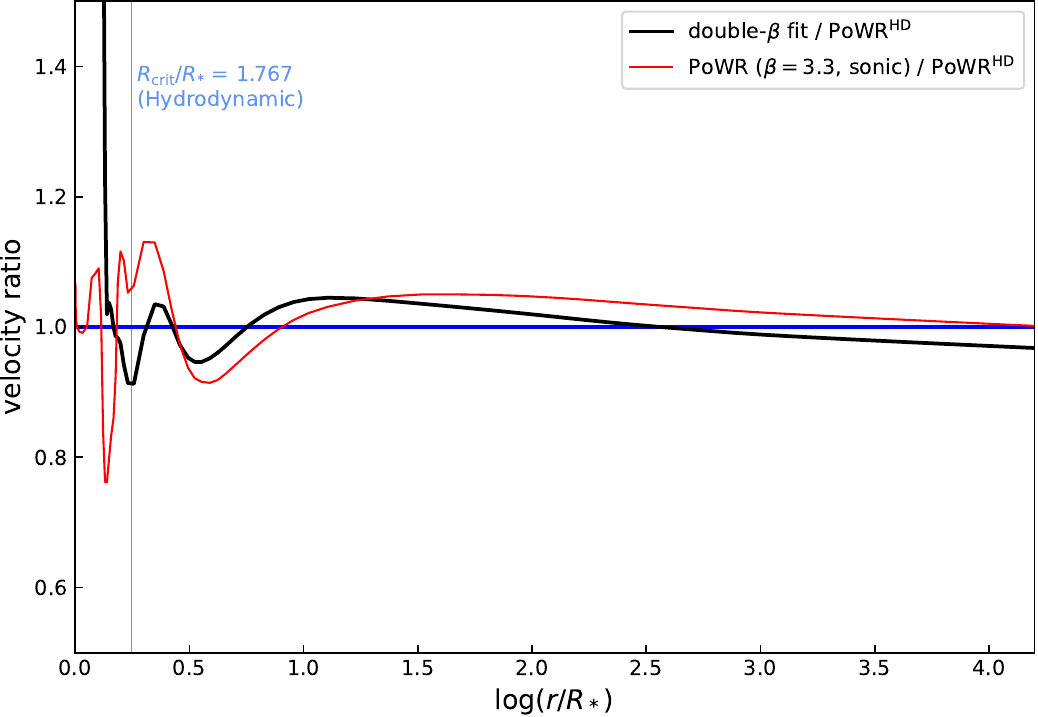}
  \caption{
    Comparison of the hydrodynamic consistent solution (blue thick line) with fits of a double $\beta$ profile (black thick line) as presented in \cite{Bjorklund+2021}. We also show for reference the best $\beta$ model from Fig.\,\ref{fig:z1sco-beta} as the red line.
  }
    \label{fig:double-beta-fit}
\end{figure}

We also tested models with different $\beta$ parameters. We do not find meaningful differences in the UV spectral appearance as most of the lines are already saturated. 
However, the lines in the optical and near-IR are significantly affected, as the velocity field around the connection between the quasi-hydrostatic and wind regime is considerably altered. High-velocity gradients ($\beta <  2.3$) in the inner wind layers produce atmospheres with weaker wind features. This underlines the need for $\beta>2.0$ for such stars. Additionally, even if adopting extremely high values of $\beta$ (e.g., $\beta> 10$), there is no good match with the wind velocity profile of the hydrodynamic model at $r \gtrsim 10 R_\ast$. One possibility to circumvent this problem is to adopt a double-$\beta$ law velocity profile, which can account for an inner acceleration different from the acceleration in the outer parts of the wind. This has for example been done for the hydrodynamical solutions in \citet{Bjorklund+2021}. Using the same approach, we obtain a parametric profile overall more similar to our hydrodynamical model up to $\sim$4\,$R_*$ (see Fig.\,\ref{fig:double-beta-fit}). 
Still, we obtain rather high values of the two parameters, namely $\beta_1=2.7$ and $\beta_2=2.0$. In contrast,  \cite{Bjorklund+2021} obtained a value of only $0.8$ for their inner $\beta_1$-parameter in their O star models describing a thinner wind regime.

As discussed for WR stars in \cite{Lefever+2023}, there may be degeneracies between the $\beta$ and effective temperature for optically thick winds. This is especially an issue if the whole spectrum is formed in the (outer) wind, which is the case for a fraction of the Galactic WR population \citep[e.g.,][]{Hamann+2019,Sander+2019} or objects like $\eta$\,Car \citep[e.g.,][]{Hillier+2001}. While also some BHGs, such as the more extreme P\,Cyg, might be affected by this degeneracy, $\zeta^1$\,Sco seems far from that regime. Therefore, any choice of the $\beta$ value should not significantly influence its temperature. Still, different $\beta$ values alter the connection regime, and thus the density structure, which would impact the many diagnostic lines formed at the base of the wind in typical BHGs as discussed above.

\cite{Venero+2024} modeled the velocity field and H$\alpha$ of late BSGs (and the BHG, HD\,80077, B2Ia+) using approximative hydrodynamical solutions within the ``m-CAK formalism''\footnote{``CAK'' refers to the theory developed by \cite{Castor+1975} while ``m-CAK'' to the modification to the theory introduced by \cite{Pauldrach+1986} and \cite{Friend-Abbott1986}.} \citep[see also][]{Cure+2011,Venero+2016,Vink2018}. 
They found a steeper velocity gradient in the inner wind compared to a $\beta$ law profile and a flatter gradient outward. Qualitatively, we find a similar behavior of the velocity field of our early BHG.
Their obtained velocity field for HD\,80077 belongs to the $\delta$-slow solution regime. In fact, they found that the BHG wind is only viable within this topological domain.

Another possible solution of the m-CAK hydrodynamics in the B star domain is the so-called $\Omega$-slow solutions \citep[see][]{Cure+2004, Cure-Araya2023}. This type of solution emerges in fast-rotating objects and is also associated with low $\varv_\infty$, as observed in BHGs. However, this type of solution would only occur for stars rotating $\gtrsim$75\% of the critical velocity\footnote{The critical rotation velocity is defined as $\varv_\mathrm{crit} = 2/3 \, \sqrt{G M/R}$ \citep{Maeder-Meynet2000}.}.
$\zeta^1$\,Sco (and BHGs in general) are not fast rotators and far from that threshold -- namely, $\zeta^1$\,Sco rotates only about 15\% critical. Moreover, the obtained PoWR$^\textsc{hd}$ velocity profile does not have a gradient in the inner wind as steep as predicted by the $\Omega$-slow solution.

\subsection{Clumping}
\label{sec:clumping}

In this work, wind inhomogeneities are treated in the ``microclumping'' approximation, assuming that clumps are optically thin. Given that the winds of BHGs can be optically thick in certain spectral lines, this approximation neglects the effects of so-called macroclumping \citep{Oskinova+2007,SundqvistPuls2018}, that is, the effect of optically thick clumps on the spectral diagnostics. So far, PoWR only treats macroclumping in the formal integral, which implies that it does not affect our derived hydrodynamic solutions. While calculations for a BSG in \citet{Bernini-Peron+2023} showed that macroclumping has an effect on the spectrum, we refrain from introducing it in this study, as we do not aim for precise reproduction of all spectral lines.

In atmosphere models without dynamical consistency, the volume filling factor $f(r)$ (cf.\,Eq. \ref{eq:volfill}) is adjusted to reconcile mass-loss rate diagnostics obtained from resonance and recombination lines. In hydrodynamically consistent models, however, changes in the clumping factor can also directly impact the derived mass-loss rate. In particular changes of $f(R_\mathrm{crit})$ affect the resulting opacity \citep[via a change in the wind density; see, e.g.,][]{Bouret+2012} and thus the inferred wind dynamics. Therefore, we investigated the impact of different clumping parameters both for wind driving and the spectral appearance of our early-BHG model.

\paragraph{Amount and onset of clumping.}

To obtain spectra that resemble the observations of $\zeta^1$\,Sco, we required a low clumping value  ($f_\mathrm{cl} = 1.5$ or $f_\infty = 0.66$) with $\varv_\mathrm{cl} = 5$. This $f_\infty$ value is similar to those inferred by \citet{RubioDiez+2022} via fitting the IR and radio spectral regions, providing additional evidence that the winds of BHGs are not very clumpy. Our findings further align well with the results from time-dependent hydrodynamical simulations by \cite{Driessen+2019}. They found that BSGs on the ``cool side'' of the bi-stability jump should show way less clumping than their O and early-B counterparts. Notably, this result differs from the spectroscopic analysis by \cite{Clark+2012}, who obtained $f_\infty = 0.06$. 

\begin{figure}
\centering
\includegraphics[width=1.0\linewidth]{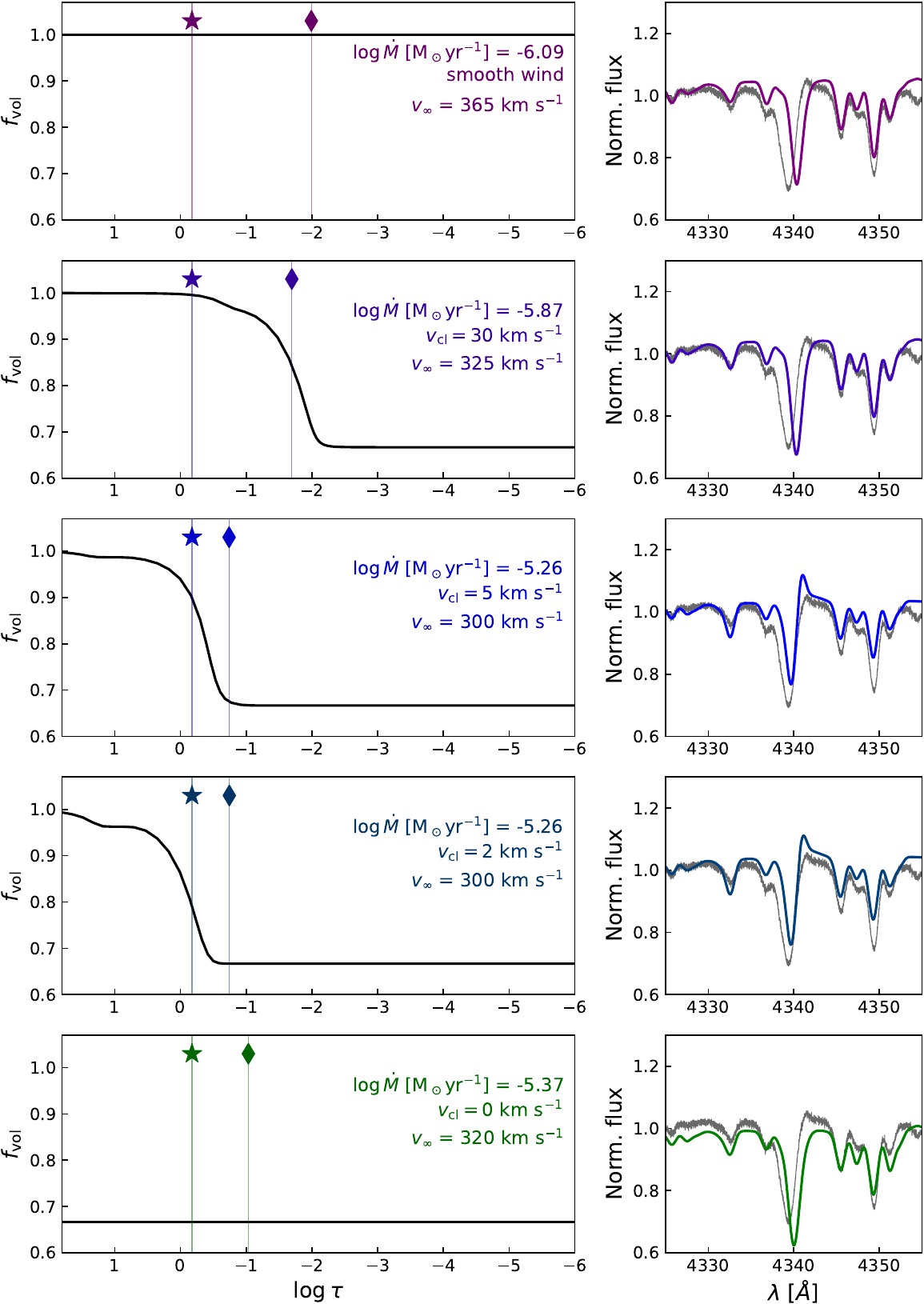     }
  \caption{
    Variation in the clumping structure and output wind properties according to different characteristic onset velocities, $\varv_\mathrm{cl}$. \textit{Left panels}: Photosphere, i.e., where the Rosseland-mean optical depth, $\tau$, is  $2/3$, as marked by the diamond symbol with its vertical line. \textit{Right panels}: Corresponding output spectra compared to the observed spectrum of $\zeta^1$\,Sco.
   }    \label{fig:clump-onset}
\end{figure}

We explored different characteristic velocities, $\varv_\mathrm{cl}$, for clumping, including subsonic values. Changing $f_\infty$ while keeping $\varv_\text{cl}$ mainly affects the amount of clumping in the outer atmosphere and thus also changes mostly the terminal velocity. However, even for the same value of $f_\infty$, different values of $\varv_\text{cl}$ affect the amount of clumping in the inner layers, especially at the critical point. In dynamically consistent models, this then impacts the wind driving as well as the spectral formation at the photosphere \citep[see also][]{Sabhahit+2025}.

In Fig.\,\ref{fig:clump-onset} we illustrate the change in the position of the critical point different $\varv_\mathrm{cl}$. Models with $\varv_\mathrm{cl} \sim 5$\,km\,s$^{-1}$ or lower produce a clumping stratification where the filling factor at the critical point $f(R_\mathrm{crit})$ is essentially identical to $ f_\infty$. On the other hand, models with higher $\varv_\mathrm{cl}$ have reduced or hardly any clumping at the critical point. Spectroscopically, these models behave more like the smooth wind model and do not predict H and He lines as P Cygni profiles, hence indicating a thinner wind.

In \cite{Clark+2012}, their best-fit model also has a clumping onset of about $\sim$60\% of the sound speed. This aligns well with the spectral analysis done by \cite{Najarro+2009} on LBVs in the Quintuplet cluster (Pistol Star and FMM362) using CMFGEN. They also obtained a low onset of clumping ($\varv_\mathrm{cl} \sim 2$\,km\,s$^{-1}$). However, both \citet[for $\zeta^1$\,Sco]{Clark+2012} and \citet[for the LBVs]{Najarro+2009} inferred a much more clumped wind with maximum volume filling factors of $0.06$ and $0.08$, respectively.

Interestingly, when we consider a model whose clumping starts from the inner boundary we find that the wind becomes again slightly optically thinner, associated with a lower $\dot{M}$. This may suggest that a variation of clumping within the subsonic regime is important to form the characteristic spectrum of $\zeta^1$\,Sco, and perhaps other early BHGs.

\paragraph{Spectral impact of clumping.}

\begin{figure}
\centering
\includegraphics[width=1.0\linewidth]{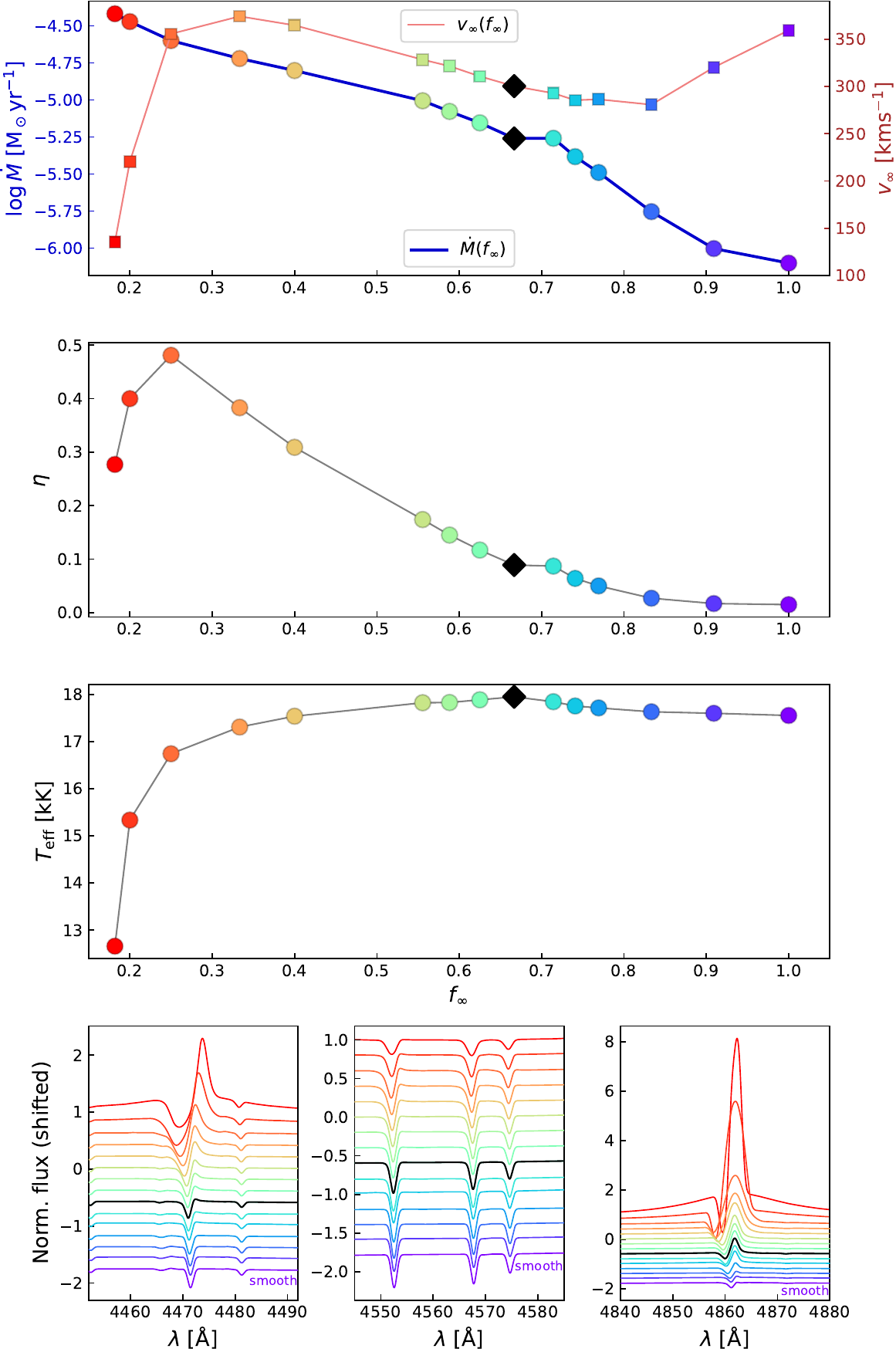}
  \caption{
  \textit{Upper panel}: Change in the mass-loss rates, $\dot{M}$, and terminal velocities, $\varv_\infty$, as a function of different values of the clumping volume filling factor, $f_\infty$. The colors trace $f_\infty$: purple indicates the homogeneous-wind model ($f_\infty$$=$$1.0$) and red is most clumped wind tested ($f_\infty$$=$$0.2$). The final model is represented by the black diamonds. 
  \textit{Middle panels}: Changes in the wind efficiency parameter ($\eta$) and $T_\mathrm{eff}$ as of function of $f_\infty$.
  \textit{Lower panels}: Different lines' spectral morphology changes as a function of different clumping values. The color scheme is the same as the upper panels, with the thick black line indicating the best-fitting model spectrum.
  }
    \label{fig:clumps}
\end{figure}

In Fig.\,\ref{fig:clumps} we compare models with different (maximum) clumping factors while leaving all other parameters untouched. More clumped winds produce spectra with more (pronounced) P-Cygni features, a consequence of the resulting higher mass-loss rates. On the other hand, if clumping is switched off, most of the optical lines remain completely in absorption, yielding a BSG-like spectrum.

\begin{figure}
\centering
\includegraphics[width=1.0\linewidth]{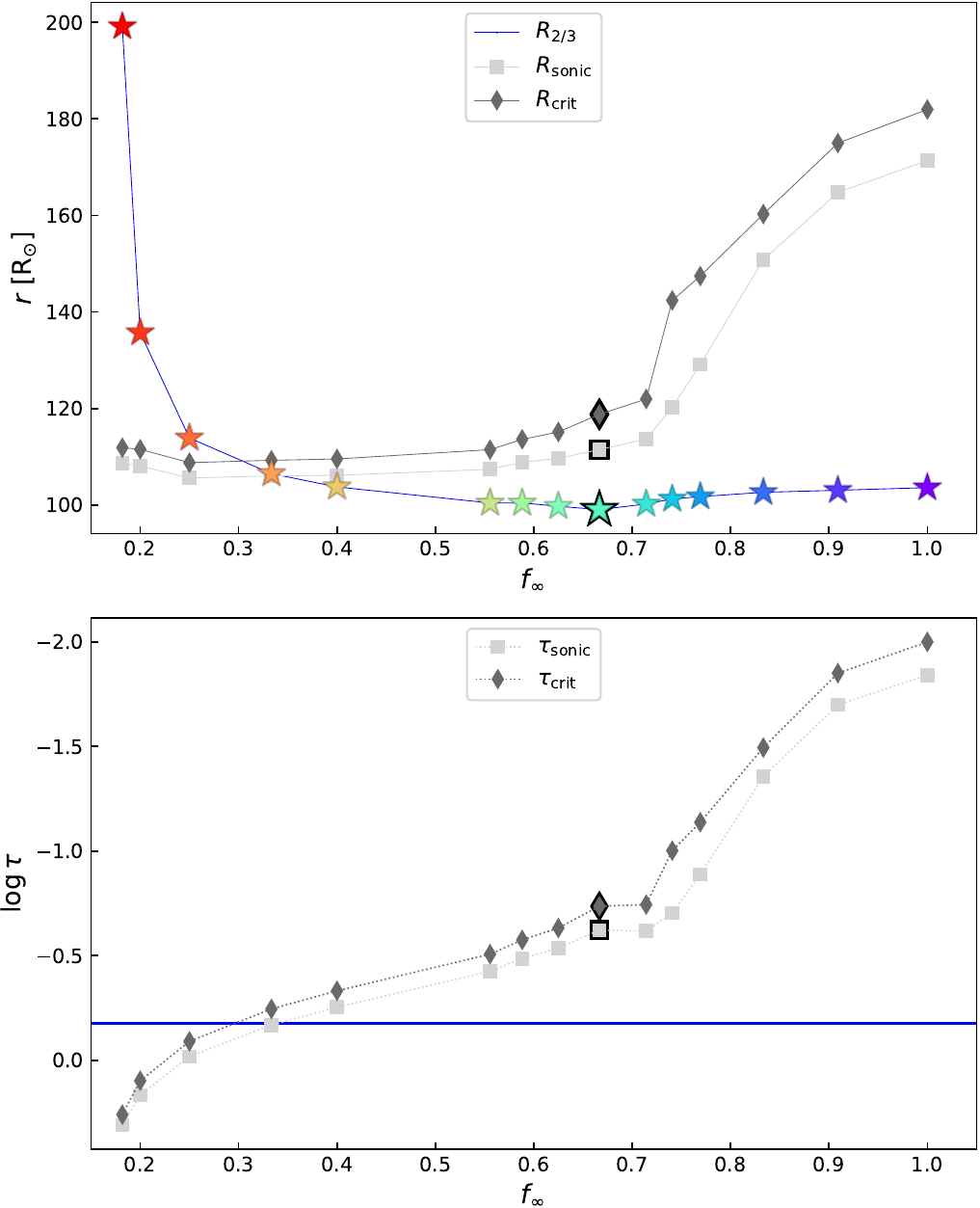}
  \caption{
  \textit{Upper panel}: Changes in the position of the critical ($R_\mathrm{crit}$; dark gray diamonds), sonic ($R_\mathrm{sonic}$; light gray squares), and photospheric ($R_{2/3}$; colored stars connected by the blue line) radii as a function of different values of the clumping volume filling factor, $f_\infty$. The bold black symbols indicate the best-fit model. \textit{Lower panel}: Same as the upper panel but showing the sonic and critical points in terms of the Rosseland-mean optical depth, $\tau$. The blue horizontal line indicates the photosphere, where $\tau = 2/3$.
  }
    \label{fig:vcl-RRR}
\end{figure}

The differences in wind density arising from different clumping are also reflected in a change of the critical point. As evident from Fig.~\ref{fig:vcl-RRR}, the models with smoother winds ($f_\infty = $ 0.73 -- 1.0) have $R_\text{crit} > R_{2/3} \equiv R(\tau=2/3)$ whereas models with higher clumping ($f_\infty = $ 0.35 -- 0.2) have $R_\text{crit} < R_{2/3}$, meaning that the wind is launched below the formation of the observed spectral continuum. Interestingly, when considering progressively more clumped winds ($f_\infty \rightarrow 0$), $\eta$ reaches a peak close to 0.5 for $f_\infty \sim 0.5$, but then decreases again.
The spectra of these models with a high clumping, however, do not look like a typical BHG anymore but instead resemble the spectral appearance of known LBVs in the quiescent phase -- also referred to as ``P-Cygni supergiants'' from a spectroscopic point of view, such as P\,Cyg, R81 and AG\,Car \citep{Clark+2012}.
The eventual decrease in $\eta$ is the result of a rapid decrease in the terminal velocity $\varv_\infty$ while $\dot{M}$ still increases (see the first panel of Fig.~\ref{fig:clumps}). Assuming the same energy budget, a denser wind would in principle move more slowly, but this reciprocal behavior between $\varv_\infty$ and $\dot{M}$ is notably only seen for very high or very low clumping in our model sequence, while for intermediate clumping values ($0.3 \lesssim$$f_\infty$$\lesssim 0.8$), $\varv_\infty$ increases with $\dot{M}$. This behavior switch should not be expected generally but depends significantly on the specific parameter regime and the choice of the clumping stratification and onset.

\paragraph{Changes in the velocity gradient.}

\begin{figure}
\centering
\includegraphics[width=1.0\linewidth]{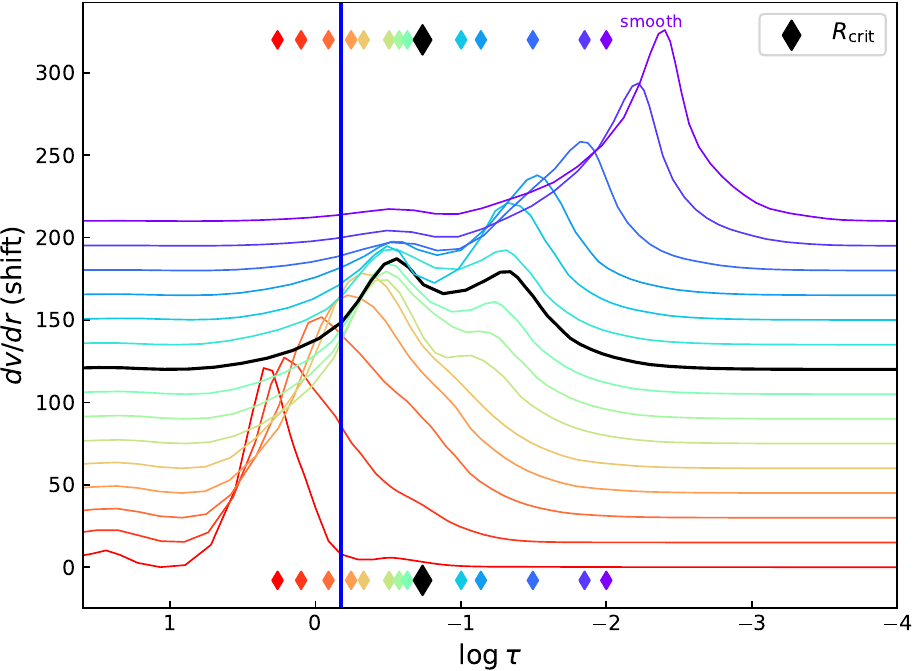}
  \caption{Velocity gradient ($d\varv/dr$) vs. Rosseland-mean optical depth ($\tau$). Each model is shifted vertically by an arbitrary amount for visualization purposes. The diamond symbols indicate the critical point, whereas the blue vertical line indicates the photosphere ($\tau = 2/3$). The thick black curve and diamonds refer to the final model ($f_\infty = 0.66$). 
  }
    \label{fig:clump-gradi}
\end{figure}

When inspecting the velocity gradient structure for the models with different clumping, it is noticeable that the models with the highest and the lowest clumping values have a single peak on their respective gradients, which correlates well with the position of their critical point. Curiously, our best model of $\zeta^1$\,Sco has a double-peak structure, which as discussed in Sect.\,\ref{sec:windvelofield} is necessary to get the proper spectral appearance of BHG winds as a transition between BSGs and quiescent LBVs.

\paragraph{Clumping-mass degeneracy.}

\citet{Sabhahit+2025} found a certain degeneracy between the clumping at the critical point and the inferred stellar masses when analyzing WNh stars of $\sim$$50\,$kK with hydrodynamically consistent models. For a given reference model, lowering (increasing) clumping at the critical point and lowering (increasing) the mass will cause the model to keep a similar spectral appearance and also yield similar wind properties. 

\begin{figure}
\centering
\includegraphics[width=1.0\linewidth]{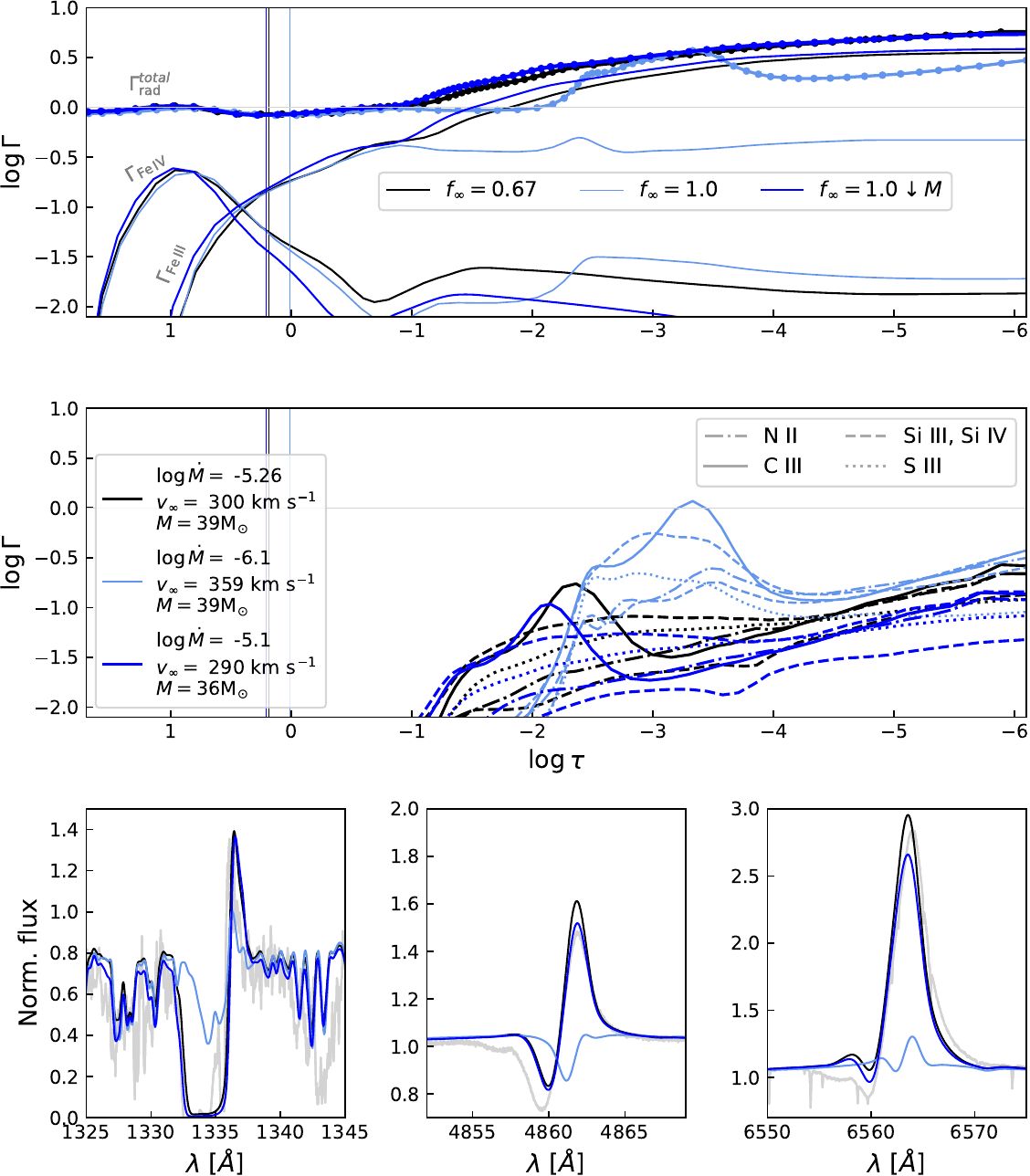}
  \caption{
  Degeneracy between clumping and stellar mass. In all panels, the black lines indicate the final model, the light blue lines the model without clumping, and the dark blue lines the model without clumping with a reduced mass. \textit{Upper panel}: Radiative acceleration vs. Rosseland optical depth. The total radiative acceleration, $\Gamma_\mathrm{rad}^{total}$, is indicated by the thick line with dots, and the thinner lines below indicate the contribution of \ion{Fe}{III} and \ion{Fe}{IV}. The thin vertical lines indicate the position of the critical points. \textit{Middle panel}: Same as the upper panes but showing the contribution of the main ions of C, N, S, and Si. \textit{Lower panels}: Comparison between the models' output spectra. Mass-loss rate units are M$_\odot$\,yr$^{-1}$.
  }
\label{fig:clump-mass}    
\end{figure}

For our studied BHG regime, we find a similar effect when comparing a clumped ($f_\infty = 0.67$, our final model) with a smooth-wind model. This comparison is illustrated in Fig.\,\ref{fig:clump-mass}, where we note a different behavior between the smooth-wind and clumped model regarding the driving and spectral appearance. Namely, the smooth-wind model has lower mass-loss rates, higher terminal velocities, and different wind-driving morphology -- all of which translates to a spectrum with fewer emission and P-Cygni lines. However, by decreasing the mass (only by 5\%) we obtain a model that is very similar to the clumped model in terms of structure, wind properties, and spectral appearance. The smooth-wind model thereby also sets a lower limit of $36\,M_\odot$ for the derived mass of $\zeta^1$\,Sco, which coincides with the value from \citet{Clark+2012}, albeit for a different luminosity.

\section{Are BHG winds in the transition regime?}
\label{sec:trans}

O stars with a spectrum dominated by P\,Cygni-type profiles in the optical regime are close to the onset of the regime of optically thick winds, where the wind onset point has an optical depth larger than unity. \citet{Vink+2012} demonstrated that in the case of $\eta \equiv \dot{M}\varv_\infty/Lc^{-1} \approx \tau_\mathrm{F} = 1$, the mass-loss rate can be directly inferred via
\begin{equation}
  \label{eq:mdottrans}
  \dot{M} = \frac{L}{\varv_\infty c} \equiv \dot{M}_\mathrm{trans} \text{.}   
\end{equation}
where the $\tau_\mathrm{F}$ is the flux-weighted optical depth, including line opacities, at the critical point.
This is called the ``transition mass-loss rate'', $\dot{M}_\text{trans}$\footnote{The transition mass-loss rate $\dot{M}_\text{trans}$ should not be mixed up with the ``transformed mass-loss rate'' $\dot{M}_\text{t}$. The latter is a quantity to compare wind strengths for stars with different luminosities and wind velocities.}. Detailed modeling with both Monte Carlo \citep{Vink+2012} and PoWR$^\textsc{HD}$ \citep{Sabhahit+2023} has shown that often $\eta = 1$ does not exactly correspond to $\tau_\mathrm{F} = 1$ and a correction factor $f$ has to be applied, typically with $f \approx 0.6$ \citep{Vink+2012,Sabhahit+2023}.

The concept of this ``transition mass-loss rate'' was successfully used to predict the $\dot{M}$ of very massive Of/WN-type stars (``slash stars''; see \citealt{Sabhahit+2022,Sabhahit+2023}). In the optical, these stars show photospheric profiles mixed with clear wind features \citep[see, e.g.,][]{Crowther+2011}. Thus, the question that may arise is whether BHGs are cooler counterparts of the slash stars. 

While BHGs are morphologically similar to the slash stars in terms of the presence of wind features contaminating the photospheric lines, our models suggest that early BHGs have a very low $\eta, < 0.1$, which is rather similar to that of BSGs. For $\tau_\mathrm{F}$, we derive a value of $0.4$ for our preferred model, meaning that $\zeta^1$\,Sco is not that close to the transition regime. To quantify this, we calculated $\dot{M}_\text{trans}$ according to Eq.\,\eqref{eq:mdottrans} using the luminosity and terminal velocity of $\zeta^1$\,Sco. The computed value of $\dot{M}_\text{trans} = 10^{-4.37}$ is almost an order of magnitude above our determined value of $10^{-5.27}$ \citep[and actually very similar to the derived value when using the recipe from][]{Vink+2001}.

When considering models with higher clumping (cf.\ Sect.\,\ref{sec:clumping}), we observe an increase in $\eta$, which approaches $\sim$0.5. However, when the wind density reaches a certain point, the terminal velocity starts to decrease, reducing $\eta$ as a consequence. While further studies with dynamically consistent atmospheres for individual BHGs and P\,Cygni-like supergiants are necessary, it seems currently unlikely that typical BHGs reach the transition regime and thus their mass-loss rates cannot be determined via the $\dot{M}_\mathrm{trans}$-formalism. 

\section{Conclusions}
\label{sec:conclu}

We used PoWR$^\textsc{hd}$ to produce the first hydrodynamically consistent atmosphere model of $\zeta^1$\,Sco -- the BHG with currently the most extensive spectroscopic coverage, considered a prototype of its kind. We successfully reproduce most of the characteristic spectral features of BHGs from the UV to the IR, which indicates that the atmosphere and wind structure are well described by our model.
From our modeling efforts and observational comparisons, we derive the following conclusions:
\begin{itemize}

    \item We derive a mass-loss rate for $\zeta^1$\,Sco of $10^{-5.27}\,M_\odot\,\mathrm{yr}^{-1}$, which is about one order of magnitude below the prediction from \citet{Vink+2001} but almost one order of magnitude higher than predicted by \citet{Bjorklund+2023} and \citet{Krticka+2024}. The radiative acceleration in the wind is dominated by \ion{Fe}{iii}. This is also true at the critical point, which is located close to, but still above, the photosphere. The derived ionization and acceleration is coherent with objects located on the cool side of the so-called bi-stability jump.

    \item Despite BHGs being spectroscopically similar to Of/WNh in terms of the presence of wind profiles in H$\beta$ (and higher-energy Balmer lines), their wind efficiency is too low to place them in the ``thick wind'' regime. This shows that for such low-temperature stars, the optical spectrum is capable of producing wind features while remaining relatively optically thin. This prevents the application of the transition mass-loss rate concept \citep{Vink+2012} in this regime.
    
    \item The radiative acceleration produces a velocity profile that differs from the widely used $\beta$ law and has a subsonic as well as a supersonic peak in the velocity gradient $d\varv/dr$. When restricted to $\beta$ laws, our derived structure can be partially reproduced by using $\beta = 3.3$ connected to a hydrostatic regime at 90\% of the sound speed.

    \item When deriving the wind dynamics, the amount of clumping and its onset play an important role in the wind launching and in shaping the optical spectrum. Lower clumping values and a subsonic onset are necessary to reproduce the main spectral features of $\zeta^1$\,Sco. On the one hand, in the absence of clumping, the BHG features are lost and the spectrum instead resembles that of a BSG. On the other hand, when clumping is increased further, the spectral appearance of quiescent LBVs is eventually reached. 
    In addition, our BHG model shows the presence of a subsonic super-Eddington layer, which will likely cause radiatively driven turbulent pressure and could also help trigger the necessary clumping.

    \item The terminal velocity obtained from the self-consistent wind solution is lower than obtained in previous quantitative spectroscopy studies with prescribed velocity fields. To reproduce the observed UV P-Cygni absorption troughs, we needed to employ a high microturbulent velocity in the formal integral. Future investigations of more targets will be required to test whether such turbulent velocities are reasonable in this regime. 
    
\end{itemize}

\begin{acknowledgements}
    We thank Joris Josiek, Wolf-Rainer Hamann, Helge Todt, and Maur\'icio G\'omez-Gonz\'alez for their insightful comments and discussions on the modeling process as well as in observational aspects of this work. We additionally thank the anonymous referee who provided insightful and constructive comments that helped to improve this manuscript.
    MBP, AACS, and RRL are supported by the German
    \emph{Deut\-sche For\-schungs\-ge\-mein\-schaft, DFG\/} in the form of an Emmy Noether Research Group -- Project-ID 445674056 (SA4064/1-1, PI Sander). GGT is supported by the German Deutsche Forschungsgemeinschaft (DFG) under Project-ID 496854903 (SA4064/2-1, PI Sander). GGT further acknowledges financial support by the Federal Ministry for Economic Affairs and Climate Action (BMWK) via the German Aerospace Center (Deutsches Zentrum f\"ur Luft- und Raumfahrt, DLR) grant 50 OR 2503 (PI: Sander).
    FN acknowledges support by PID2022-137779OB-C41 funded by MCIN/AEI/10.13039/501100011033 by ``ERDF A way of making Europe.''
    GNS and JSV are supported by STFC (Science and Technology Facilities Council) funding under grant number ST/Y001338/1.
    DP acknowledges financial support from the Deutsches Zentrum f\"ur Luft und Raumfahrt (DLR) grant FKZ 50OR2005 and the FWO junior postdoctoral fellowship No. 1256225N.    %
    This work has made use of data from the European Space Agency (ESA) mission
    {\it Gaia} (\url{https://www.cosmos.esa.int/gaia}), processed by the {\it Gaia}
    Data Processing and Analysis Consortium (DPAC,
    \url{https://www.cosmos.esa.int/web/gaia/dpac/consortium}). Funding for the DPAC
    has been provided by national institutions, in particular the institutions
    participating in the {\it Gaia} Multilateral Agreement.
    Based on data obtained from the ESO Science Archive Facility with DOIs, \url{https://doi.org/10.18727/archive/21}, \url{https://doi.org/10.18727/archive/33}, \url{https://doi.org/10.18727/archive/50}, and \url{https://doi.org/10.18727/archive/71}
      
\end{acknowledgements}

\bibliographystyle{aa}
\bibliography{biblio.bib}

\begin{thebibliography}{96}
\expandafter\ifx\csname natexlab\endcsname\relax\def\natexlab#1{#1}\fi

\bibitem[{{Asplund} {et~al.}(2009){Asplund}, {Grevesse}, {Sauval}, \& {Scott}}]{Asplund+2009}
{Asplund}, M., {Grevesse}, N., {Sauval}, A.~J., \& {Scott}, P. 2009, \araa, 47, 481

\bibitem[{{Bailer-Jones} {et~al.}(2021){Bailer-Jones}, {Rybizki}, {Fouesneau}, {Demleitner}, \& {Andrae}}]{BailerJones+2021}
{Bailer-Jones}, C.~A.~L., {Rybizki}, J., {Fouesneau}, M., {Demleitner}, M., \& {Andrae}, R. 2021, \aj, 161, 147

\bibitem[{{Barlow} \& {Cohen}(1977)}]{Barlow-Cohen1977}
{Barlow}, M.~J. \& {Cohen}, M. 1977, \apj, 213, 737

\bibitem[{{Baum} {et~al.}(1992){Baum}, {Hamann}, {Koesterke}, \& {Wessolowski}}]{Baum+1992}
{Baum}, E., {Hamann}, W.~R., {Koesterke}, L., \& {Wessolowski}, U. 1992, \aap, 266, 402

\bibitem[{{Benaglia} {et~al.}(2007){Benaglia}, {Vink}, {Mart{\'\i}}, {Ma{\'\i}z Apell{\'a}niz}, {Koribalski}, \& {Crowther}}]{Benaglia+2007}
{Benaglia}, P., {Vink}, J.~S., {Mart{\'\i}}, J., {et~al.} 2007, \aap, 467, 1265

\bibitem[{{Berghoefer} {et~al.}(1997){Berghoefer}, {Schmitt}, {Danner}, \& {Cassinelli}}]{Berghoefer+1997}
{Berghoefer}, T.~W., {Schmitt}, J.~H.~M.~M., {Danner}, R., \& {Cassinelli}, J.~P. 1997, \aap, 322, 167

\bibitem[{{Bernini-Peron} {et~al.}(2023){Bernini-Peron}, {Marcolino}, {Sander}, {Bouret}, {Ramachandran}, {Saling}, {Schneider}, {Oskinova}, \& {Najarro}}]{Bernini-Peron+2023}
{Bernini-Peron}, M., {Marcolino}, W.~L.~F., {Sander}, A.~A.~C., {et~al.} 2023, \aap, 677, A50

\bibitem[{{Bernini-Peron} {et~al.}(2024){Bernini-Peron}, {Sander}, {Ramachandran}, {Oskinova}, {Vink}, {Verhamme}, {Najarro}, {Josiek}, {Brands}, {Crowther}, {G{\'o}mez-Gonz{\'a}lez}, {Gormaz-Matamala}, {Hawcroft}, {Kuiper}, {Mahy}, {Marcolino}, {Martins}, {Mehner}, {Parsons}, {Pauli}, {Shenar}, {Schootemeijer}, {Todt}, {van Loon}, \& {XShootU Collaboration}}]{Bernini-Peron+2024}
{Bernini-Peron}, M., {Sander}, A.~A.~C., {Ramachandran}, V., {et~al.} 2024, \aap, 692, A89

\bibitem[{{Bestenlehner} {et~al.}(2025){Bestenlehner}, {Crowther}, {Hawcroft}, {Sana}, {Tramper}, {Vink}, {Brands}, \& {Sander}}]{Bestenlehner+2025}
{Bestenlehner}, J.~M., {Crowther}, P.~A., {Hawcroft}, C., {et~al.} 2025, \aap, 695, A198

\bibitem[{{Bieging} {et~al.}(1989){Bieging}, {Abbott}, \& {Churchwell}}]{Biening+1989}
{Bieging}, J.~H., {Abbott}, D.~C., \& {Churchwell}, E.~B. 1989, \apj, 340, 518

\bibitem[{{Bj{\"o}rklund} {et~al.}(2021){Bj{\"o}rklund}, {Sundqvist}, {Puls}, \& {Najarro}}]{Bjorklund+2021}
{Bj{\"o}rklund}, R., {Sundqvist}, J.~O., {Puls}, J., \& {Najarro}, F. 2021, \aap, 648, A36

\bibitem[{{Bj{\"o}rklund} {et~al.}(2023){Bj{\"o}rklund}, {Sundqvist}, {Singh}, {Puls}, \& {Najarro}}]{Bjorklund+2023}
{Bj{\"o}rklund}, R., {Sundqvist}, J.~O., {Singh}, S.~M., {Puls}, J., \& {Najarro}, F. 2023, \aap, 676, A109

\bibitem[{{Bouret} {et~al.}(2012){Bouret}, {Hillier}, {Lanz}, \& {Fullerton}}]{Bouret+2012}
{Bouret}, J.~C., {Hillier}, D.~J., {Lanz}, T., \& {Fullerton}, A.~W. 2012, \aap, 544, A67

\bibitem[{{Cardelli} {et~al.}(1989){Cardelli}, {Clayton}, \& {Mathis}}]{Cardelli+1989}
{Cardelli}, J.~A., {Clayton}, G.~C., \& {Mathis}, J.~S. 1989, \apj, 345, 245

\bibitem[{{Cassinelli} {et~al.}(1981){Cassinelli}, {Waldron}, {Sanders}, {Harnden}, {Rosner}, \& {Vaiana}}]{Cassinelli+1981}
{Cassinelli}, J.~P., {Waldron}, W.~L., {Sanders}, W.~T., {et~al.} 1981, \apj, 250, 677

\bibitem[{{Castor} {et~al.}(1975){Castor}, {Abbott}, \& {Klein}}]{Castor+1975}
{Castor}, J.~I., {Abbott}, D.~C., \& {Klein}, R.~I. 1975, \apj, 195, 157

\bibitem[{{Clark} {et~al.}(2012){Clark}, {Najarro}, {Negueruela}, {Ritchie}, {Urbaneja}, \& {Howarth}}]{Clark+2012}
{Clark}, J.~S., {Najarro}, F., {Negueruela}, I., {et~al.} 2012, \aap, 541, A145

\bibitem[{{Cohen} {et~al.}(2014){Cohen}, {Li}, {Gayley}, {Owocki}, {Sundqvist}, {Petit}, \& {Leutenegger}}]{Cohen+2014}
{Cohen}, D.~H., {Li}, Z., {Gayley}, K.~G., {et~al.} 2014, \mnras, 444, 3729

\bibitem[{{Crowther} {et~al.}(2006){Crowther}, {Lennon}, \& {Walborn}}]{Crowther+2006}
{Crowther}, P.~A., {Lennon}, D.~J., \& {Walborn}, N.~R. 2006, \aap, 446, 279

\bibitem[{{Crowther} \& {Walborn}(2011)}]{Crowther+2011}
{Crowther}, P.~A. \& {Walborn}, N.~R. 2011, \mnras, 416, 1311

\bibitem[{{Cur{\'e}} \& {Araya}(2023)}]{Cure-Araya2023}
{Cur{\'e}}, M. \& {Araya}, I. 2023, Galaxies, 11, 68

\bibitem[{{Cur{\'e}} {et~al.}(2011){Cur{\'e}}, {Cidale}, \& {Granada}}]{Cure+2011}
{Cur{\'e}}, M., {Cidale}, L., \& {Granada}, A. 2011, \apj, 737, 18

\bibitem[{{Cur{\'e}} \& {Rial}(2004)}]{Cure+2004}
{Cur{\'e}}, M. \& {Rial}, D.~F. 2004, \aap, 428, 545

\bibitem[{{Cutri} {et~al.}(2013){Cutri}, {Wright}, {Conrow}, {Fowler}, {Eisenhardt}, {Grillmair}, {Kirkpatrick}, {Masci}, {McCallon}, {Wheelock}, {Fajardo-Acosta}, {Yan}, {Benford}, {Harbut}, {Jarrett}, {Lake}, {Leisawitz}, {Ressler}, {Stanford}, {Tsai}, {Liu}, {Helou}, {Mainzer}, {Gettings}, {Gonzalez}, {Hoffman}, {Marsh}, {Padgett}, {Skrutskie}, {Beck}, {Papin}, \& {Wittman}}]{Cutri+2013}
{Cutri}, R.~M., {Wright}, E.~L., {Conrow}, T., {et~al.} 2013, {Explanatory Supplement to the AllWISE Data Release Products}, Explanatory Supplement to the AllWISE Data Release Products, by R. M. Cutri et al.

\bibitem[{{Debnath} {et~al.}(2024){Debnath}, {Sundqvist}, {Moens}, {Van der Sijpt}, {Verhamme}, \& {Poniatowski}}]{Debnath+2024}
{Debnath}, D., {Sundqvist}, J.~O., {Moens}, N., {et~al.} 2024, \aap, 684, A177

\bibitem[{{Driessen} {et~al.}(2019){Driessen}, {Sundqvist}, \& {Kee}}]{Driessen+2019}
{Driessen}, F.~A., {Sundqvist}, J.~O., \& {Kee}, N.~D. 2019, \aap, 631, A172

\bibitem[{{Evans} {et~al.}(2015){Evans}, {Kennedy}, {Dufton}, {Howarth}, {Walborn}, {Markova}, {Clark}, {de Mink}, {de Koter}, {Dunstall}, {H{\'e}nault-Brunet}, {Ma{\'\i}z Apell{\'a}niz}, {McEvoy}, {Sana}, {Sim{\'o}n-D{\'\i}az}, {Taylor}, \& {Vink}}]{Evans+2015}
{Evans}, C.~J., {Kennedy}, M.~B., {Dufton}, P.~L., {et~al.} 2015, \aap, 574, A13

\bibitem[{{Fitzpatrick}(1991)}]{Fitzpatrick-Edward1991}
{Fitzpatrick}, E.~L. 1991, \pasp, 103, 1123

\bibitem[{{Friend} \& {Abbott}(1986)}]{Friend-Abbott1986}
{Friend}, D.~B. \& {Abbott}, D.~C. 1986, \apj, 311, 701

\bibitem[{{Gaia Collaboration} {et~al.}(2016){Gaia Collaboration}, {Prusti}, {de Bruijne}, {Brown}, {Vallenari}, {Babusiaux}, {Bailer-Jones}, {Bastian}, {Biermann}, {Evans}, {Eyer}, {Jansen}, {Jordi}, {Klioner}, {Lammers}, {Lindegren}, {Luri}, {Mignard}, {Milligan}, {Panem}, {Poinsignon}, {Pourbaix}, {Randich}, {Sarri}, {Sartoretti}, {Siddiqui}, {Soubiran}, {Valette}, {van Leeuwen}, {Walton}, {Aerts}, {Arenou}, {Cropper}, {Drimmel}, {H{\o}g}, {Katz}, {Lattanzi}, {O'Mullane}, {Grebel}, {Holland}, {Huc}, {Passot}, {Bramante}, {Cacciari}, {Casta{\~n}eda}, {Chaoul}, {Cheek}, {De Angeli}, {Fabricius}, {Guerra}, {Hern{\'a}ndez}, {Jean-Antoine-Piccolo}, {Masana}, {Messineo}, {Mowlavi}, {Nienartowicz}, {Ord{\'o}{\~n}ez-Blanco}, {Panuzzo}, {Portell}, {Richards}, {Riello}, {Seabroke}, {Tanga}, {Th{\'e}venin}, {Torra}, {Els}, {Gracia-Abril}, {Comoretto}, {Garcia-Reinaldos}, {Lock}, {Mercier}, {Altmann}, {Andrae}, {Astraatmadja}, {Bellas-Velidis}, {Benson}, {Berthier}, {Blomme}, {Busso}, {Carry}, {Cellino}, {Clementini},
  {Cowell}, {Creevey}, {Cuypers}, {Davidson}, {De Ridder}, {de Torres}, {Delchambre}, {Dell'Oro}, {Ducourant}, {Fr{\'e}mat}, {Garc{\'\i}a-Torres}, {Gosset}, {Halbwachs}, {Hambly}, {Harrison}, {Hauser}, {Hestroffer}, {Hodgkin}, {Huckle}, {Hutton}, {Jasniewicz}, {Jordan}, {Kontizas}, {Korn}, {Lanzafame}, {Manteiga}, {Moitinho}, {Muinonen}, {Osinde}, {Pancino}, {Pauwels}, {Petit}, {Recio-Blanco}, {Robin}, {Sarro}, {Siopis}, {Smith}, {Smith}, {Sozzetti}, {Thuillot}, {van Reeven}, {Viala}, {Abbas}, {Abreu Aramburu}, {Accart}, {Aguado}, {Allan}, {Allasia}, {Altavilla}, {{\'A}lvarez}, {Alves}, {Anderson}, {Andrei}, {Anglada Varela}, {Antiche}, {Antoja}, {Ant{\'o}n}, {Arcay}, {Atzei}, {Ayache}, {Bach}, {Baker}, {Balaguer-N{\'u}{\~n}ez}, {Barache}, {Barata}, {Barbier}, {Barblan}, {Baroni}, {Barrado y Navascu{\'e}s}, {Barros}, {Barstow}, {Becciani}, {Bellazzini}, {Bellei}, {Bello Garc{\'\i}a}, {Belokurov}, {Bendjoya}, {Berihuete}, {Bianchi}, {Bienaym{\'e}}, {Billebaud}, {Blagorodnova}, {Blanco-Cuaresma}, {Boch},
  {Bombrun}, {Borrachero}, {Bouquillon}, {Bourda}, {Bouy}, {Bragaglia}, {Breddels}, {Brouillet}, {Br{\"u}semeister}, {Bucciarelli}, {Budnik}, {Burgess}, {Burgon}, {Burlacu}, {Busonero}, {Buzzi}, {Caffau}, {Cambras}, {Campbell}, {Cancelliere}, {Cantat-Gaudin}, {Carlucci}, {Carrasco}, {Castellani}, {Charlot}, {Charnas}, {Charvet}, {Chassat}, {Chiavassa}, {Clotet}, {Cocozza}, {Collins}, {Collins}, {Costigan}, {Crifo}, {Cross}, {Crosta}, {Crowley}, {Dafonte}, {Damerdji}, {Dapergolas}, {David}, {David}, {De Cat}, {de Felice}, {de Laverny}, {De Luise}, {De March}, {de Martino}, {de Souza}, {Debosscher}, {del Pozo}, {Delbo}, {Delgado}, {Delgado}, {di Marco}, {Di Matteo}, {Diakite}, {Distefano}, {Dolding}, {Dos Anjos}, {Drazinos}, {Dur{\'a}n}, {Dzigan}, {Ecale}, {Edvardsson}, {Enke}, {Erdmann}, {Escolar}, {Espina}, {Evans}, {Eynard Bontemps}, {Fabre}, {Fabrizio}, {Faigler}, {Falc{\~a}o}, {Farr{\`a}s Casas}, {Faye}, {Federici}, {Fedorets}, {Fern{\'a}ndez-Hern{\'a}ndez}, {Fernique}, {Fienga}, {Figueras}, {Filippi},
  {Findeisen}, {Fonti}, {Fouesneau}, {Fraile}, {Fraser}, {Fuchs}, {Furnell}, {Gai}, {Galleti}, {Galluccio}, {Garabato}, {Garc{\'\i}a-Sedano}, {Gar{\'e}}, {Garofalo}, {Garralda}, {Gavras}, {Gerssen}, {Geyer}, {Gilmore}, {Girona}, {Giuffrida}, {Gomes}, {Gonz{\'a}lez-Marcos}, {Gonz{\'a}lez-N{\'u}{\~n}ez}, {Gonz{\'a}lez-Vidal}, {Granvik}, {Guerrier}, {Guillout}, {Guiraud}, {G{\'u}rpide}, {Guti{\'e}rrez-S{\'a}nchez}, {Guy}, {Haigron}, {Hatzidimitriou}, {Haywood}, {Heiter}, {Helmi}, {Hobbs}, {Hofmann}, {Holl}, {Holland}, {Hunt}, {Hypki}, {Icardi}, {Irwin}, {Jevardat de Fombelle}, {Jofr{\'e}}, {Jonker}, {Jorissen}, {Julbe}, {Karampelas}, {Kochoska}, {Kohley}, {Kolenberg}, {Kontizas}, {Koposov}, {Kordopatis}, {Koubsky}, {Kowalczyk}, {Krone-Martins}, {Kudryashova}, {Kull}, {Bachchan}, {Lacoste-Seris}, {Lanza}, {Lavigne}, {Le Poncin-Lafitte}, {Lebreton}, {Lebzelter}, {Leccia}, {Leclerc}, {Lecoeur-Taibi}, {Lemaitre}, {Lenhardt}, {Leroux}, {Liao}, {Licata}, {Lindstr{\o}m}, {Lister}, {Livanou}, {Lobel}, {L{\"o}ffler},
  {L{\'o}pez}, {Lopez-Lozano}, {Lorenz}, {Loureiro}, {MacDonald}, {Magalh{\~a}es Fernandes}, {Managau}, {Mann}, {Mantelet}, {Marchal}, {Marchant}, {Marconi}, {Marie}, {Marinoni}, {Marrese}, {Marschalk{\'o}}, {Marshall}, {Mart{\'\i}n-Fleitas}, {Martino}, {Mary}, {Matijevi{\v{c}}}, {Mazeh}, {McMillan}, {Messina}, {Mestre}, {Michalik}, {Millar}, {Miranda}, {Molina}, {Molinaro}, {Molinaro}, {Moln{\'a}r}, {Moniez}, {Montegriffo}, {Monteiro}, {Mor}, {Mora}, {Morbidelli}, {Morel}, {Morgenthaler}, {Morley}, {Morris}, {Mulone}, {Muraveva}, {Musella}, {Narbonne}, {Nelemans}, {Nicastro}, {Noval}, {Ord{\'e}novic}, {Ordieres-Mer{\'e}}, {Osborne}, {Pagani}, {Pagano}, {Pailler}, {Palacin}, {Palaversa}, {Parsons}, {Paulsen}, {Pecoraro}, {Pedrosa}, {Pentik{\"a}inen}, {Pereira}, {Pichon}, {Piersimoni}, {Pineau}, {Plachy}, {Plum}, {Poujoulet}, {Pr{\v{s}}a}, {Pulone}, {Ragaini}, {Rago}, {Rambaux}, {Ramos-Lerate}, {Ranalli}, {Rauw}, {Read}, {Regibo}, {Renk}, {Reyl{\'e}}, {Ribeiro}, {Rimoldini}, {Ripepi}, {Riva}, {Rixon},
  {Roelens}, {Romero-G{\'o}mez}, {Rowell}, {Royer}, {Rudolph}, {Ruiz-Dern}, {Sadowski}, {Sagrist{\`a} Sell{\'e}s}, {Sahlmann}, {Salgado}, {Salguero}, {Sarasso}, {Savietto}, {Schnorhk}, {Schultheis}, {Sciacca}, {Segol}, {Segovia}, {Segransan}, {Serpell}, {Shih}, {Smareglia}, {Smart}, {Smith}, {Solano}, {Solitro}, {Sordo}, {Soria Nieto}, {Souchay}, {Spagna}, {Spoto}, {Stampa}, {Steele}, {Steidelm{\"u}ller}, {Stephenson}, {Stoev}, {Suess}, {S{\"u}veges}, {Surdej}, {Szabados}, {Szegedi-Elek}, {Tapiador}, {Taris}, {Tauran}, {Taylor}, {Teixeira}, {Terrett}, {Tingley}, {Trager}, {Turon}, {Ulla}, {Utrilla}, {Valentini}, {van Elteren}, {Van Hemelryck}, {van Leeuwen}, {Varadi}, {Vecchiato}, {Veljanoski}, {Via}, {Vicente}, {Vogt}, {Voss}, {Votruba}, {Voutsinas}, {Walmsley}, {Weiler}, {Weingrill}, {Werner}, {Wevers}, {Whitehead}, {Wyrzykowski}, {Yoldas}, {{\v{Z}}erjal}, {Zucker}, {Zurbach}, {Zwitter}, {Alecu}, {Allen}, {Allende Prieto}, {Amorim}, {Anglada-Escud{\'e}}, {Arsenijevic}, {Azaz}, {Balm}, {Beck}, {Bernstein},
  {Bigot}, {Bijaoui}, {Blasco}, {Bonfigli}, {Bono}, {Boudreault}, {Bressan}, {Brown}, {Brunet}, {Bunclark}, {Buonanno}, {Butkevich}, {Carret}, {Carrion}, {Chemin}, {Ch{\'e}reau}, {Corcione}, {Darmigny}, {de Boer}, {de Teodoro}, {de Zeeuw}, {Delle Luche}, {Domingues}, {Dubath}, {Fodor}, {Fr{\'e}zouls}, {Fries}, {Fustes}, {Fyfe}, {Gallardo}, {Gallegos}, {Gardiol}, {Gebran}, {Gomboc}, {G{\'o}mez}, {Grux}, {Gueguen}, {Heyrovsky}, {Hoar}, {Iannicola}, {Isasi Parache}, {Janotto}, {Joliet}, {Jonckheere}, {Keil}, {Kim}, {Klagyivik}, {Klar}, {Knude}, {Kochukhov}, {Kolka}, {Kos}, {Kutka}, {Lainey}, {LeBouquin}, {Liu}, {Loreggia}, {Makarov}, {Marseille}, {Martayan}, {Martinez-Rubi}, {Massart}, {Meynadier}, {Mignot}, {Munari}, {Nguyen}, {Nordlander}, {Ocvirk}, {O'Flaherty}, {Olias Sanz}, {Ortiz}, {Osorio}, {Oszkiewicz}, {Ouzounis}, {Palmer}, {Park}, {Pasquato}, {Peltzer}, {Peralta}, {P{\'e}turaud}, {Pieniluoma}, {Pigozzi}, {Poels}, {Prat}, {Prod'homme}, {Raison}, {Rebordao}, {Risquez}, {Rocca-Volmerange}, {Rosen},
  {Ruiz-Fuertes}, {Russo}, {Sembay}, {Serraller Vizcaino}, {Short}, {Siebert}, {Silva}, {Sinachopoulos}, {Slezak}, {Soffel}, {Sosnowska}, {Strai{\v{z}}ys}, {ter Linden}, {Terrell}, {Theil}, {Tiede}, {Troisi}, {Tsalmantza}, {Tur}, {Vaccari}, {Vachier}, {Valles}, {Van Hamme}, {Veltz}, {Virtanen}, {Wallut}, {Wichmann}, {Wilkinson}, {Ziaeepour}, \& {Zschocke}}]{GaiaCollab+2016}
{Gaia Collaboration}, {Prusti}, T., {de Bruijne}, J.~H.~J., {et~al.} 2016, \aap, 595, A1

\bibitem[{{Gaia Collaboration} {et~al.}(2023){Gaia Collaboration}, {Vallenari}, {Brown}, {Prusti}, {de Bruijne}, {Arenou}, {Babusiaux}, {Biermann}, {Creevey}, {Ducourant}, {Evans}, {Eyer}, {Guerra}, {Hutton}, {Jordi}, {Klioner}, {Lammers}, {Lindegren}, {Luri}, {Mignard}, {Panem}, {Pourbaix}, {Randich}, {Sartoretti}, {Soubiran}, {Tanga}, {Walton}, {Bailer-Jones}, {Bastian}, {Drimmel}, {Jansen}, {Katz}, {Lattanzi}, {van Leeuwen}, {Bakker}, {Cacciari}, {Casta{\~n}eda}, {De Angeli}, {Fabricius}, {Fouesneau}, {Fr{\'e}mat}, {Galluccio}, {Guerrier}, {Heiter}, {Masana}, {Messineo}, {Mowlavi}, {Nicolas}, {Nienartowicz}, {Pailler}, {Panuzzo}, {Riclet}, {Roux}, {Seabroke}, {Sordo}, {Th{\'e}venin}, {Gracia-Abril}, {Portell}, {Teyssier}, {Altmann}, {Andrae}, {Audard}, {Bellas-Velidis}, {Benson}, {Berthier}, {Blomme}, {Burgess}, {Busonero}, {Busso}, {C{\'a}novas}, {Carry}, {Cellino}, {Cheek}, {Clementini}, {Damerdji}, {Davidson}, {de Teodoro}, {Nu{\~n}ez Campos}, {Delchambre}, {Dell'Oro}, {Esquej},
  {Fern{\'a}ndez-Hern{\'a}ndez}, {Fraile}, {Garabato}, {Garc{\'\i}a-Lario}, {Gosset}, {Haigron}, {Halbwachs}, {Hambly}, {Harrison}, {Hern{\'a}ndez}, {Hestroffer}, {Hodgkin}, {Holl}, {Jan{\ss}en}, {Jevardat de Fombelle}, {Jordan}, {Krone-Martins}, {Lanzafame}, {L{\"o}ffler}, {Marchal}, {Marrese}, {Moitinho}, {Muinonen}, {Osborne}, {Pancino}, {Pauwels}, {Recio-Blanco}, {Reyl{\'e}}, {Riello}, {Rimoldini}, {Roegiers}, {Rybizki}, {Sarro}, {Siopis}, {Smith}, {Sozzetti}, {Utrilla}, {van Leeuwen}, {Abbas}, {{\'A}brah{\'a}m}, {Abreu Aramburu}, {Aerts}, {Aguado}, {Ajaj}, {Aldea-Montero}, {Altavilla}, {{\'A}lvarez}, {Alves}, {Anders}, {Anderson}, {Anglada Varela}, {Antoja}, {Baines}, {Baker}, {Balaguer-N{\'u}{\~n}ez}, {Balbinot}, {Balog}, {Barache}, {Barbato}, {Barros}, {Barstow}, {Bartolom{\'e}}, {Bassilana}, {Bauchet}, {Becciani}, {Bellazzini}, {Berihuete}, {Bernet}, {Bertone}, {Bianchi}, {Binnenfeld}, {Blanco-Cuaresma}, {Blazere}, {Boch}, {Bombrun}, {Bossini}, {Bouquillon}, {Bragaglia}, {Bramante}, {Breedt},
  {Bressan}, {Brouillet}, {Brugaletta}, {Bucciarelli}, {Burlacu}, {Butkevich}, {Buzzi}, {Caffau}, {Cancelliere}, {Cantat-Gaudin}, {Carballo}, {Carlucci}, {Carnerero}, {Carrasco}, {Casamiquela}, {Castellani}, {Castro-Ginard}, {Chaoul}, {Charlot}, {Chemin}, {Chiaramida}, {Chiavassa}, {Chornay}, {Comoretto}, {Contursi}, {Cooper}, {Cornez}, {Cowell}, {Crifo}, {Cropper}, {Crosta}, {Crowley}, {Dafonte}, {Dapergolas}, {David}, {David}, {de Laverny}, {De Luise}, {De March}, {De Ridder}, {de Souza}, {de Torres}, {del Peloso}, {del Pozo}, {Delbo}, {Delgado}, {Delisle}, {Demouchy}, {Dharmawardena}, {Di Matteo}, {Diakite}, {Diener}, {Distefano}, {Dolding}, {Edvardsson}, {Enke}, {Fabre}, {Fabrizio}, {Faigler}, {Fedorets}, {Fernique}, {Fienga}, {Figueras}, {Fournier}, {Fouron}, {Fragkoudi}, {Gai}, {Garcia-Gutierrez}, {Garcia-Reinaldos}, {Garc{\'\i}a-Torres}, {Garofalo}, {Gavel}, {Gavras}, {Gerlach}, {Geyer}, {Giacobbe}, {Gilmore}, {Girona}, {Giuffrida}, {Gomel}, {Gomez}, {Gonz{\'a}lez-N{\'u}{\~n}ez},
  {Gonz{\'a}lez-Santamar{\'\i}a}, {Gonz{\'a}lez-Vidal}, {Granvik}, {Guillout}, {Guiraud}, {Guti{\'e}rrez-S{\'a}nchez}, {Guy}, {Hatzidimitriou}, {Hauser}, {Haywood}, {Helmer}, {Helmi}, {Sarmiento}, {Hidalgo}, {Hilger}, {H{\l}adczuk}, {Hobbs}, {Holland}, {Huckle}, {Jardine}, {Jasniewicz}, {Jean-Antoine Piccolo}, {Jim{\'e}nez-Arranz}, {Jorissen}, {Juaristi Campillo}, {Julbe}, {Karbevska}, {Kervella}, {Khanna}, {Kontizas}, {Kordopatis}, {Korn}, {K{\'o}sp{\'a}l}, {Kostrzewa-Rutkowska}, {Kruszy{\'n}ska}, {Kun}, {Laizeau}, {Lambert}, {Lanza}, {Lasne}, {Le Campion}, {Lebreton}, {Lebzelter}, {Leccia}, {Leclerc}, {Lecoeur-Taibi}, {Liao}, {Licata}, {Lindstr{\o}m}, {Lister}, {Livanou}, {Lobel}, {Lorca}, {Loup}, {Madrero Pardo}, {Magdaleno Romeo}, {Managau}, {Mann}, {Manteiga}, {Marchant}, {Marconi}, {Marcos}, {Marcos Santos}, {Mar{\'\i}n Pina}, {Marinoni}, {Marocco}, {Marshall}, {Martin Polo}, {Mart{\'\i}n-Fleitas}, {Marton}, {Mary}, {Masip}, {Massari}, {Mastrobuono-Battisti}, {Mazeh}, {McMillan}, {Messina}, {Michalik},
  {Millar}, {Mints}, {Molina}, {Molinaro}, {Moln{\'a}r}, {Monari}, {Mongui{\'o}}, {Montegriffo}, {Montero}, {Mor}, {Mora}, {Morbidelli}, {Morel}, {Morris}, {Muraveva}, {Murphy}, {Musella}, {Nagy}, {Noval}, {Oca{\~n}a}, {Ogden}, {Ordenovic}, {Osinde}, {Pagani}, {Pagano}, {Palaversa}, {Palicio}, {Pallas-Quintela}, {Panahi}, {Payne-Wardenaar}, {Pe{\~n}alosa Esteller}, {Penttil{\"a}}, {Pichon}, {Piersimoni}, {Pineau}, {Plachy}, {Plum}, {Poggio}, {Pr{\v{s}}a}, {Pulone}, {Racero}, {Ragaini}, {Rainer}, {Raiteri}, {Rambaux}, {Ramos}, {Ramos-Lerate}, {Re Fiorentin}, {Regibo}, {Richards}, {Rios Diaz}, {Ripepi}, {Riva}, {Rix}, {Rixon}, {Robichon}, {Robin}, {Robin}, {Roelens}, {Rogues}, {Rohrbasser}, {Romero-G{\'o}mez}, {Rowell}, {Royer}, {Ruz Mieres}, {Rybicki}, {Sadowski}, {S{\'a}ez N{\'u}{\~n}ez}, {Sagrist{\`a} Sell{\'e}s}, {Sahlmann}, {Salguero}, {Samaras}, {Sanchez Gimenez}, {Sanna}, {Santove{\~n}a}, {Sarasso}, {Schultheis}, {Sciacca}, {Segol}, {Segovia}, {S{\'e}gransan}, {Semeux}, {Shahaf}, {Siddiqui}, {Siebert},
  {Siltala}, {Silvelo}, {Slezak}, {Slezak}, {Smart}, {Snaith}, {Solano}, {Solitro}, {Souami}, {Souchay}, {Spagna}, {Spina}, {Spoto}, {Steele}, {Steidelm{\"u}ller}, {Stephenson}, {S{\"u}veges}, {Surdej}, {Szabados}, {Szegedi-Elek}, {Taris}, {Taylor}, {Teixeira}, {Tolomei}, {Tonello}, {Torra}, {Torra}, {Torralba Elipe}, {Trabucchi}, {Tsounis}, {Turon}, {Ulla}, {Unger}, {Vaillant}, {van Dillen}, {van Reeven}, {Vanel}, {Vecchiato}, {Viala}, {Vicente}, {Voutsinas}, {Weiler}, {Wevers}, {Wyrzykowski}, {Yoldas}, {Yvard}, {Zhao}, {Zorec}, {Zucker}, \& {Zwitter}}]{GaiaCollab+2023}
{Gaia Collaboration}, {Vallenari}, A., {Brown}, A.~G.~A., {et~al.} 2023, \aap, 674, A1

\bibitem[{{Gonz{\'a}lez-Tor{\`a}} {et~al.}(2025){Gonz{\'a}lez-Tor{\`a}}, {Sander}, {Sundqvist}, {Debnath}, {Delbroek}, {Josiek}, {Lefever}, {Moens}, {Van der Sijpt}, \& {Verhamme}}]{Gonzalez-Tora+2025}
{Gonz{\'a}lez-Tor{\`a}}, G., {Sander}, A.~A.~C., {Sundqvist}, J.~O., {et~al.} 2025, \aap, 694, A269

\bibitem[{{Gr{\"a}fener} \& {Hamann}(2005)}]{Graefener-Hamann2005}
{Gr{\"a}fener}, G. \& {Hamann}, W.~R. 2005, \aap, 432, 633

\bibitem[{{Gr{\"a}fener} {et~al.}(2002){Gr{\"a}fener}, {Koesterke}, \& {Hamann}}]{Graefener+2002}
{Gr{\"a}fener}, G., {Koesterke}, L., \& {Hamann}, W.~R. 2002, \aap, 387, 244

\bibitem[{{Gr{\"a}fener} \& {Vink}(2013)}]{Grafener-Vink2013}
{Gr{\"a}fener}, G. \& {Vink}, J.~S. 2013, \aap, 560, A6

\bibitem[{{Hamann} {et~al.}(2019){Hamann}, {Gr{\"a}fener}, {Liermann}, {Hainich}, {Sander}, {Shenar}, {Ramachandran}, {Todt}, \& {Oskinova}}]{Hamann+2019}
{Hamann}, W.~R., {Gr{\"a}fener}, G., {Liermann}, A., {et~al.} 2019, \aap, 625, A57

\bibitem[{{Haucke} {et~al.}(2018){Haucke}, {Cidale}, {Venero}, {Cur{\'e}}, {Kraus}, {Kanaan}, \& {Arcos}}]{Haucke+2018}
{Haucke}, M., {Cidale}, L.~S., {Venero}, R.~O.~J., {et~al.} 2018, \aap, 614, A91

\bibitem[{{Hillier} {et~al.}(2001){Hillier}, {Davidson}, {Ishibashi}, \& {Gull}}]{Hillier+2001}
{Hillier}, D.~J., {Davidson}, K., {Ishibashi}, K., \& {Gull}, T. 2001, \apj, 553, 837

\bibitem[{{Hillier} {et~al.}(2003){Hillier}, {Lanz}, {Heap}, {Hubeny}, {Smith}, {Evans}, {Lennon}, \& {Bouret}}]{Hillier+2003}
{Hillier}, D.~J., {Lanz}, T., {Heap}, S.~R., {et~al.} 2003, \apj, 588, 1039

\bibitem[{{Hillier} \& {Miller}(1998)}]{Hillier-Miller1998}
{Hillier}, D.~J. \& {Miller}, D.~L. 1998, \apj, 496, 407

\bibitem[{{Humphreys} \& {Davidson}(1979)}]{Humphreys-Davidson1979}
{Humphreys}, R.~M. \& {Davidson}, K. 1979, \apj, 232, 409

\bibitem[{{Kostenkov} {et~al.}(2020){Kostenkov}, {Fabrika}, {Sholukhova}, {Sarkisyan}, \& {Bizyaev}}]{Kostenkov+2020}
{Kostenkov}, A., {Fabrika}, S., {Sholukhova}, O., {Sarkisyan}, A., \& {Bizyaev}, D. 2020, \mnras, 496, 5455

\bibitem[{{Krti{\v{c}}ka} \& {Kub{\'a}t}(2016)}]{Krticka-Kubat2016}
{Krti{\v{c}}ka}, J. \& {Kub{\'a}t}, J. 2016, Advances in Space Research, 58, 710

\bibitem[{{Krti{\v{c}}ka} \& {Kub{\'a}t}(2017)}]{Krticka-Kubat2017}
{Krti{\v{c}}ka}, J. \& {Kub{\'a}t}, J. 2017, \aap, 606, A31

\bibitem[{{Krti{\v{c}}ka} {et~al.}(2021){Krti{\v{c}}ka}, {Kub{\'a}t}, \& {Krti{\v{c}}kov{\'a}}}]{Krticka+2021}
{Krti{\v{c}}ka}, J., {Kub{\'a}t}, J., \& {Krti{\v{c}}kov{\'a}}, I. 2021, \aap, 647, A28

\bibitem[{{Krti{\v{c}}ka} {et~al.}(2024){Krti{\v{c}}ka}, {Kub{\'a}t}, \& {Krti{\v{c}}kov{\'a}}}]{Krticka+2024}
{Krti{\v{c}}ka}, J., {Kub{\'a}t}, J., \& {Krti{\v{c}}kov{\'a}}, I. 2024, \aap, 681, A29

\bibitem[{{Lamers} {et~al.}(1995){Lamers}, {Snow}, \& {Lindholm}}]{Lamers+1995}
{Lamers}, H. J.~G.~L.~M., {Snow}, T.~P., \& {Lindholm}, D.~M. 1995, \apj, 455, 269

\bibitem[{{Lefever} {et~al.}(2023){Lefever}, {Sander}, {Shenar}, {Poniatowski}, {Dsilva}, \& {Todt}}]{Lefever+2023}
{Lefever}, R.~R., {Sander}, A.~A.~C., {Shenar}, T., {et~al.} 2023, \mnras, 521, 1374

\bibitem[{{Leitherer} \& {Robert}(1991)}]{Leitherer-Robert1991}
{Leitherer}, C. \& {Robert}, C. 1991, \apj, 377, 629

\bibitem[{{Lennon} {et~al.}(1992){Lennon}, {Dufton}, \& {Fitzsimmons}}]{Lennon+1992}
{Lennon}, D.~J., {Dufton}, P.~L., \& {Fitzsimmons}, A. 1992, \aaps, 94, 569

\bibitem[{{Maeder} \& {Meynet}(2000)}]{Maeder-Meynet2000}
{Maeder}, A. \& {Meynet}, G. 2000, \aap, 361, 159

\bibitem[{{Mahy} {et~al.}(2016){Mahy}, {Hutsem{\'e}kers}, {Royer}, \& {Waelkens}}]{Mahy+2016}
{Mahy}, L., {Hutsem{\'e}kers}, D., {Royer}, P., \& {Waelkens}, C. 2016, \aap, 594, A94

\bibitem[{{Markova} \& {Puls}(2008)}]{Markova_Puls2008}
{Markova}, N. \& {Puls}, J. 2008, \aap, 478, 823

\bibitem[{{Maryeva} {et~al.}(2022){Maryeva}, {Karpov}, {Kniazev}, \& {Gvaramadze}}]{Maryeva+2022}
{Maryeva}, O.~V., {Karpov}, S.~V., {Kniazev}, A.~Y., \& {Gvaramadze}, V.~V. 2022, \mnras, 513, 5752

\bibitem[{{Moens} {et~al.}(2022){Moens}, {Poniatowski}, {Hennicker}, {Sundqvist}, {El Mellah}, \& {Kee}}]{Moens+2022}
{Moens}, N., {Poniatowski}, L.~G., {Hennicker}, L., {et~al.} 2022, \aap, 665, A42

\bibitem[{{Najarro} {et~al.}(2009){Najarro}, {Figer}, {Hillier}, {Geballe}, \& {Kudritzki}}]{Najarro+2009}
{Najarro}, F., {Figer}, D.~F., {Hillier}, D.~J., {Geballe}, T.~R., \& {Kudritzki}, R.~P. 2009, \apj, 691, 1816

\bibitem[{{Najarro} {et~al.}(2011){Najarro}, {Hanson}, \& {Puls}}]{Najarro+2011}
{Najarro}, F., {Hanson}, M.~M., \& {Puls}, J. 2011, \aap, 535, A32

\bibitem[{{Najarro} {et~al.}(1996){Najarro}, {Kudritzki}, {Cassinelli}, {Stahl}, \& {Hillier}}]{Najarro+1996}
{Najarro}, F., {Kudritzki}, R.~P., {Cassinelli}, J.~P., {Stahl}, O., \& {Hillier}, D.~J. 1996, \aap, 306, 892

\bibitem[{{Odegard} \& {Cassinelli}(1982)}]{Odegard-Cassinelli1982}
{Odegard}, N. \& {Cassinelli}, J.~P. 1982, \apj, 256, 568

\bibitem[{{Oskinova} {et~al.}(2007){Oskinova}, {Hamann}, \& {Feldmeier}}]{Oskinova+2007}
{Oskinova}, L.~M., {Hamann}, W.~R., \& {Feldmeier}, A. 2007, \aap, 476, 1331

\bibitem[{{Pauldrach} {et~al.}(1986){Pauldrach}, {Puls}, \& {Kudritzki}}]{Pauldrach+1986}
{Pauldrach}, A., {Puls}, J., \& {Kudritzki}, R.~P. 1986, \aap, 164, 86

\bibitem[{{Pauldrach} \& {Puls}(1990)}]{Pauldrach_Puls1990}
{Pauldrach}, A.~W.~A. \& {Puls}, J. 1990, \aap, 237, 409

\bibitem[{{Paunzen}(2022)}]{Paunzen+2022}
{Paunzen}, E. 2022, \aap, 661, A89

\bibitem[{{Petrov} {et~al.}(2016){Petrov}, {Vink}, \& {Gr{\"a}fener}}]{Petrov+2016}
{Petrov}, B., {Vink}, J.~S., \& {Gr{\"a}fener}, G. 2016, \mnras, 458, 1999

\bibitem[{{Poniatowski} {et~al.}(2021){Poniatowski}, {Sundqvist}, {Kee}, {Owocki}, {Marchant}, {Decin}, {de Koter}, {Mahy}, \& {Sana}}]{Poniatowski+2021}
{Poniatowski}, L.~G., {Sundqvist}, J.~O., {Kee}, N.~D., {et~al.} 2021, \aap, 647, A151

\bibitem[{{Rubio-D{\'\i}ez} {et~al.}(2022){Rubio-D{\'\i}ez}, {Sundqvist}, {Najarro}, {Traficante}, {Puls}, {Calzoletti}, \& {Figer}}]{RubioDiez+2022}
{Rubio-D{\'\i}ez}, M.~M., {Sundqvist}, J.~O., {Najarro}, F., {et~al.} 2022, \aap, 658, A61

\bibitem[{{Sabhahit} {et~al.}(2022){Sabhahit}, {Vink}, {Higgins}, \& {Sander}}]{Sabhahit+2022}
{Sabhahit}, G.~N., {Vink}, J.~S., {Higgins}, E.~R., \& {Sander}, A. A.~C. 2022, \mnras, 514, 3736

\bibitem[{{Sabhahit} {et~al.}(2025){Sabhahit}, {Vink}, {Sander}, {Bernini-Peron}, {Crowther}, {Lefever}, \& {Shenar}}]{Sabhahit+2025}
{Sabhahit}, G.~N., {Vink}, J.~S., {Sander}, A. A.~C., {et~al.} 2025, arXiv e-prints, arXiv:2502.14957

\bibitem[{{Sabhahit} {et~al.}(2023){Sabhahit}, {Vink}, {Sander}, \& {Higgins}}]{Sabhahit+2023}
{Sabhahit}, G.~N., {Vink}, J.~S., {Sander}, A. A.~C., \& {Higgins}, E.~R. 2023, \mnras, 524, 1529

\bibitem[{{Sander} {et~al.}(2015){Sander}, {Shenar}, {Hainich}, {G{\'\i}menez-Garc{\'\i}a}, {Todt}, \& {Hamann}}]{Sander+2015}
{Sander}, A., {Shenar}, T., {Hainich}, R., {et~al.} 2015, \aap, 577, A13

\bibitem[{{Sander} {et~al.}(2024){Sander}, {Bouret}, {Bernini-Peron}, {Puls}, {Backs}, {Berlanas}, {Bestenlehner}, {Brands}, {Herrero}, {Martins}, {Maryeva}, {Pauli}, {Ramachandran}, {Crowther}, {G{\'o}mez-Gonz{\'a}lez}, {Gormaz-Matamala}, {Hamann}, {Hillier}, {Kuiper}, {Larkin}, {Lefever}, {Mehner}, {Najarro}, {Oskinova}, {Sch{\"o}sser}, {Shenar}, {Todt}, {ud-Doula}, \& {Vink}}]{Sander+2024}
{Sander}, A.~A.~C., {Bouret}, J.~C., {Bernini-Peron}, M., {et~al.} 2024, \aap, 689, A30

\bibitem[{{Sander} {et~al.}(2018){Sander}, {F{\"u}rst}, {Kretschmar}, {Oskinova}, {Todt}, {Hainich}, {Shenar}, \& {Hamann}}]{Sander+2018}
{Sander}, A.~A.~C., {F{\"u}rst}, F., {Kretschmar}, P., {et~al.} 2018, \aap, 610, A60

\bibitem[{{Sander} {et~al.}(2017){Sander}, {Hamann}, {Todt}, {Hainich}, \& {Shenar}}]{Sander+2017}
{Sander}, A.~A.~C., {Hamann}, W.~R., {Todt}, H., {Hainich}, R., \& {Shenar}, T. 2017, \aap, 603, A86

\bibitem[{{Sander} {et~al.}(2019){Sander}, {Hamann}, {Todt}, {Hainich}, {Shenar}, {Ramachandran}, \& {Oskinova}}]{Sander+2019}
{Sander}, A.~A.~C., {Hamann}, W.~R., {Todt}, H., {et~al.} 2019, \aap, 621, A92

\bibitem[{{Sander} {et~al.}(2023){Sander}, {Lefever}, {Poniatowski}, {Ramachandran}, {Sabhahit}, \& {Vink}}]{Sander+2023}
{Sander}, A.~A.~C., {Lefever}, R.~R., {Poniatowski}, L.~G., {et~al.} 2023, \aap, 670, A83

\bibitem[{{Sander} \& {Vink}(2020)}]{Sander+2020}
{Sander}, A. A.~C. \& {Vink}, J.~S. 2020, \mnras, 499, 873

\bibitem[{{Shenar} {et~al.}(2024){Shenar}, {Bodensteiner}, {Sana}, {Crowther}, {Lennon}, {Abdul-Masih}, {Almeida}, {Backs}, {Berlanas}, {Bernini-Peron}, {Bestenlehner}, {Bowman}, {Bronner}, {Britavskiy}, {de Koter}, {de Mink}, {Deshmukh}, {Evans}, {Fabry}, {Gieles}, {Gilkis}, {Gonz{\'a}lez-Tor{\`a}}, {Gr{\"a}fener}, {G{\"o}tberg}, {Hawcroft}, {H{\'e}nault-Brunet}, {Herrero}, {Holgado}, {Janssens}, {Johnston}, {Josiek}, {Justham}, {Kalari}, {Katabi}, {Keszthelyi}, {Klencki}, {Kub{\'a}t}, {Kub{\'a}tov{\'a}}, {Langer}, {Lefever}, {Ludwig}, {Mackey}, {Mahy}, {Ma{\'\i}z Apell{\'a}niz}, {Mandel}, {Maravelias}, {Marchant}, {Menon}, {Najarro}, {Oskinova}, {O'Grady}, {Ovadia}, {Patrick}, {Pauli}, {Pawlak}, {Ramachandran}, {Renzo}, {Rocha}, {Sander}, {Sayada}, {Schneider}, {Schootemeijer}, {Sch{\"o}sser}, {Sch{\"u}rmann}, {Sen}, {Shahaf}, {Sim{\'o}n-D{\'\i}az}, {Stoop}, {Toonen}, {Tramper}, {van Loon}, {Valli}, {van Son}, {Vigna-G{\'o}mez}, {Villase{\~n}or}, {Vink}, {Wang}, \& {Willcox}}]{Shenar+2024}
{Shenar}, T., {Bodensteiner}, J., {Sana}, H., {et~al.} 2024, \aap, 690, A289

\bibitem[{{Sloan} {et~al.}(2003){Sloan}, {Kraemer}, {Price}, \& {Shipman}}]{Sloan+2003}
{Sloan}, G.~C., {Kraemer}, K.~E., {Price}, S.~D., \& {Shipman}, R.~F. 2003, \apjs, 147, 379

\bibitem[{{Sundqvist} {et~al.}(2019){Sundqvist}, {Bj{\"o}rklund}, {Puls}, \& {Najarro}}]{Sundqvist+2019}
{Sundqvist}, J.~O., {Bj{\"o}rklund}, R., {Puls}, J., \& {Najarro}, F. 2019, \aap, 632, A126

\bibitem[{{Sundqvist} \& {Puls}(2018)}]{SundqvistPuls2018}
{Sundqvist}, J.~O. \& {Puls}, J. 2018, \aap, 619, A59

\bibitem[{{Urbaneja} {et~al.}(2005){Urbaneja}, {Herrero}, {Kudritzki}, {Najarro}, {Smartt}, {Puls}, {Lennon}, \& {Corral}}]{Urbaneja+2005}
{Urbaneja}, M.~A., {Herrero}, A., {Kudritzki}, R.~P., {et~al.} 2005, \apj, 635, 311

\bibitem[{{van Genderen} {et~al.}(1982){van Genderen}, {van Leeuwen}, \& {Brand}}]{vanGenderen+1982}
{van Genderen}, A.~M., {van Leeuwen}, F., \& {Brand}, J. 1982, \aaps, 47, 591

\bibitem[{{Venero} {et~al.}(2016){Venero}, {Cur{\'e}}, {Cidale}, \& {Araya}}]{Venero+2016}
{Venero}, R.~O.~J., {Cur{\'e}}, M., {Cidale}, L.~S., \& {Araya}, I. 2016, \apj, 822, 28

\bibitem[{{Venero} {et~al.}(2024){Venero}, {Cur{\'e}}, {Puls}, {Cidale}, {Haucke}, {Araya}, {Gormaz-Matamala}, \& {Arcos}}]{Venero+2024}
{Venero}, R.~O.~J., {Cur{\'e}}, M., {Puls}, J., {et~al.} 2024, \mnras, 527, 93

\bibitem[{{Verhamme} {et~al.}(2024){Verhamme}, {Sundqvist}, {de Koter}, {Sana}, {Backs}, {Brands}, {Najarro}, {Puls}, {Vink}, {Crowther}, {Kub{\'a}tov{\'a}}, {Sander}, {Bernini-Peron}, {Kuiper}, {Prinja}, {Schillemans}, {Shenar}, {van Loon}, \& {XShootu collaboration}}]{Verhamme+2024}
{Verhamme}, O., {Sundqvist}, J., {de Koter}, A., {et~al.} 2024, \aap, 692, A91

\bibitem[{{Vink}(2018)}]{Vink2018}
{Vink}, J.~S. 2018, \aap, 619, A54

\bibitem[{{Vink} {et~al.}(1999){Vink}, {de Koter}, \& {Lamers}}]{Vink+1999}
{Vink}, J.~S., {de Koter}, A., \& {Lamers}, H.~J.~G.~L.~M. 1999, \aap, 350, 181

\bibitem[{{Vink} {et~al.}(2001){Vink}, {de Koter}, \& {Lamers}}]{Vink+2001}
{Vink}, J.~S., {de Koter}, A., \& {Lamers}, H.~J.~G.~L.~M. 2001, \aap, 369, 574

\bibitem[{{Vink} \& {Gr{\"a}fener}(2012)}]{Vink+2012}
{Vink}, J.~S. \& {Gr{\"a}fener}, G. 2012, \apjl, 751, L34

\bibitem[{{Vink} {et~al.}(2023){Vink}, {Mehner}, {Crowther}, {Fullerton}, {Garcia}, {Martins}, {Morrell}, {Oskinova}, {St-Louis}, {ud-Doula}, {Sander}, {Sana}, {Bouret}, {Kub{\'a}tov{\'a}}, {Marchant}, {Martins}, {Wofford}, {van Loon}, {Grace Telford}, {G{\"o}tberg}, {Bowman}, {Erba}, {Kalari}, {Abdul-Masih}, {Alkousa}, {Backs}, {Barbosa}, {Berlanas}, {Bernini-Peron}, {Bestenlehner}, {Blomme}, {Bodensteiner}, {Brands}, {Evans}, {David-Uraz}, {Driessen}, {Dsilva}, {Geen}, {G{\'o}mez-Gonz{\'a}lez}, {Grassitelli}, {Hamann}, {Hawcroft}, {Herrero}, {Higgins}, {John Hillier}, {Ignace}, {Istrate}, {Kaper}, {Kee}, {Kehrig}, {Keszthelyi}, {Klencki}, {de Koter}, {Kuiper}, {Laplace}, {Larkin}, {Lefever}, {Leitherer}, {Lennon}, {Mahy}, {Ma{\'\i}z Apell{\'a}niz}, {Maravelias}, {Marcolino}, {McLeod}, {de Mink}, {Najarro}, {Oey}, {Parsons}, {Pauli}, {Pedersen}, {Prinja}, {Ramachandran}, {Ram{\'\i}rez-Tannus}, {Sabhahit}, {Schootemeijer}, {Reyero Serantes}, {Shenar}, {Stringfellow}, {Sudnik}, {Tramper}, \&
  {Wang}}]{Vink+2023}
{Vink}, J.~S., {Mehner}, A., {Crowther}, P.~A., {et~al.} 2023, \aap, 675, A154

\bibitem[{{Vink} \& {Sander}(2021)}]{Vink-Sander2021}
{Vink}, J.~S. \& {Sander}, A. A.~C. 2021, \mnras, 504, 2051

\bibitem[{{Walborn} {et~al.}(2016){Walborn}, {Morrell}, {Barb{\'a}}, \& {Sota}}]{Walborn+2016}
{Walborn}, N.~R., {Morrell}, N.~I., {Barb{\'a}}, R.~H., \& {Sota}, A. 2016, \aj, 151, 91

\bibitem[{{Walborn} \& {Nichols-Bohlin}(1987)}]{Walborn+1987}
{Walborn}, N.~R. \& {Nichols-Bohlin}, J. 1987, \pasp, 99, 40

\bibitem[{{Walborn} {et~al.}(2015){Walborn}, {Sana}, {Evans}, {Taylor}, {Sabbi}, {Barb{\'a}}, {Morrell}, {Ma{\'\i}z Apell{\'a}niz}, {Sota}, {Dufton}, {McEvoy}, {Clark}, {Markova}, \& {Ulaczyk}}]{Walborn+2015}
{Walborn}, N.~R., {Sana}, H., {Evans}, C.~J., {et~al.} 2015, \apj, 809, 109

\bibitem[{{Welch} {et~al.}(2022){Welch}, {Coe}, {Zackrisson}, {de Mink}, {Ravindranath}, {Anderson}, {Brammer}, {Bradley}, {Yoon}, {Kelly}, {Diego}, {Windhorst}, {Zitrin}, {Dimauro}, {Jim{\'e}nez-Teja}, {Abdurro'uf}, {Nonino}, {Acebron}, {Andrade-Santos}, {Avila}, {Bayliss}, {Ben{\'\i}tez}, {Broadhurst}, {Bhatawdekar}, {Brada{\v{c}}}, {Caminha}, {Chen}, {Eldridge}, {Farag}, {Florian}, {Frye}, {Fujimoto}, {Gomez}, {Henry}, {Hsiao}, {Hutchison}, {James}, {Joyce}, {Jung}, {Khullar}, {Larson}, {Mahler}, {Mandelker}, {McCandliss}, {Morishita}, {Newshore}, {Norman}, {O'Connor}, {Oesch}, {Oguri}, {Ouchi}, {Postman}, {Rigby}, {Ryan}, {Sharma}, {Sharon}, {Strait}, {Strolger}, {Timmes}, {Toft}, {Trenti}, {Vanzella}, \& {Vikaeus}}]{Welch+2022}
{Welch}, B., {Coe}, D., {Zackrisson}, E., {et~al.} 2022, \apjl, 940, L1

\bibitem[{{Zacharias} {et~al.}(2004){Zacharias}, {Monet}, {Levine}, {Urban}, {Gaume}, \& {Wycoff}}]{Zacharias+2005}
{Zacharias}, N., {Monet}, D.~G., {Levine}, S.~E., {et~al.} 2004, in American Astronomical Society Meeting Abstracts, Vol. 205, American Astronomical Society Meeting Abstracts, 48.15

\end{thebibliography}

\begin{appendix}
\section{ OB hypergiants in the Local Group}

In Table\,\ref{tab:all_BHGs} we provide a list of stars classified as BHGs (and O hypergiants) in the Local Group. In total, we counted 35 objects. The list also include objects that did not formally receive the ``Ia+'' but show the spectroscopic characteristics. In this table, we sought to exclude stars confirmed LBVs such as P\,Cyg and HD\,168607.

\begin{table}[]
\caption{\label{tab:all_BHGs}OB hypergiants in the local group.}
\centering
\begin{tabular}{lcr}
\hline
Star                          & Sp.\,type              & Reference       \\
\hline\hline
\multicolumn{3}{c}{Milky Way}             \\
\hline
HD105056                     & O9.7Iae          & {Wa2016}  \\
HD173010                     & O9.7Ia+/B0Ia+    & {Wa2016} \\
BP\,Cru                      & B1 Ia+           & {Cl2012}    \\
HD169454                     & B1 Ia+           & {Cl2012}    \\
$\zeta^1$\,Sco               & B1.5 Ia+         & {Cl2012}    \\
HD190603                     & B1.5 Ia+         & {Cl2012}    \\
HD80077                      & B2.5 Ia+         & {Cl2012}    \\
Cyg\,OB2\,\#12               & B3-4 Ia+         & {Cl2012}    \\
Wd1-5                        & WNL/B Ia+        & {Cl2012}    \\
Wd1-13                       & WNL/B Ia+        & {Cl2012}    \\
Wd1-7                        & B5 Ia+           & {Cl2012}    \\
Wd1-33                       & B5 Ia+           & {Cl2012}    \\
HD183143                     & B7 Iae           & {Cl2012}    \\
HD168625                     & B8 Ia+           & {Cl2012}    \\
HD199478                     & B9 Iae           & {Cl2012}    \\
Wd1-42a                      & B9 Ia+           & {Cl2012}    \\
\hline\hline
\multicolumn{3}{c}{M\,33 / Triangulum}          \\
\hline
HS80\,110A                     & B1 Ia+             & Ur2005 \\
\hline\hline
\multicolumn{3}{c}{LMC}                    \\
\hline
Sk -66 169                    & O9.7 Ia+         & FE1991  \\
SK -68 135                    & ON9.7 Ia+        & Vi2023           \\
Sk -69 279                    & O9.5 Ia+         & Be2025           \\
VFTS 430                      & B0.5 Ia+((n))Nwk & Ev2015           \\
VFTS 003                      & B1 Ia+            & Ev2015           \\
Sk -68 114                    & B1 Iae           & vG1982           \\
SK -69 224                     & B1 Ia+           & Vi2023           \\
Sk -67 2                      & B1.5 Ia+          & Be2025           \\
VFTS 533                      & B1.5 Ia+p Nwk    & Ev2015           \\
Sk -68 63                     & B2.5 Ieq         & vG1982           \\
Sk -68 8                      & B5 Ia+           & Be2025           \\
VFTS 458                      & B5 Ia+p          & Ev2015           \\
Sk -66 50                     & B7 Ia+            & Bl2025           \\
Sk -67 17                     & B9 Iae           & Vi2023           \\
VFTS 424                      & B9 Ia+p          & Ev2015           \\

\hline\hline
\multicolumn{3}{c}{SMC}                     \\
\hline
AzV 78                        & B1 Ia+           & Be2025           \\
AzV 393                       & B3 Ia+            & Be2025           \\
AzV 415                       & B8 Ie            & vG1982          \\
\hline\hline
\end{tabular}
\tablefoot{
Cl2012 stands for \cite{Clark+2012}, Ur2005 for \cite{Urbaneja+2005}, Be2025 for \cite{Bestenlehner+2025}, FE1991 for \cite{Fitzpatrick-Edward1991}, Wa2016 for \cite{Walborn+2016}, Ev2015 for \cite{Evans+2015}, vG1982 for \cite{vanGenderen+1982}, Vi2023 for \cite{Vink+2023},and Sh2024 for \cite{Shenar+2024}.
}
\end{table}

\section{Atomic data}

\begin{table}[]
\caption{\label{tab:atomic} Atoms and levels considered in the models.}
\begin{tabular}{lccc|lccr}
\hline\hline
Elem. & ion & levels & lines & Elem. & ion & levels & lines \\
\hline
H  & I   & 30 & 435  & Cl & III  & 1  & 0    \\
H  & II  & 1  & 0    & Cl & IV   & 24 & 276  \\
He & I   & 45 & 990  & Cl & V    & 18 & 153  \\
He & II  & 30 & 435  & Cl & VI   & 23 & 253  \\
He & III & 1  & 0    & Cl & VII  & 1  & 0    \\
N  & I   & 10 & 45   & Ar & I    & 75 & 2775 \\
N  & II  & 38 & 703  & Ar & II   & 99 & 4851 \\
N  & III & 85 & 3570 & Ar & III  & 30 & 435  \\
N  & IV  & 38 & 703  & Ar & IV   & 13 & 78   \\
N  & V   & 20 & 190  & Ar & V    & 10 & 45   \\
N  & VI  & 14 & 91   & Ar & VI   & 9  & 36   \\
C  & I   & 15 & 105  & Ar & VII  & 20 & 190  \\
C  & II  & 32 & 496  & Ar & VIII & 1  & 0    \\
C  & III & 40 & 780  & K  & I    & 62 & 1891 \\
C  & IV  & 25 & 300  & K  & II   & 20 & 190  \\
C  & V   & 29 & 406  & K  & III  & 20 & 190  \\
C  & VI  & 1  & 0    & K  & IV   & 23 & 253  \\
O  & I   & 13 & 78   & K  & V    & 19 & 171  \\
O  & II  & 37 & 666  & K  & VI   & 1  & 0    \\
O  & III & 33 & 528  & Ca & I    & 20 & 190  \\
O  & IV  & 29 & 406  & Ca & II   & 20 & 190  \\
O  & V   & 54 & 1431 & Ca & III  & 14 & 91   \\
O  & VI  & 35 & 595  & Ca & IV   & 24 & 276  \\
O  & VII & 15 & 105  & Ca & V    & 15 & 105  \\
S  & II  & 32 & 496  & Ca & VI   & 15 & 105  \\
S  & III & 23 & 253  & Ca & VII  & 20 & 190  \\
S  & IV  & 25 & 300  & Ca & VIII & 1  & 0    \\
S  & V   & 20 & 190  & Ne & I    & 30 & 435  \\
S  & VI  & 22 & 231  & Ne & II   & 42 & 861  \\
Mg & I   & 1  & 0    & Ne & III  & 18 & 153  \\
Mg & II  & 32 & 496  & Ne & IV   & 35 & 595  \\
Mg & III & 43 & 903  & Ne & V    & 20 & 190  \\
Mg & IV  & 17 & 136  & Ne & VI   & 20 & 190  \\
Mg & V   & 20 & 190  & Ne & VII  & 1  & 0    \\
Si & I   & 20 & 190  & Fe & I    & 13 & 40   \\
Si & II  & 20 & 190  & Fe & II   & 14 & 48   \\
Si & III & 24 & 276  & Fe & III  & 13 & 40   \\
Si & IV  & 55 & 1485 & Fe & IV   & 18 & 77   \\
Si & V   & 52 & 1326 & Fe & V    & 22 & 107  \\
P  & IV  & 12 & 66   & Fe & VI   & 29 & 194  \\
P  & V   & 11 & 55   & Fe & VII  & 19 & 87   \\
P  & VI  & 1  & 0    & Fe & VIII & 14 & 49   \\
Al & I   & 1  & 0    & Fe & IX   & 15 & 56   \\
Al & II  & 10 & 45   & Fe & X    & 1  & 0    \\
Al & III & 10 & 45   &    &      &    &      \\
Al & IV  & 1  & 0    &    &      &    &    \\
\hline
\end{tabular}
\end{table}

In Table\,\ref{tab:atomic} we list the elements and ions with the number of levels and line transitions considered in our atmosphere models. For Fe, the levels and line numbers refer to superlevels and superlines \citep[see][for more details]{Graefener+2002}.

\section{Obtaining the final model}
\label{app:diagnostics}

We started our hydrodynamically consistent PoWR$^\textsc{hd}$ modeling with a model based on the properties of the best-fit CMFGEN model from \cite{Clark+2012}. While their study currently marks the most precise spectral fitting of $\zeta^1$\,Sco, employing their basic parameters in a dynamically consistent model does not automatically yield the desired spectroscopic similarity. This was expected because not only are there  differences between PoWR and CMFGEN, but there is an inherent coupling of stellar and wind properties in PoWR$^\textsc{hd}$, which affects the derived spectrum.

Consequently, we varied the stellar properties aiming to alter the wind parameters and improve the similarity with the observed UV, optical, and IR spectra. The primary focus is to reproduce the main BHGs spectral signatures \citep{vanGenderen+1982}, namely the P-Cygni profiles (full or partial) of H$\beta$, H$\gamma$, H$\delta$, and \ion{He}{I} lines. We further sought to reproduce metal lines in pure absorption as they inform us about the star's more fundamental properties and adequate boundary conditions (e.g., temperature, mass, and photospheric turbulence).

\begin{figure}
\centering
\includegraphics[width=1.0\linewidth]{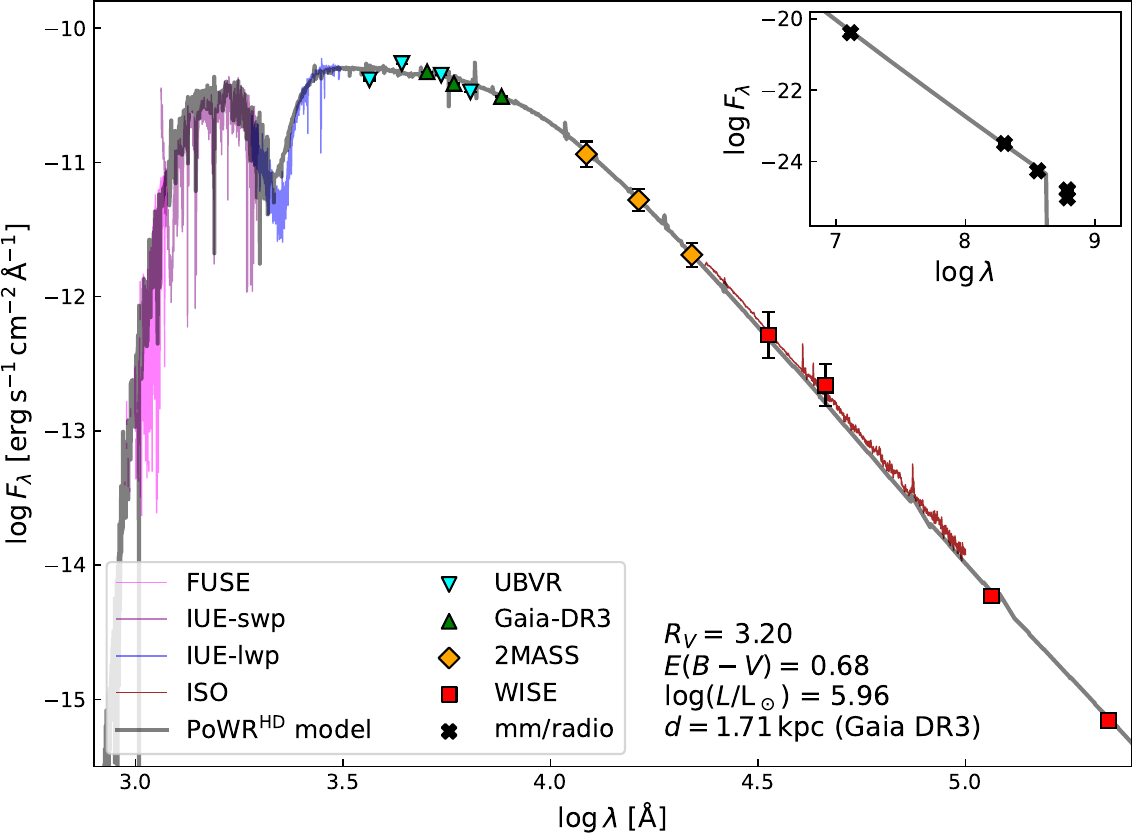}
  \caption{
  Comparison of the hydrodynamically consistent model of $\zeta^1$ Sco (gray line) with the star's SED and flux-calibrated UV and mid-IR spectra. The UV spectra are retrieved from FUSE and IUE (fuchsia, purple, and blue lines). The mid-IR spectrum was obtained with ISO (brown line). The cyan triangles are the flux points of the UBVI photometry. The orange diamonds are from the 2MASS JHK photometry, and the red squares are from WISE mid- and far-IR photometry. The inlet plot shows the millimeter/radio regime; the black crosses are the flux points obtained with the Very Large Array (VLA). The truncation in the synthetic spectrum is an artifact of the code output. See Appendix \ref{app:magnitudes-SED}.
  }
\label{fig:Z1Sco-SED}
\end{figure}

To obtain the luminosity, we aimed to reproduce the SED from the far-UV to the millimeter (or radio) regime. We applied the extinction law by \cite{Cardelli+1989} considering initially the reddening and extinction parameters from \cite{Clark+2012}, namely $E(B-V) = 0.66$ and $R_V = 3.3$. To obtain a slightly better fit, these were adjusted to the slightly higher $E(B-V) = 0.70$ and a ``standard'' $R_V = 3.1$. In Fig.\,\ref{fig:Z1Sco-SED} we show the resulting fit.

The final model nicely reproduces the observed SED across the whole considered range, indicating a well-constrained luminosity. However, there are remaining offsets of about 0.1\,dex in the far-UV regime (IUE-sws) and $\sim$0.2\,dex, in the near-UV regime (IUE-lws). While this could potentially imply an underestimation of the extinction, higher values did not improve the overall SED fit.

\begin{figure}
\centering
\includegraphics[width=1.0\linewidth]{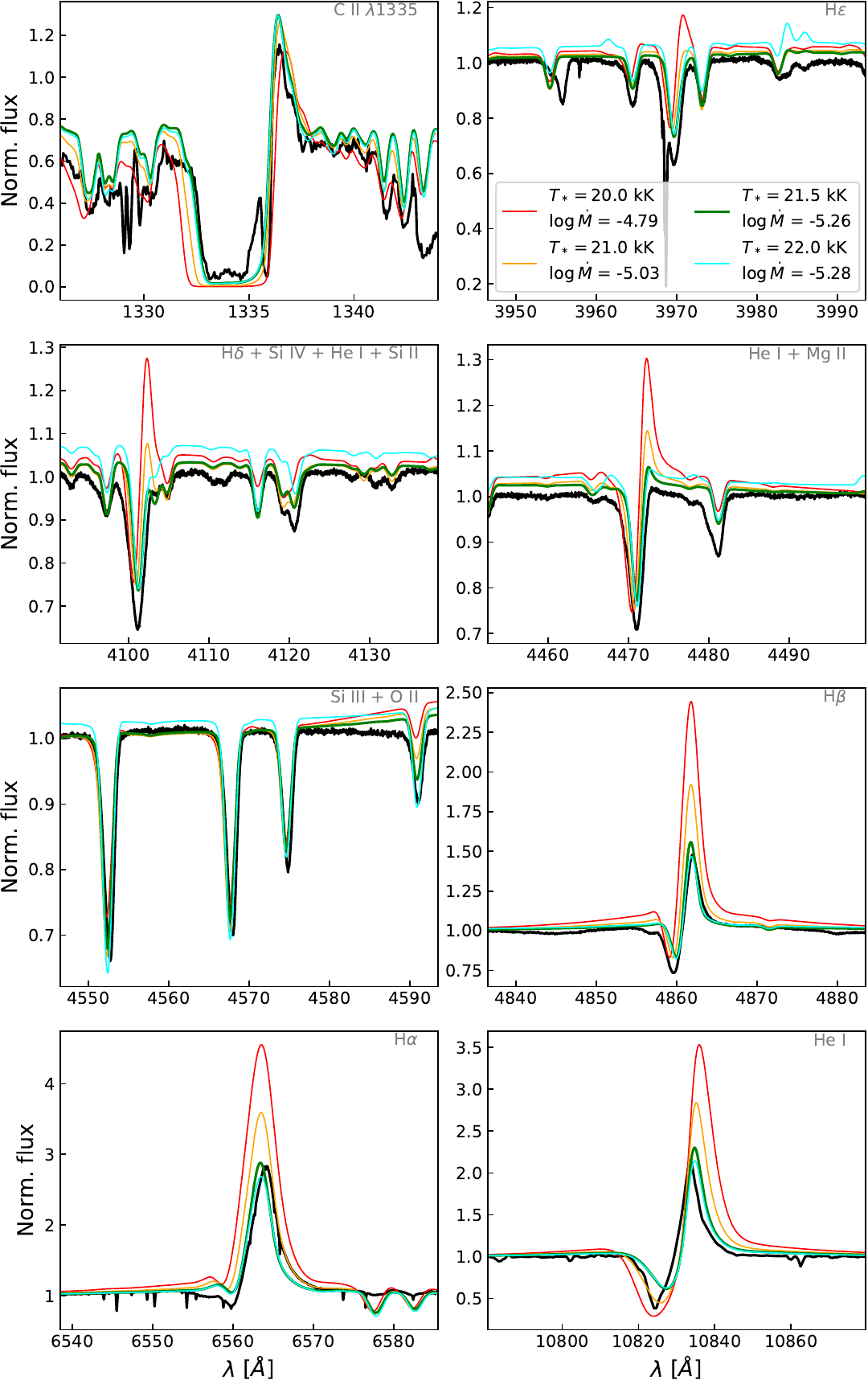    }
  \caption{Resulting variation in the spectral features and wind properties when varying the $T_\ast$ input for PoWR$^\textsc{HD}$ calculations. The observed spectrum is shown in black.\label{fig:Z1Sco-T-seq}}
\end{figure}

To determine the effective temperature $T_\mathrm{eff}$, we sought to find the best compromise between 
\ion{Si}{II}\,$\lambda$4128,32, \ion{Si}{III}\,$\lambda$4550,69,76, \ion{Si}{IV}\,$\lambda$4090,4116, and \ion{Mg}{II}\,$\lambda$4481. Other lines such as \ion{O}{I}\,$\lambda$7772, and \ion{He}{I} lines were used as secondary diagnostics.
We notice that while \ion{Si}{III}~$\lambda$4550, 65, and 75 are well fitted, \ion{Si}{IV}~$\lambda$4090,4116 is overestimated while \ion{Si}{II}~$\lambda$4128,32 is underestimated, hinting to an overestimation of $T_\mathrm{eff}$. We also observe a similar result for the ionization balance of \ion{O}{II} and \ion{O}{I}, where \ion{O}{II} lines (e.g., \ion{O}{II}~$\lambda$4070 and \ion{O}{II}~$\lambda$4590,95) seem slightly overestimated, but overall well reproduced, while \ion{O}{I}~$\lambda$7772 is underestimated. Yet, as illustrated in Fig.\,\ref{fig:Z1Sco-T-seq}, lower choices of $T_\ast$ would have worsened the spectral reproduction due to the consequences on the hydrodynamic wind parameters coupled to the stellar parameters. As depicted, a slight change in $T_\ast$ ($\Delta T_\mathrm{eff} \sim 1$ kK) causes the $\dot{M}$ to increase by a factor of 2. As $\log L$ and $M$ are kept fixed, decreasing(increasing) $T_\ast$ is equivalent to increasing(decreasing) the stellar radius -- via the Stefan-Boltzmann relation --, thereby reducing the surface gravity. 

To fix the surface gravity, we relied on H$\epsilon$ and H$\zeta$ as they were the cleanest Balmer lines without meaningful wind features -- even though models with higher mass-loss rates eventually show wind features in this line as also illustrated in Fig.\,\ref{fig:Z1Sco-T-seq}. In general, we could not obtain models with the wings as broad as the observed spectra, which indicates that the obtained surface gravity, and likely mass, are still too low for this particular star. Due to the inherent coupling between wind and stellar parameters, increasing the mass would spoil the agreement of the main characteristics of $\zeta^1$\,Sco.

\begin{figure}
\centering
\includegraphics[width=1.0\linewidth]{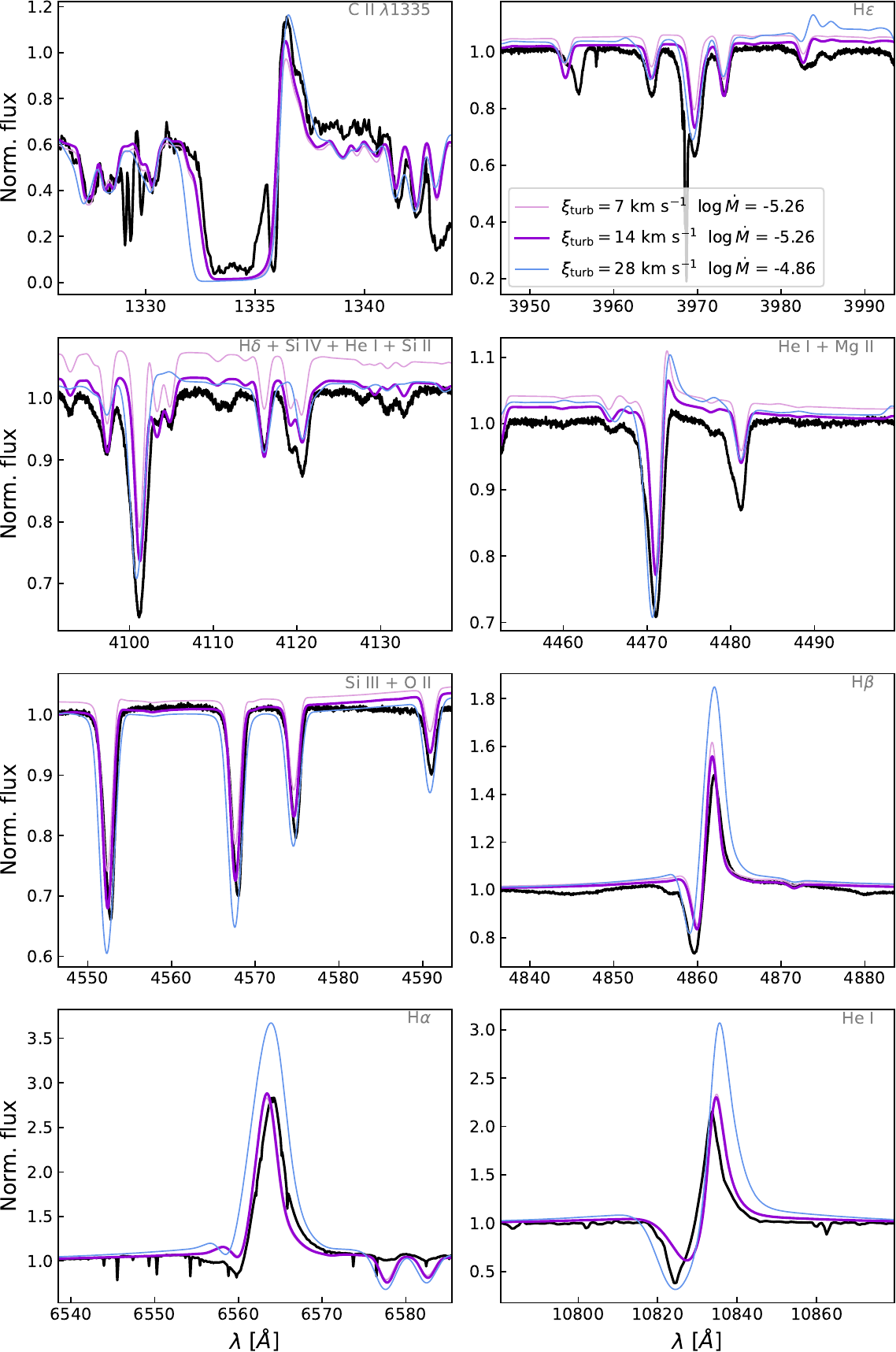    }
  \caption{Resulting variation in the spectral features and wind properties when varying the $\varv_\mathrm{turb}$ input for PoWR$^\textsc{HD}$ calculations. The observed spectrum is shown in black.\label{fig:Z1Sco-Vmic-seq}}
\end{figure}

In Fig.\,\ref{fig:Z1Sco-Vmic-seq} we illustrate the change in the model's properties and spectrum by varying the $\varv_\mathrm{turb}$. Values lower than $\varv_\mathrm{turb} \sim 14$\,km\,s$^{-1}$ yield similar wind properties as output, but weaken the absorption lines, including temperature and gravity diagnostics (e.g., Si and H lines). 
Since in general our final model, whose $\varv_\mathrm{turb} = 14$\,km\,s$^{-1}$, underestimates the absorption features of most of the spectral lines, it is possible that $\varv_\mathrm{turb} \sim 15$\,km\,s$^{-1}$ underestimates the microturbulence in the star. Moreover, the presence of super-Eddington layers and results from multidimensional simulations of hot stars \citep[albeit in the context of O stars;][]{Debnath+2024} point in that direction.

Still, an increase in the $\varv_\mathrm{turb}$ beyond $\sim$25\,km\,s$^{-1}$ would cause meaningful increase in the $\dot{M}$ and $\varv_\infty$, as well as an increase in the absorption features. This value of $\varv_\mathrm{turb}$ is higher than microturbulence values usually used in classical modeling of BSGs \citep[e.g.,][]{Crowther+2006, Haucke+2018}. We also notice that models with high turbulence have lower $T_\mathrm{eff}$ (linked to a higher $\dot{M}$), and regarding many wind lines behave similarly to cooler models, despite having the putative temperature diagnostics \ion{Si}{III} and \ion{Si}{IV} lines stronger. This illustrates some degeneracies that may arise between certain parameters, which also highlights how challenging it is to obtain final spectral fitting based on the stellar properties within the hydrodynamically consistent framework. In that context, having a consistent treatment of the turbulence would lift such a degeneracy, as $\varv_\mathrm{turb}$ would also emerge from the input physics.

\section{Spectral variability}
\label{app:specvar}

\begin{figure*}
\centering
\includegraphics[width=1.0\linewidth]{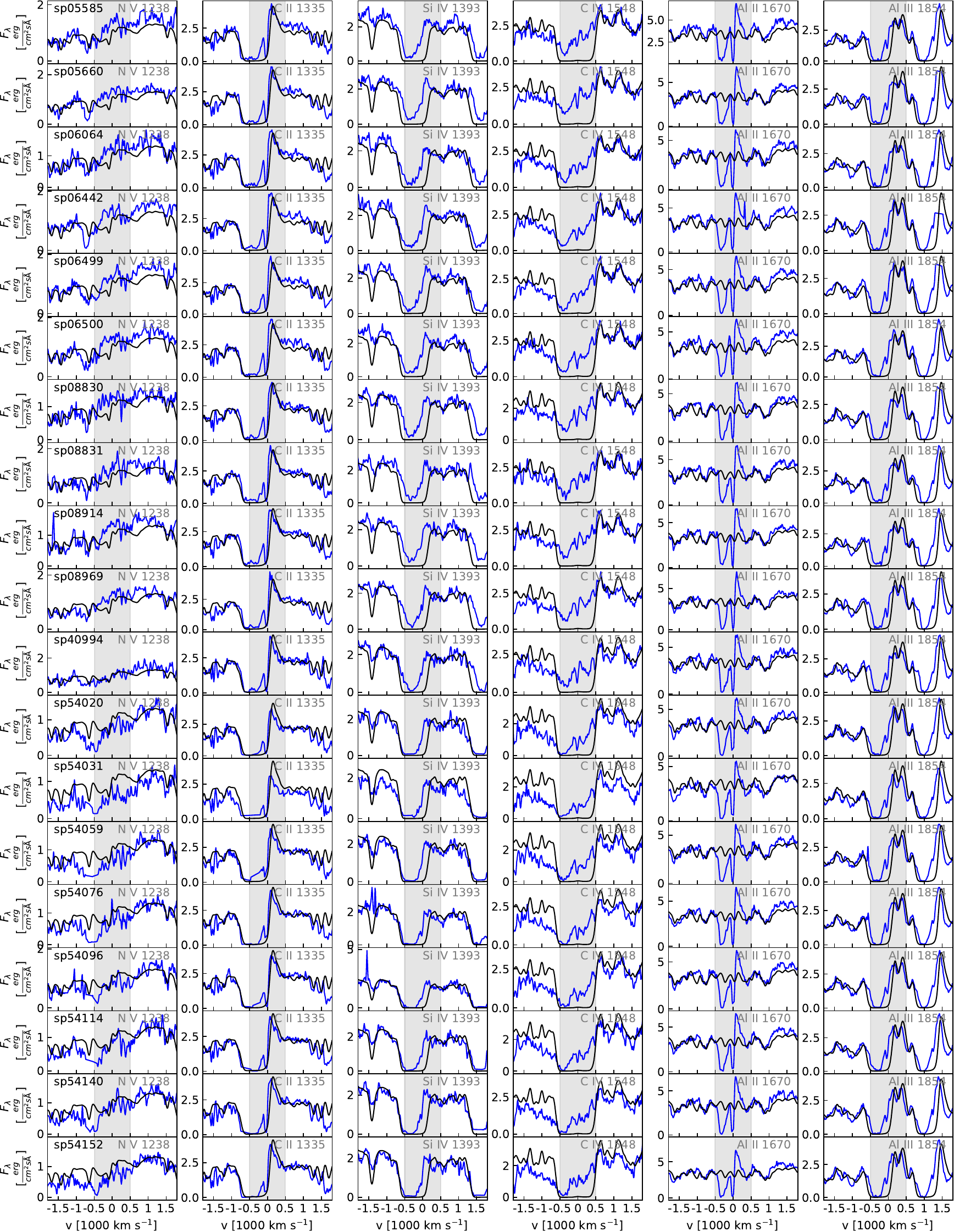    }
  \caption{IUE spectra of $\zeta^1$\,Sco. The blue line is the observation, and the black line is our final PoWR$\textsc{hd}$ model. The shaded area marks the interval of $\pm$500\,km\,s$^{-1}$ around the reference wavelength.
   }    \label{fig:iue-var}
\end{figure*}

\begin{figure*}
\centering
\includegraphics[width=1.0\linewidth]{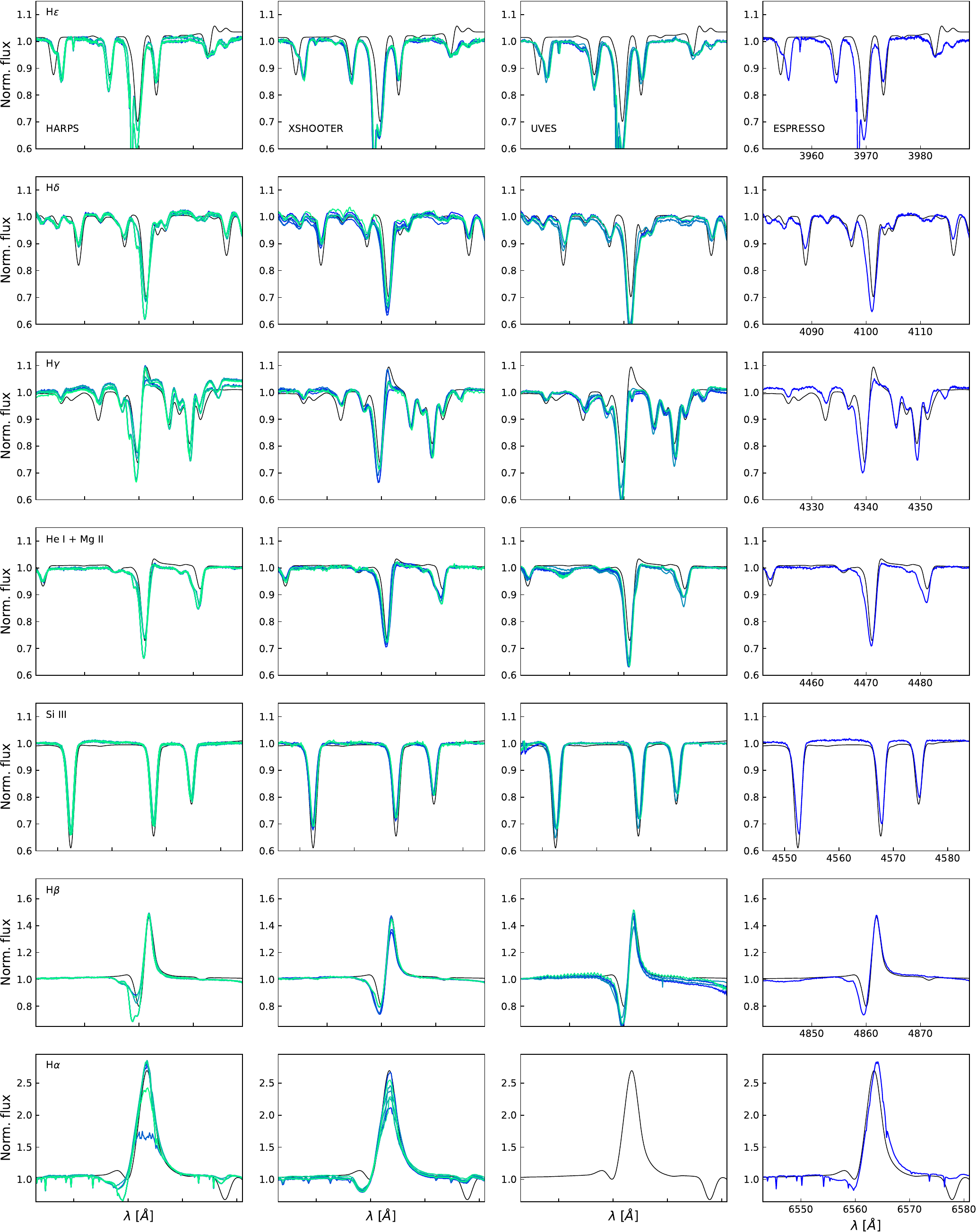    }
  \caption{Variability in the spectral features in the optical of $\zeta^1$\,Sco. Each column refers to a different instrument.
   }    \label{fig:opt-var}
\end{figure*}

\begin{figure}
\centering
\includegraphics[width=1.0\linewidth]{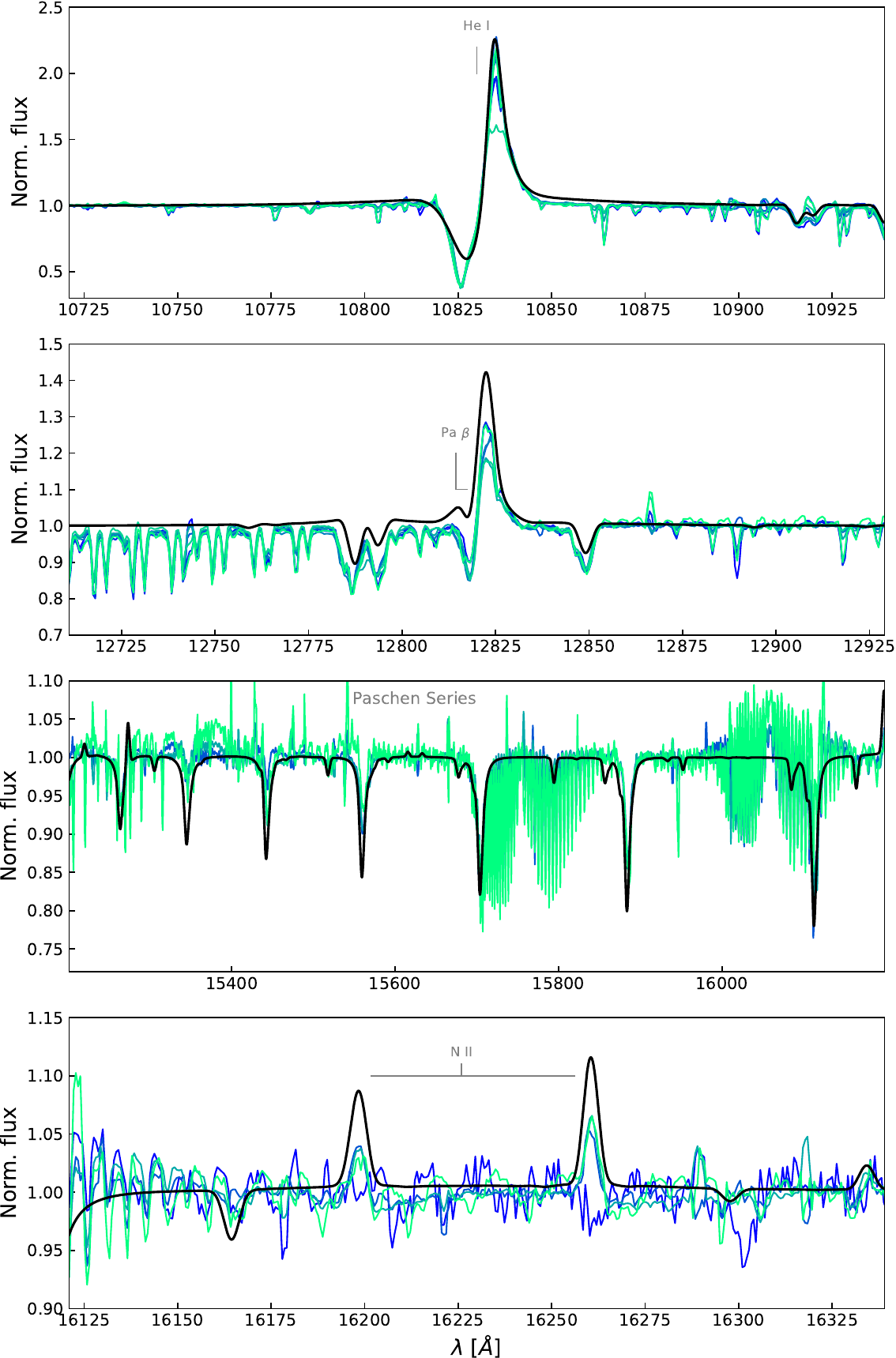    }
  \caption{Comparison of multi-epoch X-Shooter IR observations of $\zeta^1$\,Sco.}
  \label{fig:nir-var}
\end{figure}

A thorough examination of the spectral variability of $\zeta^1$\,Sco, and their causes is out of the scope of this study. Nonetheless, to give a general impression of the observed variability, we present the most relevant diagnostic lines in the UV (Fig.\,\ref{fig:iue-var}), optical (Fig.\,\ref{fig:opt-var}) and IR (Fig.\,\ref{fig:nir-var}). For reference, we also compare with our final PoWR$^\textsc{hd}$ model. More information on the observational dataset can be found in the supplement material at Zenodo\footnote{Downloadable at: \url{https://zenodo.org/records/15050256}.}.

In the UV range, we notice only mild variability in the P-Cygni profiles. \ion{N}{V}\,$\lambda$1240 is the line that presents the most variable behavior.  In the optical, we notice the bulk of the variability happening in the emission component of the P-Cygni lines, especially the Balmer lines. However, we noticed a distinctly strong absorption with a secondary component in the HARPS spectra from the year 2006. The metal lines do not suffer much variation.
In the X-Shooter IR spectra, which in general are much noisier than the optical, we do not observe meaningful variation, except for the emission components of the \ion{He}{I}\,$\lambda$10830 and Paschen-H$\beta$. We also see variations on the \ion{N}{II}\,$\lambda$16164,256, which switches between emission (reproduced by our PoWR$^\textsc{hd}$ model) and absorption.

\section{Magnitudes used in the SED}
\label{app:magnitudes-SED}

\begin{table}
\caption{\label{tab:phot} Magnitudes used to compute the SED.}
\centering
\begin{tabular}{lcr}
\hline\hline
Filter & Mag./\textit{Flux} & Source\\
\hline
U  & 4.526 & \cite{Paunzen+2022} \\
B  & 4.84  & \cite{Zacharias+2005} \\
V  & 4.032 & \cite{Zacharias+2005} \\
R  & 2.997 & \cite{Zacharias+2005} \\
Gb & 2.61 & \cite{GaiaCollab+2023} \\
G  & 2.612 & \cite{GaiaCollab+2023} \\
Gr & 2.157 & \cite{GaiaCollab+2023} \\
J  & 3.592 & \cite{Cutri+2013} \\
H  & 3.341 & \cite{Cutri+2013} \\
K  & 3.217 & \cite{Cutri+2013} \\
W1 & 5.162 & \cite{Cutri+2013} \\
W2 & 4.776 & \cite{Cutri+2013} \\
W3 & 4.53  & \cite{Cutri+2013} \\
W4 & 5.015 & \cite{Cutri+2013} \\
0.13cm & \textit{23.2}\,mJy & \cite{Leitherer-Robert1991} \\
2.0cm & \textit{4.3}\,mJy & \cite{Biening+1989} \\
3.6cm & \textit{2.4}\,mJy & \cite{Benaglia+2007} \\
6.1cm & \textit{2.0}\,mJy & \cite{Biening+1989} \\\hline
\end{tabular}
\tablefoot{
Values in italic in the millimeter/radio regime indicate fluxes in mJy.
}
\end{table}

Table\,\ref{tab:phot} provides an overview of all the photometry used to derive the observed SED for $\zeta^1$\,Sco. For each magnitude, the source is listed as well.

\end{appendix}

\end{document}